%% file: 6dBootstrap.tex
\RequirePackage{pdf14}

\documentclass[letterpaper]{article}
\usepackage[usenames,dvipsnames,table]{xcolor}
\usepackage{graphicx}
\usepackage{caption}
\usepackage{subcaption}
\usepackage{array,setspace,mathrsfs,amsthm,amsfonts,amsmath,yfonts,dsfont,bbm,colonequals,mathtools}
\usepackage{./aux/tikz-cd}
\usepackage{./aux/jheppub}
\usepackage{afterpage}
\usepackage{xspace}
\usepackage[normalem]{ulem}

\input{./aux/topmatter}

\title{The \texorpdfstring{$(2,0)$}{(2,0)} superconformal bootstrap}
\author[1]{Christopher Beem,}
\author[2]{Madalena Lemos,}
\author[2]{Leonardo Rastelli,}
\author[3]{Balt C. van Rees}

\affiliation[1]{Institute for Advanced Study, Einstein Drive, Princeton, NJ 08540, USA}
\affiliation[2]{C.~N.~Yang Institute for Theoretical Physics, Stony Brook University, Stony Brook, NY 11794-3840, USA}
\affiliation[3]{Theory Group, Physics Department, CERN, CH-1211 Geneva 23, Switzerland}

\preprint{CERN-PH-TH-2015-165, YITP-SB-15-25}

\bigskip
\abstract{We develop the conformal bootstrap program for six-dimensional conformal field theories with $(2,0)$ supersymmetry, focusing on the universal four-point function of stress tensor multiplets. We review the solution of the superconformal Ward identities and describe the superconformal block decomposition of this correlator. We apply numerical bootstrap techniques to derive bounds on OPE coefficients and scaling dimensions from the constraints of crossing symmetry and unitarity. We also derive analytic results for the large spin spectrum using the lightcone expansion of the crossing equation. Our principal result is strong evidence that the $A_1$ theory realizes the minimal allowed central charge $(c=25)$ for any interacting $(2,0)$ theory. This implies that the full stress tensor four-point function of the $A_1$ theory is the unique unitary solution to the crossing symmetry equation at $c=25$. For this theory, we estimate the scaling dimensions of the lightest unprotected operators appearing in the stress tensor operator product expansion. We also find rigorous upper bounds for dimensions and OPE coefficients for a general interacting $(2,0)$ theory of central charge $c$. For large $c$, our bounds appear to be saturated by the holographic predictions obtained from eleven-dimensional supergravity.
}


\begin{document}
\setcounter{tocdepth}{2}
\maketitle
\setcounter{page}{1}

\input{sections/1_introduction}
\input{sections/2_overview}
\input{sections/3_correlator}
\input{sections/4_block_decomposition_and_crossing}
\input{sections/5_numerics}
\input{sections/6_results}
\input{sections/acknowledgments}

\appendix

\input{appendices/A_representations}
\input{appendices/B_blocks}
\input{appendices/C_lightcone}

\bibliography{./aux/biblio}
\bibliographystyle{./aux/JHEP}

\end{document}

%% file: aux/topmatter.tex
\newcommand\eea{\end{eqnarray}}
\newcommand\ee{\end{equation}}

\newcommand\mf{\mathfrak}

\newcommand\nn{\nonumber}
\newcommand\ph{\phantom}

\newcommand\eg{{\it e.g.}}
\newcommand\ie{{\it i.e.}}
\newcommand\cf{{\it cf.}}

\newtheorem*{result*}{Result}

\newtheorem*{fact*}{Fact}

\newtheorem*{conj*}{Conjecture}
\newtheorem*{spec*}{Speculation}

\newtheorem*{cor*}{Corollary}

\newcommand\tr{\text{Tr}}

\newcommand\del{\partial}

\newcommand\hf{\frac{1}{2}}

\newcommand\Cb{\mathbb{C}}

\newcommand\Rb{\mathbb{R}}

\renewcommand\AA{\mathcal{A}}

\newcommand\BB{\mathcal{B}}
\newcommand\CC{\mathcal{C}}
\newcommand\DD{\mathcal{D}}

\newcommand\FF{\mathcal{F}}
\newcommand\GG{\mathcal{G}}
\newcommand\HH{\mathcal{H}}
\newcommand\II{\mathcal{I}}

\newcommand\LL{\mathcal{L}}
\newcommand\MM{\mathcal{M}}
\newcommand\NN{\mathcal{N}}
\newcommand\OO{\mathcal{O}}

\newcommand\QQ{\mathcal{Q}}
\newcommand\RR{\mathcal{R}}

\newcommand\WW{\mathcal{W}}
\newcommand\XX{\mathcal{X}}

\newcommand\zb{{\bar z}}

\newcommand\qq{\mathbbmtt{Q}}

\newcommand\sof{{\mf{so}}}
\newcommand\suf{{\mf{su}}}
\newcommand\ospf{{\mf{osp}}}
\newcommand\uspf{{\mf{usp}}}
\newcommand\slf{{\mf{sl}}}

\newcommand\gf{{\mf{g}}}

\newcommand{\vev}[1]{\ensuremath{\langle #1 \rangle}\xspace}
\global \long \def \a{\alpha}
\global \long \def \b{\beta}
\global\long\global\long\def\D{\Delta}

\newcommand\restr[2]{{
  \left.\kern-\nulldelimiterspace 
  #1 
  \vphantom{\big|} 
  \right|_{#2} 
  }}
\newcommand\fverb{\setbox\fverbbox=\hbox\bgroup\verb}
\newcommand\fverbdo{\egroup\medskip\noindent%
			\fbox{\unhbox\fverbbox}\ }
\newcommand\fverbit{\egroup\item[\fbox{\unhbox\fverbbox}]}
\newbox\fverbbox

%% file: sections/1_introduction.tex

\section{Introduction and summary}
\label{sec:intro}

In this work we introduce and develop the modern conformal bootstrap program for $(2,0)$ superconformal theories in six dimensions. These theories provide a powerful organizing principle for lower-dimensional supersymmetric dynamics. From their existence one can infer a vast landscape of supersymmetric field theories in various dimensions and rationalize many deep, nonperturbative dualities that act within this landscape (see, \eg, \cite{Gaiotto:2009we, Gaiotto:2009hg, Dimofte:2011ju, Bah:2012dg, Gadde:2013sca}). Despite their increasingly central role, the $(2,0)$ theories have proved stubbornly resistant to study by traditional quantum field theory techniques. This situation, coupled with the high degree of symmetry and conjectured uniqueness of these theories, suggests that the $(2,0)$ theories may be a prime target for the conformal bootstrap approach.

\subsection*{\texorpdfstring{$(2,0)$}{(2,0)} theories}

It has been known since the work of Nahm \cite{Nahm:1977tg} that superconformal algebras can only be defined for spacetime dimension less than or equal to six. In six dimensions, the possible superconformal algebras are the $(\NN,0)$ algebras $\ospf(8^\star|2\NN)$. The maximum six-dimensional superconformal algebra for which there can exist a stress tensor multiplet is the $(2,0)$ algebra $\ospf(8^*|4)$ \cite{Cordova_banff}. Thus the six-dimensional $(2,0)$ theories are singled out as the maximally supersymmetric local CFTs in the maximum number of dimensions. 

While it is easy to identify a free field theory that realizes $(2,0)$ superconformal symmetry -- namely the abelian tensor model -- the existence of \emph{interacting} $(2,0)$ theories was only inferred in the mid 1990s on the basis of string theory constructions \cite{Witten:1995zh,Strominger:1995ac}. A decoupling limit of Type IIB string theory on ALE spaces predicts the existence of a list of interacting $(2,0)$ theories labelled by the simple and simply-laced Lie algebras $(A_{n \geqslant1}, D_{n \geqslant 4}, E_6, E_7, E_8)$ \cite{Witten:1995zh}. The $A_{N-1}$ model can also be realized as the low-energy limit of the worldvolume theory of $N$ coincident M5 branes in M-theory \cite{Strominger:1995ac}. Two important properties of these theories can be deduced almost immediately from their string/M-theory constructions:
\begin{itemize}
\item[(i)] On $\Rb^6$, the $(2,0)$ theory of type $\gf$ has a moduli space of vacua given by the orbifold
\begin{equation} \label{Mg}
{\MM}_{\gf} \colonequals (\Rb^5)^{r_{\gf}} / W_{\gf}~,
\end{equation}
where $r_{\gf}$ is the rank of the simply laced Lie algebra ${\gf}$ and $W_{\gf}$ its Weyl group. At a generic point in the moduli space, the long distance physics is described by a collection of $r_{\gf}$ decoupled free tensor multiplets.
\item[(ii)] Upon supersymmetric compactification on $\Rb^5 \times S^1$, the $(2,0)$ theory of type $\gf$ reduces at a low energies to a maximally supersymmetric five-dimensional Yang-Mills theory with gauge algebra $\gf$, whose dimensionful gauge coupling is controlled by the $S^1$ radius, $g_{\rm YM}^2 \sim R$.
\end{itemize}
The string/M-theory constructions give physically compelling evidence for the existence of the interacting $(2,0)$ theories, but the evidence is indirect and requires an extensive conceptual superstructure. Moreover, at present these constructions do not provide tools for computing anything in the conformal phase of the theory beyond a very limited set of robust observables, such as anomalies.\footnote{
The 't Hooft anomalies for the $R$-symmetry, as well as the gravitational and mixed anomalies, are encoded in an eight-form anomaly polynomial. First obtained for the $A_n$ and $D_n$ theories by anomaly inflow arguments in M-theory \cite{Duff:1995wd,Witten:1996hc,Freed:1998tg,Harvey:1998bx,Yi:2001bz}, the anomaly polynomial can also be reproduced by purely field-theoretic reasoning \cite{Intriligator:2000eq,Ohmori:2014kda}, relying only on properties (i) and (ii). A similar field-theoretic derivation has recently been performed for the $a$-type Weyl anomaly \cite{Cordova:2015vwa}, \ie, the coefficient in front of the Euler density in the trace anomaly. Finally, the $c$-type anomalies, \ie, the coefficients in front of the the three independent Weyl invariants in the trace anomaly, are proportional to each other and related by supersymmetry to the two-point function of the canonically normalized stress tensor, see \eqref{c(g)} below.}

At large $n$, the $A_{n}$ theories can be described holographically in terms of eleven-dimensional supergravity on ${\rm AdS}_7 \times S^4$ \cite{Maldacena:1997re}.\footnote{There is an analogous conjecture for the $D_n$ theories, relating them to supergravity on ${\rm AdS}_7 \times \mathbb{RP}_4$ \cite{Witten:1998xy,Aharony:1998rm}.} The AdS/CFT correspondence then renders these large $n$ theories extremely tractable. However, the extension to finite $n$ is presently only possible at leading order. Higher order corrections require a method for computing quantum corrections in M-theory. An intrinsic \emph{field theoretic} construction of the $(2,0)$ theories would therefore be indispensable. Turning this logic around, such an independent definition would offer a window into quantum M-theory.

However, most standard field theory methods are inadequate for describing the $(2,0)$ theories. The mere existence of unitary, interacting quantum field theories in $d > 4$ appears surprising from effective field theory reasoning -- conventional local Lagrangians are ruled out by power counting. The $(2,0)$ theories are isolated, intrinsically quantum mechanical conformal field theories, which cannot be reached as infrared fixed points of local RG flows starting from a gaussian fixed point. This is in sharp contrast to more familiar examples of isolated CFTs in lower dimensions, such as the critical Ising model in three dimensions. It is unclear whether an unconventional (non-local?) Lagrangian for the interacting $(2,0)$ theories could be written down (see, \eg, \cite{Lambert:2010wm,Ho:2011ni,Chu:2012um,Bonetti:2012st,Samtleben:2012fb,Saemann:2013pca} for a partial list of attempts), but in any event it would be unlikely to lend itself to a semiclassical approximation. 

There have been several attempts at field-theoretic definitions of the $(2,0)$ theories. The most concrete proposal \cite{Aharony:1997th,Aharony:1997an} relates the Discrete Light Cone Quantization (DLCQ) of the $A_n$ theory (with $k$ units of lightcone momentum) to supersymmetric quantum mechanics on the moduli space of $k$ $SU(n+1)$ instantons -- to recover the $(2,0)$ theory on $\Rb^6$ one is instructed to send $k \to \infty$. This moduli space is singular due to small instanton singularities. A specific regularization procedure has been advocated in \cite{Aharony:1997an}, but its conceptual status seems somewhat unclear, and one may be concerned that important features of the theory might be hidden in the details of the UV regulator. The DLCQ proposal remains largely unexplored due to the inherent difficulty of performing calculations in a strongly coupled quantum mechanical model, and the need to take the large $k$ limit. To the best of our knowledge, the only explicit result obtained in this framework is the calculation of the half BPS spectrum of the $A_n$ theory, carried out in \cite{Aharony:1997an}. According to another little-explored proposal, the $A_n$ $(2,0)$ theory on $\Rb^4 \times T^2$ (with a finite-size $T^2$) can be ``deconstructed'' in terms of four-dimensional ${\NN} = 2$ quiver gauge theories \cite{ArkaniHamed:2001ie}. Finally there is the suggestion \cite{Douglas:2010iu,Lambert:2010iw} that five-dimensional maximally supersymmetric Yang-Mills theory is a consistent quantum field theory at the nonperturbative level, without additional UV degrees of freedom, and that it gives a complete definition of the $(2,0)$ theory on $S^1$. The conceptual similarities shared by these three proposals have been emphasized in \cite{Lambert:2012qy}. It would be of great interest to develop them further, ideally to the point where quantitative information for the non-protected operator spectrum could be derived and compared to the bootstrap results obtained here.

\subsection*{Bootstrap approach} 

In the present work, we will eschew the problem of identifying ``the fundamental degrees of freedom'' of the $(2,0)$ theories. Indeed, we are not certain that this question makes sense. Instead, we will do our best to rely exclusively on symmetry. We regard the $(2,0)$ theories as abstract conformal field theories (CFTs), to be constrained -- ideally, completely determined -- by bootstrap methods. The conformal bootstrap program was formulated in the pioneering papers \cite{Ferrara:1973yt,Ferrara:1973vz,Polyakov:1974gs} and has undergone a modern renaissance starting with \cite{Rattazzi:2008pe}. The algebra of local operators, which is defined by the spectrum of local operators and their operator product expansion (OPE) coefficients, is taken as the primary object. The conformal bootstrap aims to fix these data by relying exclusively on symmetries and general consistency requirements, such as associativity and unitarity.%
\footnote{
In principle, the bootstrap for local operators could be enlarged by also including non-local operators (such as defects of various codimensions), or by considering non-local observables (such as partition functions in non-trivial geometries), or both. Apart from the calculation of new interesting observables, such an enlarged framework may yield additional constraints on the local operator algebra itself. This is familiar in two-dimensional CFT, where modular invariance imposes additional strong restrictions on the operator spectrum. Constraints arising from non-trivial geometries are however much less transparent in higher-dimensional CFTs, and their incorporation in the bootstrap program remains an open problem. Refs \cite{Liendo:2012hy,Gaiotto:2013nva,Gliozzi:2015qsa} contain some work on the bootstrap in the presence of defects.}
Intuitively, we expect this approach to be especially powerful for theories that are uniquely determined by their symmetries and perhaps a small amount of additional data, such as central charges. The $(2,0)$ theories, with their conjectured ADE classification and absence of exactly marginal deformations, are an attractive target.

We are making the fundamental assumption that the $(2,0)$ theories admit a local operator algebra satisfying the usual properties. This may warrant some discussion, in light of the fact that the $(2,0)$ theories require a slight generalization of the usual axioms of quantum field theory \cite{Witten:1996hc,Aharony:1998qu,Witten:1998wy,Moore:2004jv,Freed:2006yc,Witten:2009at,Henningson:2010rc,Freed:2012bs}. In contrast to a standard QFT, each $(2,0)$ theory does not have a well-defined partition function on a manifold of non-trivial topology (with non-trivial three-cycles), but it yields instead a {\it vector} worth of partition functions. This subtlety does not affect the correlation functions of local operators on $\Rb^6$ where there are no interesting three-cycles to be found. That the $(2,0)$ theories of type $A_n$ and $D_n$ have a conventional local operator algebra is manifest from AdS/CFT, at least for large $n$. The extrapolation of this property to finite $n$ is a very plausible conjecture. Ultimately, in the absence of an alternative calculable definition of the $(2,0)$ theories, we are going to take the existence of a local operator algebra as axiomatic. Our work will give new compelling evidence that this is a consistent hypothesis.

In this paper, we focus on the crossing symmetry constraints that arise from the four-point function of stress tensor multiplets. This is a natural starting point for the bootstrap program, since the stress tensor is the one non-trivial operator that we know for certain must exist in a $(2,0)$ theory. By superconformal representation theory, the stress tensor belongs to a half BPS multiplet, whose superconformal primary (highest weight state) is a dimension four scalar operator $\Phi$ in the two-index symmetric traceless representation of $\sof(5)_R$. It is equivalent but technically simpler to focus on the four-point function of $\Phi$, which contains the same information as the stress tensor four-point function. (Indeed, the two-, three- and four-point functions of all half BPS multiplets are known to admit a unique structure in superspace \cite{Eden:2001wg,Arutyunov:2002ff,Dolan:2004mu}.)

The two- and three-point functions of the stress tensor supermultiplet are uniquely determined in terms of a single parameter, the central charge $c$. Normalizing $c$ to be one for the free tensor multiplet, the central charge of the $(2,0)$ theory of type $\gf$ is given by \cite{Beem:2014kka}\footnote{This is the unique expression compatible with the structure of the eight-form anomaly polynomial and with the explicit knowledge of $c$ for the $A_n$ theories at large $n$, which is available from a holographic calculation \cite{Henningson:1998gx,Henningson:1998ey}.}
\begin{equation}
\label{c(g)}
c(\gf) = 4d_{\gf} h^\vee_{\gf} + r_{\gf}~, 
\end{equation}
where $d_{\gf}$, $h^\vee_{\gf}$ and $r_{\gf}$ are the dimension, dual Coxeter number, and rank of $\gf$, respectively. For example, this gives $c(A_{N-1}) = 4 N^3 - 3N -1$, which exhibits the famous $O(N^3)$ growth of degrees of freedom. Equation \eqref{c(g)} gives the value of the central charge for all the \emph{known} $(2,0)$ theories, but we do not wish to make any such \emph{a priori} assumptions about the theories we are studying. We will therefore treat the central charge as an arbitrary parameter, imposing only the unitarity requirement that it is real and positive.
 
Building on previous work \cite{Ferrara:2001uj,Eden:2001wg,Arutyunov:2002ff,Dolan:2004mu,Heslop:2004du}, we impose the constraints of superconformal invariance on the four-point function of $\Phi$ and decompose it in a double OPE expansion into an infinite sum of superconformal blocks. Schematically we have
\begin{equation}
\label{schematicexp}
\langle\Phi\,\Phi\,\Phi\,\Phi\rangle = \sum_{\OO} f^2_{\Phi \Phi {\OO} } \; G^\Phi_{{\OO}}~.
\end{equation}
The sum is over all the superconformal multiplets allowed by selection rules to appear in the $\Phi \times\Phi$ OPE. Each multiplet is labelled by the corresponding superconformal primary ${\OO}$, with associated superconformal block $G^\Phi_{{\OO} }$, a known function of the two independent conformal cross ratios. Finally $f_{\Phi \Phi {\OO} } \in \Rb$ denotes the OPE coefficient. We further assume that no conserved currents of spin $\ell >2$ 
appear in this expansion. Very generally, the presence of higher-spin conserved currents in a CFT implies that the theory contains a free decoupled subsector \cite{Maldacena:2011jn}, while we wish to focus on interacting $(2,0)$ theories. There are three classes of supermultiplets that contribute in the double OPE expansion \eqref{schematicexp}:
\begin{itemize}
\item[(i)] An infinite set $\{{\OO}_\chi\}$ of BPS multiplets, whose quantum numbers are known from shortening conditions and whose OPE coefficients $f_{\Phi \Phi {\OO}_\chi}$ can be determined in closed form using crossing symmetry, as functions of the central charge $c$. For this particular four-point function, this follows from elementary algebraic manipulations, but there is a deeper structure underlying this analytic result. The operator algebra of any $(2,0)$ theory admits a closed subalgebra, isomorphic to a two-dimensional chiral algebra \cite{Beem:2013sza,Beem:2014kka}. Certain protected contributions to four-point functions of BPS operators of the $(2,0)$ theory are entirely captured by this chiral algebra. In the case at hand, the operator $\Phi$ corresponds to the holomorphic stress tensor $T$ of the chiral algebra, with central charge $c_{2d} = c$, while the operators $\{ {\OO}_\chi \}$ map to products of derivatives of $T$. 
\item[(ii)] Another infinite tower of BPS multiplets $\{ \DD, \BB_1, \BB_3\, , \BB_5 \, , \dots \}$.
The $\BB_\ell$ operators have spin $\ell$, while the single $\DD$ operator is a scalar.\footnote{Here we use an abbreviated notation for the supermultiplets, dropping their $R$-charge quantum numbers. The translation to the precise notation introduced in section 2 is as follows: 
 $\DD \colonequals \DD[0,4]$, $\BB_\ell \colonequals \BB[0, 2]_\ell$, $\LL_{\Delta, \ell} \colonequals \LL[0, 0]_{\Delta, \ell}$.
 }
All their quantum numbers (including the conformal dimension) are known from shortening conditions, but their contribution to the four-point function is {\it not} captured by the chiral algebra.
\item[(iii)] An infinite set of non-BPS operators $\{ \LL_{\Delta, \ell} \}$, labelled by the conformal dimension $\Delta$ and spin $\ell$. Only even $\ell$ contribute because of Bose symmetry. Both the set of $\{ (\Delta, \ell) \}$ entering the sum and the OPE coefficients are {\it a priori} unknown. Unitarity gives the bound $\Delta \geqslant \ell + 6$. It turns out that 
\begin{eqnarray}
\lim_{\Delta \to \ell + 6} G^\Phi_{\LL_{\Delta, \ell}} & = & G^\Phi_{\BB_{\ell-1} }~, \qquad  \; \ell = 2, 4, \dots~,\\
\lim_{\Delta \to 6} G^\Phi_{\LL_{\Delta, 0}} & = & G^\Phi_{\DD}~, \nonumber
\end{eqnarray}
so we can view the BPS contributions of type (ii) as a limiting case of the non-BPS contributions.
\end{itemize}
We then analyze the constraints of crossing symmetry and unitarity on the unknown CFT data $\{ (\Delta, \ell), f_{\Phi \Phi \LL_{\Delta, \ell} },f_{\Phi\Phi\DD},f_{\Phi\Phi\BB_\ell}\}$. Some partial analytic results can be derived by taking the Lorentzian lightcone limit of the four-point function. As shown in \cite{Fitzpatrick:2012yx,Komargodski:2012ek}, crossing symmetry relates the leading singularity in one channel with the large spin asymptotics in the crossed channel. By this route we demonstrate that the BPS operators $\BB_\ell$ are necessarily present -- at least for large $\ell$ -- and compute the asymptotic behavior of the OPE coefficients $f_{\Phi \Phi \BB_{ \ell }}$ as $\ell \to \infty$. To proceed further we must resort to numerics. There is by now a standard suite of numerical techniques to derive rigorous inequalities in the space of CFT data, following the blueprint of \cite{Rattazzi:2008pe}. The numerical algorithm requires us to choose a finite-dimensional space of linear functionals that act on functions of the conformal cross ratios. We parametrize the space of functionals by an integer $\Lambda$. Greater $\Lambda$ corresponds to a bigger space of functionals, and hence more stringent bounds.

\subsection*{Summary of results}

The most fundamental bound is for the central charge $c$ itself. We derive a rigorous lower bound $c > 21.45$. This is the bound for the maximal value of $\Lambda$ allowed by our numerical resources. However, it turns out that the lower bound on $c$ has a very regular dependence on the cut-off $\Lambda$, see Fig. \ref{Fig:cbound}. This leads to the compelling conjecture that it converges exactly to $c_{\rm min} = 25$ as $\Lambda \to \infty$. This is precisely the central charge for the $A_1$ theory, the smallest central charge of all the known interacting $(2,0)$ theories.\footnote{
Recall that $c=1$ for the free tensor theory, but we are excluding free theories in our ansatz by not allowing for higher spin conserved currents in the operator algebra.} 

This result rules out the existence of exotic $(2,0)$ theories with a central charge smaller than that of the $A_1$ theory. A much stronger conclusion follows if one accepts the standard bootstrap wisdom \cite{Poland:2010wg,ElShowk:2012hu} that the crossing equation has a {\it unique} unitarity solution whenever a bound is saturated. We are then making the precise mathematical conjecture that for $c=25$ the CFT data contained in \eqref{schematicexp} are completely determined by the bootstrap equation. We test this conjecture by extrapolating various observables to $\Lambda \to \infty$ using different schemes, always finding a consistent picture. As an example, we determine numerically the dimension $\Delta_0$ of the leading-twist scalar non-BPS operator, $6.387 < \Delta_0 < 6.443$, see Fig. \ref{Fig:extrapol_l0}. (The range of values reflects our estimate of the uncertainty in the $\Lambda \to \infty$ extrapolation.) One is tempted to further speculate that all other crossing equations will also have unique solutions, \ie, that \emph{the $A_1$ theory can be completely bootstrapped}.

An important check of our claim that for $c = c_{\rm min} \to 25$ we are bootstrapping the $A_1$ theory follows from examining the $c$ dependence of the OPE coefficient $f_{\Phi \Phi \DD}$. We find that it is zero for $c = c_{\rm min}$. In fact, it is precisely the vanishing of this OPE coefficient that is responsible for the existence of a lower bound on $c$. For $c < c_{\rm min}$, the squared OPE coefficient becomes negative, violating unitarity. Now the $\DD$ operator is a chiral ring operator, and the chiral spectrum of the $(2,0)$ theory of type ${\gf}$ has been computed \cite{Bhattacharyya:2007sa} assuming \eqref{Mg} and a standard folk theorem relating chiral operators to holomorphic functions on the moduli space. This analysis reveals that $\DD$ is \emph{absent} in the $A_1$ theory, in nice agreement with the vanishing of $f_{\Phi \Phi \DD}$ for $c = c_{\rm min}$. 

We also derive bounds on operator dimensions and OPE coefficients in the entire range $c \in [c_{\rm min}, \infty)$. For $c \to \infty$, our bounds must be compatible with the existence of a known unitary solution of the crossing equation, given by the holographic four-point correlator of $\Phi$ evaluated from ${\rm AdS}_7 \times S^4$ supergravity. For strictly infinite central charge, this is simply the ``generalized free field theory'' answer, for which the four-point function factorizes into products of two-point functions. The supergravity answer gives a non-trivial $1/c$ correction to the disconnected result. We find compelling evidence that the holographic answer -- including the $1/c$ correction -- saturates the best possible numerical bounds of this type. The same phenomenon has been observed for ${\NN}=4$ SCFTs in four dimensions \cite{Beem:2013qxa}.

In summary, we find strong evidence that both for small and for large central charge the bootstrap bounds are saturated by actual SCFTs -- the $A_1$ theory for $c = c_{\rm min} \to 25$ and the $A_{n \to \infty}$ theory for $c \to \infty$. It is natural to conjecture that all $A_n$ theories saturate the bounds at the appropriate value \eqref{c(g)} of the central charge. The bounds depend smoothly on $c$, and when they are saturated one expects to find a unique unitary solution of this particular crossing equation. Presumably, only the discrete values \eqref{c(g)} of the central charge will turn out to be compatible with the remaining, infinite set of bootstrap equations. 

\subsection*{Outlook}

The bootstrap results derived from the $\Phi$ four-point function are completely universal. The only input is the existence of $\Phi$ itself, which is tantamount to the existence of a stress tensor. An obvious direction for future work is to make additional spectral assumptions, leveraging what is conjecturally known about the $(2,0)$ theories. A relevant additional piece of data is the half BPS spectrum, which is easily deduced from \eqref{Mg}. In the $(2,0)$ theory of type $\gf$, the half BPS ring is generated by $r_{\gf}$ operators. These generators are in one-to-one correspondence with the Casimir invariants of $\gf$ and have conformal dimension $\Delta = 2 k$, where $k$ is the order of the Casimir invariant. The operator $\Phi$ corresponds to the quadratic Casimir, it is always the lowest-dimensional generator and is in fact the unique generator for the $A_1$ theory. 

The natural next step in the $(2,0)$ bootstrap program is then to consider four-point correlators of higher-dimensional half BPS operators -- both individual correlators and systems of multiple correlators.\footnote{The study of multiple correlators has proved extremely fruitful in the bootstrap of $3d$ CFTs. For example, they have led the world's most precise determination of critical exponents for the $3d$ critical Ising model, with rigorous error bars \cite{Kos:2014bka,Simmons-Duffin:2015qma}.} The protected chiral algebra associated to the $(2,0)$ theory of type $\gf$ has been identified with the $\WW_{\gf}$ algebra \cite{Beem:2014kka} and will be an essential tool in the analysis of these more general correlators. The chiral algebra controls an infinite amount of CFT data, which would be very difficult to obtain otherwise. For the $\Phi$ four-point function only the universal subalgebra generated by the holomorphic stress tensor is needed, but more complicated correlators make use of the very non-trivial structure constants of the $\WW_{\gf}$ algebra. We have seen that there is a sense in which the $A_1$ theory is uniquely cornered by the vanishing of the OPE coefficient $f_{\Phi \Phi \DD}$, which reflects the chiral ring relation that sets to zero the quarter BPS operator $\DD$. The higher-rank theories admit analogous chiral ring relations, which imply certain relations amongst the OPE coefficients appearing in a suitable system of multiple correlators. At least in principle, this gives a strategy to bootstrap the general $(2,0)$ theory of type~$\gf$.
 
\medskip

The remainder of this paper is organized as follows. In section \ref{sec:strategy} we provide some useful background on the six-dimensional $(2,0)$ theories and discuss how to formulate the corresponding bootstrap program in full generality. Sections \ref{sec:structure_of_correlator}, \ref{Sec:blockdecomp}, and \ref{sec:numerics} contain the nuts and bolts of the bootstrap setup considered in this paper: they contain, respectively, the detailed structure of the $\langle\Phi\,\Phi\,\Phi\,\Phi\rangle$ correlator, its superconformal block decomposition, and a review of the numerical approach to the bootstrap. The results of our numerical analysis are then presented in section \ref{sec:results}, and supplementary material can be found in the appendices. Casual readers may limit themselves to section \ref{sec:strategy} and the discussion surrounding Figs.~\ref{Fig:cbound}, \ref{Fig:OPE_D_bound}, and \ref{Fig:extrapol_l0} in section \ref{sec:results}.

%% file: sections/2_overview.tex

\section{The bootstrap program for \texorpdfstring{$(2,0)$}{(2,0)} theories}
\label{sec:strategy}

A great virtue of the bootstrap approach to conformal field theory is its generality. Indeed, this will be the reason that we can make progress in studying the conformal phase of $(2,0)$ SCFTs despite the absence of a conventional definition. Thus in broad terms this work will mirror many recent bootstrap studies \cite{Rychkov:2009ij,Vichi:2009zz,Caracciolo:2009bx,Poland:2010wg,Rattazzi:2010gj,Rattazzi:2010yc,Vichi:2011ux,Poland:2011ey,ElShowk:2012ht,Liendo:2012hy,ElShowk:2012hu,Beem:2013qxa,Gliozzi:2013ysa,Kos:2013tga,El-Showk:2013nia,Alday:2013opa,Gaiotto:2013nva,Berkooz:2014yda,El-Showk:2014dwa,Gliozzi:2014jsa,Nakayama:2014lva,Nakayama:2014yia,Alday:2014qfa,Chester:2014fya,Chester:2014mea,Kos:2014bka,Caracciolo:2014cxa,Paulos:2014vya,Bae:2014hia,Simmons-Duffin:2015qma,Gliozzi:2015qsa,Bobev:2015vsa,Bobev:2015jxa,Kos:2015mba}. We will not review the basic philosophy in any detail here. Instead, the purpose of this section is to describe in fairly general terms how $(2,0)$ supersymmetry affects the bootstrap problem, and also to review some aspects of the known $(2,0)$ theories that are relevant for this chapter of the bootstrap program. In subsequent sections we will provide a more detailed account of the specific crossing symmetry problem we are studying, culminating in a bootstrap equation that can be fruitfully analyzed.

\subsection{Local operators}
\label{subsec:local_op_review}

The basic objects in the bootstrap approach to CFT are the local operators, which are organized into representations of the conformal algebra. The local operators in a unitary $(2,0)$ SCFT must further organize into unitary representations of the $\ospf(8^\star|4)$ superconformal algebra. A unitary representation of $\ospf(8^\star|4)$ is a highest weight representation and is completely determined by the transformations of its highest weight state (the superconformal primary state) under a maximal abelian subalgebra. For generators of the maximal abelian subalgebra we take the generators $H_{1,2,3}$ of rotations in three orthogonal planes in $\Rb^6$, generators $R_1$ and $R_2$ of a Cartan subalgebra of $\sof(5)_R$, and the dilatation generator $D$. We define the quantum numbers of a state with respect to these generators as follows,\footnote{See Appendix \ref{App:representations} for our naming conventions and more details about these representations.}
\begin{eqnarray}
\label{eq:eigenvalue_conventions}
H_i|\psi\rangle&=&h_i|\psi\rangle~,\nn\\
R_i|\psi\rangle&=&d_i|\psi\rangle~,\\
D|\psi\rangle  &=&\Delta|\psi\rangle~.\nn
\end{eqnarray}
There are five families of unitary representations, each admitting various special cases. These families are characterized by linear relations obeyed by the quantum numbers of the superconformal primary state \cite{Dobrev:2002dt,Minwalla:1997ka}:
\begin{alignat}{3}
&\LL~&:&~ \Delta>h_1+h_2-h_3+2(d_1+d_2)+6~, \qquad && h_1\geqslant h_2\geqslant h_3~,\nn\\
&\AA~&:&~ \Delta=h_1+h_2-h_3+2(d_1+d_2)+6~, \qquad && h_1\geqslant h_2\geqslant h_3~,\nn\\
&\BB~&:&~ \Delta=h_1+2(d_1+d_2)+4~,                && h_1\geqslant h_2=h_3~,\label{eq:rep_families}\\
&\CC~&:&~ \Delta=h_1+2(d_1+d_2)+2~,                && h_1=h_2=h_3~,\nn\\
&\DD~&:&~ \Delta=2(d_1+d_2)~,	                   && h_1=h_2=h_3=0~.\nn
\end{alignat}
Representations of type $\LL$ are called long or generic representations, and their scaling dimension can be any real number consistent with the inequality in \eqref{eq:rep_families}. The other families of representations are short representations.\footnote{In the literature a distinction is sometimes drawn between short and semi-short representations. We make no such distinction here.} These representations have additional null states appearing in the Verma module built on the superconformal primary by the action of raising operators in $\ospf(8^\star|4)$. These representations are also sometimes called protected representations or -- in an abuse of terminology -- BPS representations.

\bigskip

\noindent Now let us review what is understood about the spectrum of local operators in the known $(2,0)$ theories.

\subsubsection*{BPS operators and chiral rings}

Perhaps the most familiar short representations are those of type $\DD$. The highest weight states for these representations are scalars that are half BPS (annihilated by two full spinorial supercharges) if $d_2=0$, and they are one quarter BPS (annihilated by one full spinorial supercharge) otherwise. As an example, the $\DD[1,0]$ multiplet is just the abelian tensor multiplet, whose primary is a free scalar field (with scaling dimension $\Delta=2$) transforming in the $\bf 5$ of $\sof(5)_R$. A more important example in the present paper is the $\DD[2,0]$ multiplet. The superconformal primary in this multiplet is a scalar operator $\Phi^{AB}$ of dimension four that transforms in the $\bf 14$ of $\sof(5)_R$. This multiplet also contains the $R$-symmetry currents, supercurrents, and the stress tensor for the theory, so such a multiplet should always be present in the spectrum of a local $(2,0)$ theory.

The BPS operators form two commutative rings -- the half and quarter BPS chiral rings. The OPE of BPS operators is non-singular, and multiplication in the chiral rings can be defined by taking the short distance limit of the OPE of these BPS operators.\footnote{Alternatively, one may define these rings cohomologically by passing to the cohomology of the relevant supercharges.} In the known $(2,0)$ theories of type $\gf$, these rings can be identified as the coordinate rings of certain complex subspaces of the moduli space of vacua -- they take relatively simple forms \cite{Bhattacharyya:2007sa}:
\begin{equation}
\label{eq:chiralrings}
\RR_{\frac12}={\Cb[z_1,\ldots,z_{r_{\gf}}]}/W_{\gf}~,\qquad \RR_{\frac14}=\Cb[z_1,\ldots,z_{r_{\gf}};w_1,\ldots,w_{r_{\gf}}]/W_{\gf}~,
\end{equation}
where $r_{\gf}$ is the rank of $\gf$ and $W_{\gf}$ is the Weyl group acting in the natural way. 

Knowing these rings for a given $(2,0)$ theory determines the full spectrum of $\DD$-type multiplets in said theory.\footnote{It isn't quite the case that the ring elements are in one-to-one correspondence with the $\DD$ multiplets. This is because half BPS operators have $\sof(5)_R$ descendants that are quarter BPS.} Note, however, that the ring structure for these operators does not determine numerically the value of any OPE coefficients since the normalizations of the BPS operators corresponding to particular holomorphic functions on the moduli space are unknown. To put it another way, if we demand that all BPS operators be canonically normalized, then the structure constants of the chiral ring are no longer known.

\subsubsection*{Chiral algebra operators}

There is a larger class of protected representations that participate in a more elaborate algebraic structure than the chiral rings -- namely the protected chiral algebra introduced in \cite{Beem:2014kka} (extending the analogous story for $\NN=2$ SCFTs in four dimensions \cite{Beem:2013sza}). Each of the following representations contains operators that act as two-dimensional meromorphic operators in the protected chiral algebra:\footnote{Here we are reverting to the conventions for $\sof(6)$ quantum numbers used in Appendix \ref{App:representations}. The $c_i$ are linearly related to the $h_i$ above in \eqref{eq:so6_change_of_basis}.}
\begin{alignat}{3}
\label{eq:reps_in_chiral_algebra}
&\DD[0,0,0;d,0]_{\ph{[h_1,0,0]}}\qquad&&~ \DD[0,0,0;d,\,1] \qquad&& \DD[0,0,0;d,\,2\,]\nn\\
&\CC\,[c_1,0,0;d,\,0\,]\qquad&&~ \CC\,[c_1,0,0;d,\,1\,]\qquad&& \\
&\BB[c_1,c_2,0;d,\,0\,]\qquad&&~ \qquad&& \nn
\end{alignat}
We observe that all half BPS operators and certain quarter BPS operators make an appearance in both the chiral rings and the chiral algebra, but the chiral algebra also knows about an infinite number of $\BB$-type multiplets that are probably less familiar.

In \cite{Beem:2014kka} the chiral algebras of the known $(2,0)$ theories were identified as the affine $\WW$-algebras of type $\gf$, where $\gf$ is the same simply laced Lie algebra that labels the $(2,0)$ theory, so the spectrum of the above multiplets is known.\footnote{This is the same chiral algebra that appears in the AGT correspondence in connection with the $(2,0)$ theory of type $\gf$ \cite{Alday:2009aq,Wyllard:2009hg}. The precise connection between these two appearances of the same chiral algebra remains somewhat perplexing.} We will have more to say about the information encoded in the chiral algebra below.

\subsubsection*{General short representations and the superconformal index}

The full spectrum of short representations is encoded in the superconformal index up to cancellations between representations that can recombine (group theoretically) to form a long representation \cite{Bhattacharya:2008zy}. The allowed recombinations are reviewed in Appendix \ref{App:representations}. This means that the full index unambiguously encodes the number of representations of the following types:
\begin{alignat}{4}
\label{eq:non_recombining_reps}
&\DD[0,0,0;d,0]_{\ph{[h_1,0,0]}}\qquad&&~ \DD[0,0,0;d,\,1] \qquad&&~ \DD[0,0,0;d,\,2\,]\qquad&&~ \DD[0,0,0;d,\,3\,]\nn\\
&\CC\,[c_1,0,0;d,\,0\,]\qquad&&~ \CC\,[c_1,0,0;d,\,1\,]\qquad&&\CC\,[c_1,0,0;d,\,2\,]\qquad&& \\
&\BB[c_1,c_2,0;d,\,0\,]\qquad&&~ \BB[c_1,c_2,0;d,\,1\,]\qquad\qquad&& \nn
\end{alignat}
It also provides lower bounds for the number of operators transforming in any short representation that appears in a recombination rule. 

A proposal has been made for the full superconformal index of the $(2,0)$ theories in \cite{Kim:2012qf,Kim:2013nva,Lockhart:2012vp}. To our knowledge, this proposal has not yet been systematically developed to the point where it will produce the unambiguous spectral data mentioned above. For our purposes, we will not need the full superconformal index, but for future generalizations of our bootstrap approach it would be very helpful to develop the technology to such a point.\footnote{Refs.~\cite{Kim:2012ava,Bullimore:2014upa} contain proposals for the index in an unrefined limit, where the enhanced supersymmetry leads to a simpler expression.}

\subsubsection*{Generic representations}

The situation for generic representations is much worse than that for short representations. Namely, in the known $(2,0)$ theories, the spectrum of long multiplets is almost completely mysterious. Outside of the the holographic regime, we are not aware of a single result concerning the spectrum of such operators. This is precisely the kind of information that one hopes will be attainable using bootstrap methods.

In the large $n$ limit of the $A_n$ theories, the full spectrum of local operators is known from AdS/CFT. Local operators are in one-to-one correspondence with  single- and multi-graviton states of the bulk  supergravity theory. Single-graviton states are the Kaluza-Klein modes of eleven-dimensional supergravity on ${\rm AdS}_7 \times S^4$~\cite{Gunaydin:1984wc} and correspond to ``single-trace'' operators of the boundary  theory. They can be organized into an infinite tower of half BPS representations of the $(2, 0)$ superconformal algebra. Similarly, multi-graviton states in the bulk are dual to ``multi-trace'' operators of the boundary $(2, 0)$ theory. They can be organized into a list of (generically long) multiplets of the superconformal algebra. At strictly infinite $n$, the bulk supergravity is free so the energy of a multi-graviton state is the sum of the energies of its single-graviton constituents. This translates into an analogous statement for the conformal dimension of the dual multi-trace operator. The first finite $n$ correction to the conformal dimensions of these operators can be computed from tree-level gravitational interactions in the bulk.

Of particular interest to us will be the double-trace operators that are constructed from the superconformal primaries of the stress tensor multiplet:
\begin{equation}
\label{eq:generalized_double_trace}
\OO_{m,\ell} = \left[\Phi\,(\partial^2)^m\partial_{\mu_1}\!\cdots\partial_{\mu_\ell}\,\Phi\right]_{[0,0]}~.
\end{equation}
At large $n$ these are the leading twist long multiplets. The scaling dimensions of the first few operators of this type at large $n$ are given by \cite{Arutyunov:2002ff,Heslop:2004du}
\begin{eqnarray} \label{DTsugra}
\Delta[\OO_{0,\ell=0}]&=&8-\frac{24}{n^3}+ \dots ~,\nn\\
\Delta[\OO_{0,\ell=2}]&=&10-\frac{30}{11 n^3}+ \dots ~,\\
\Delta[\OO_{0,\ell=4}]&=&12-\frac{72}{91n^3}+\dots ~.\nn
\end{eqnarray}
Similar results can be obtained for more general double trace operators at large $n$ -- see \cite{Heslop:2004du}. However, it should be noted that because the holographic dual is realized in M-theory, there is at present no method -- even in principle -- to generate the higher order corrections. This stands in contrast to the case of theories with string theory duals, where there is at least a framework for describing higher order corrections at large central charge.

\subsection{OPE coefficients}
\label{subsec:OPE_coeff_review}

The OPE coefficients of the known $(2,0)$ theories generally appear even more difficult to access than the spectrum of operators. Indeed, until recently the only three-point functions that were known for the finite rank $(2,0)$ theories were those that were fixed directly by conformal symmetry, \ie, those encoding the OPE of a conserved current with a charged operator.

\subsubsection*{Selection rules}

Some information is available in the form of selection rules that dictate which superconformal multiplets can appear in the OPE of members of two other multiplets. These selection rules are nontrivial to derive, and provide useful simplifications when studying, \eg, the conformal block decomposition of four-point functions.

We are not aware of a complete catalogue of selection rules for the $(2,0)$ superconformal multiplets. However, an algebraic algorithm has been developed in \cite{Heslop:2004du} based on writing three-point functions in analytic superspace, and this should be sufficient to determine the selection rules in all cases. For some special cases the selection rules have been determined explicitly, \cf\ \cite{Ferrara:2001uj,Eden:2001wg,Arutyunov:2002ff,Heslop:2004du}. We will use a particular case of this below in our discussion of half BPS operators.

\subsubsection*{OPE coefficients from the chiral algebra}

Going beyond selection rules, the numerical determination of some OPE coefficients (aside from those mentioned above) has recently become possible as a consequence of the identification of the protected chiral algebras of the $(2,0)$ theories \cite{Beem:2014kka}. In particular, up to choice of normalization, the three-point couplings between three chiral-algebra-type operators in a $(2,0)$ theory will be equal to the three-point coupling of the corresponding meromorphic operators in the associated chiral algebra. The structure constants of the $\WW_{\gf}$ algebras are completely fixed in terms of the Virasoro central charge, which is related by general arguments to the $c$-type Weyl anomaly coefficient of the corresponding $(2,0)$ theory, \cf\ \eqref{c(g)}.

Aside from these, the three-point functions of other protected operators are generally unknown -- it would be very interesting if it were possible to determine, \eg, the three-point functions of quarter BPS operators not described by the chiral algebra, perhaps using some argument related to supersymmetric localization.

\subsubsection*{Undetermined short operators and generic representations}

The situation for generic representations is again much worse, and aside from the OPEs encoding the charges of operators under global symmetries, we are not aware of any results for any OPE coefficients involving long multiplets outside of the holographic limit. At large $n$ many OPE coefficients can be computed at leading order in the $1/n$ expansion -- see \cite{Heslop:2004du} for example. We will be particularly interested later in this paper in the three-point functions coupling two stress tensor multiplets and certain generalized double traces that turn out to realize the $\DD[0,4]$ and $\BB[0,2]_{\ell}$ multiplets mentioned above. In particular, we have
\begin{eqnarray}
\label{eq:double_trace_BPS}
\OO_{\DD[0,4]}&=& \left[\Phi\Phi\right]_{[0,4]}~,\nn\\
\OO_{\BB[0,2]_{\ell}}&=&\left[\Phi\,\partial_{\mu_1}\ldots\partial_{\mu_{\ell}}\Phi\right]_{[0,2]}~,\qquad \ell=1,3,\ldots
\end{eqnarray}
Let us introduce the slightly awkward convention that $\OO_\ell:=\OO_{\BB[0,2]_{\ell}}$ and $\OO_{\ell=-1}=\OO_{\DD[0,4]}$. Then the values of the three-point couplings (squared) for these double traces take the following form:
\begin{equation}
\label{eq:OPE_large_k}
\lambda^2_{\Phi\Phi\OO_{\ell}}=\frac{2^{\ell+1}(\ell+2)(\ell+5)!(\ell+6)!}{(2\ell+9)!}\left(\frac{(\ell+3)(\ell+8)(\ell+9)}{36}-\frac{10}{n^3}\frac{(\ell^2+11\ell+27)}{(\ell+2)(\ell+4)(\ell+7)}+\ldots\right)~.
\end{equation}
In particular, for the first few low values of $\ell$ these are given by
\begin{eqnarray}
\label{eq:OPE_large_k_small_l}
\lambda^2_{\Phi\Phi\OO_{\DD[0,4]}}   &=&\frac{16}{9}-\frac{340}{63n^3}+\ldots~,\nn\\
\lambda^2_{\Phi\Phi\OO_{\BB[0,2]_1}} &=&\frac{120}{11} - \frac{39}{11n^3}+\ldots~,\\ 
\lambda^2_{\Phi\Phi\OO_{\BB[0,2]_3}} &=&\frac{256}{13}-\frac{8832}{5005n^3}+\ldots~.\nn
\end{eqnarray}

\subsection{Four-point functions of half BPS operators}
\label{subsec:half_BPS_crossing}

In this paper we will be focusing on the four-point function of stress tensor multiplets. However, many nice features of the bootstrap problem for stress tensor multiplets occur more generally in studying the four-point function of arbitrary half BPS operators, and ultimately the approach developed in this paper should be extendable to this more general class of correlators without great conceptual difficulty. Therefore let us first give a schematic description of the crossing symmetry problem in this more general context, before specializing to the specific case of interest.

\subsubsection*{Three-point functions}

It is a convenient fact that the full superspace structure of a three-point function involving two half BPS multiplets and any third multiplet is completely determined by the three-point function of superconformal primary operators \cite{Eden:2001wg}. Since the superconformal primaries of half BPS representations are spacetime scalars, it follows that each such superspace three-point function is determined by a single numerical coefficient.

This simplicity of three-point functions (or equivalently of the OPE between half BPS operators) in superspace has pleasant consequences for the conformal block expansion of four-point functions and the associated crossing symmetry equation. In particular, it means that the \emph{superconformal block} associated to the exchange of all operators in a superconformal multiplet is fixed up to a single overall coefficient, even though the exchanged operators could live in several conformal multiplets and transform in different representations of $\sof(5)_R$.

Thus, the conformal block decomposition will take the form of a sum over superconformal multiplets of a single real coefficient times a superconformal block. Schematically this takes the following form
\begin{equation} 
\label{eq:schematic_conformal_block_expansion}
\langle  \OO_1^A \, \OO_2^B\,  \OO_3^C \, \OO_4^D \, \rangle ~~\sim~~ \sum_{\XX} \lambda_{12\XX}\lambda_{34\XX}\left(\sum_{R\in\XX}P^{ABCD}_R G^{1234}_{\XX,R}(z,\zb)\right)~.
\end{equation}
Here $\XX$ runs over the irreducible representations appearing in both the $\OO_1\times\OO_2$ and $\OO_3\times\OO_4$ selection rules, $R$ runs over the different $\sof(5)_R$ representations appearing in the supermultiplet $\XX$, and we have introduced projection tensors $P^{ABCD}_{R}$. The precise form of these projectors is not particularly important -- in the technical analysis of Section \ref{Sec:blockdecomp} we will introduce some additional structure in the form of complex $R$-symmetry polarization vectors to simplify manipulations of these superconformal blocks. 

The superconformal blocks $G^{1234}_{\XX,R}(z,\zb)$ for each $R\in\XX$ are functions of conformal cross ratios, and their form is fixed in terms of the representations of the four external operators and $\XX$. These functions also have fixed relative normalizations. This is a manifestation of the simplification mentioned above, and it means that there is only a single free numerical parameter that determines the contribution of a full supermultiplet to the four-point function.

\subsubsection*{Selection Rules}

The representations $\XX$ that can appear in the sum on the right hand side of \eqref{eq:schematic_conformal_block_expansion} are constrained by the superconformal selection rules mentioned previously. For the particular case of the OPE of two half BPS operators, these selection rules have been studied starting with the work of \cite{Eden:2001wg,Ferrara:2001uj,Arutyunov:2002ff}, with the complete answer being given in \cite{Heslop:2004du}. The results are as follows:
\begin{eqnarray}
\label{eq:selection_rules}
\DD[d,0] \times \DD[\tilde{d},0] &~~=~~& \sum_{k=0}^{\scriptscriptstyle\min(d,\tilde{d})}\sum_{i=0}^{\scriptscriptstyle\min(d,\tilde{d})-k}\DD[d+\tilde{d}-2k-2i,2i]~+\nn\\
&+& \sum_{k=1}^{\scriptscriptstyle\min(d,\tilde{d})}\sum_{i=0}^{\scriptscriptstyle\min(d,\tilde{d})-k}\sum_{s=0,1,2,\ldots}\BB[d+\tilde{d}-2k-2i,2i]_{\ell}~+\\
&+& \sum_{k=2}^{\scriptscriptstyle\min(d,\tilde{d})}\sum_{i=0}^{\scriptscriptstyle\min(d,\tilde{d})-k}\sum_{\overset{~~s\,=\,0,1,2,\ldots}{\scriptscriptstyle\Delta>6+\ell~~}}\LL_{\Delta}[d+\tilde{d}-2k-2i,2i]_{\ell}~.\nn
\end{eqnarray}
An interesting point that will become relevant in a moment is that every short representation appearing on the right hand side of \eqref{eq:selection_rules} is of one of two types:
\begin{itemize}
\item[(A)] Representations that appear in the decomposition a long multiplet in \eqref{eq:selection_rules} at the appropriate unitarity bound.
\item[(B)] Representations from the list \eqref{eq:reps_in_chiral_algebra} that include chiral algebra operators.
\end{itemize}
Notice in particular that for long multiplets that decompose at the unitarity bound (\cf\ Appendix \ref{App:representations}), only one of their irreducible components is allowed by the selection rules. This implies that the superconformal blocks for those short multiplets of type (A) above will simply be obtained as the limit of a long superconformal block when its scaling dimension is set to the unitarity bound.\footnote{A consequence of this general structure is that the number of independent, unknown functions of conformal cross ratios appearing in the superconformal block expansion of a four-point function of half BPS operators is just equal to the number of different $R$-symmetry representations in which the long multiplets in \eqref{eq:selection_rules} transform.}

\subsubsection*{Fixed and unfixed BPS contributions}

The two types of short operators mentioned above will participate in the crossing symmetry problem very differently. Operators of type (B) will have their three-point functions with the external half BPS operators determined by the chiral algebra, so subject to identification of the chiral algebra their contribution to the four-point function will be completely fixed. For the known $(2,0)$ theories the chiral algebra has been identified, so for a general four-point function of half BPS operators we can put in some fairly intricate data about the theory we wish to study by fixing the chiral algebra part of the correlator.

Operators of type (A) on the other hand are not described by their chiral algebra and we cannot \emph{a priori} fix their contribution to the four-point function. Indeed, we can see that as the dimension of the $\sof(5)_R$-symmetry representation of the external half BPS operators is increased, an ever larger number of short operators that are not connected to the chiral algebra will appear in the OPE. This situation stands in contrast to analogous bootstrap problems in four dimensions \cite{Beem:2013qxa,Beem:2014zpa}, where the entirety of the short operator spectrum contributing to the desired four-point functions is constrained by the chiral algebra.

From the point of view of the conformal block decomposition, since these unfixed short multiplets occur in the decomposition of long multiplets at threshold, their conformal blocks will be limits of long conformal blocks and so they do not introduce any truly new ingredients into the crossing symmetry problem. However, since these are protected operators there is a chance that we may know something about their spectrum. We will explain below that precisely such a situation can occur for the four-point function of stress tensors multiplets.

\subsection{Specialization to stress tensor multiplets}
\label{subsec:stress_tensor_crossing}

Let us now restrict our attention to the special case of interest, which is the four-point function of stress tensor multiplets, \ie, $\DD[2,0]$ multiplets. This leads to some simplifications in the structure outlined above. We will see these simplifications in much greater detail in the coming sections.

\medskip 

The sums in \eqref{eq:selection_rules} truncate fairly early when $d=\tilde{d}=2$. Additionally some representations are ruled out by the requirement that the OPE be symmetric under the exchange of the two identical operators. This leaves the following selection rules:
\begin{eqnarray}
\label{eq:stress_tensor_selection_rules}
\DD[2,0] \times \DD[2,0] &~~=~~& \mathbf 1 + \DD[4,0]+\DD[2,0]+\DD[0,4]+\nn\\[1.5ex]
&+& \sum_{\ell=0,2,\ldots}\left(\BB[2,0]_{\ell}+\BB[0,2]_{\ell+1}+\text{\sout{$\BB[0,0]_{\ell}$}}\right)~+\\
&+& \sum_{\ell=0,2,\ldots}\sum_{\Delta>6+\ell}\LL[0,0]_{\Delta,\ell}~.\nn
\end{eqnarray}
The representations that are struck out contain higher spin conserved currents, and so will be absent in interacting theories. 

Referring back to \eqref{eq:reps_in_chiral_algebra}, we see that almost all of the short representations allowed by these selection rules are included in the chiral algebra. Their OPE coefficients will consequently be determined by the corresponding chiral algebra correlator. An important feature of this special case is that the chiral algebra correlator that is related to this particular four-point function is the four-point function of holomorphic stress tensors. This is important because the four-point function of holomorphic stress tensors is determined uniquely up to a single constant -- the Virasoro central charge. Thus the chiral algebra contribution to this four-point function can be completely characterized in terms of the central charge, with no additional dependence on the theory being studied. This is in contrast to the case of general four-point functions, where the chiral algebra correlator is some four-point function in the $\WW_{\gf}$ algebra that may in principle look rather different for different choices of $\gf$.

Thus the contributions of chiral algebra operators to the four-point function will be fixed in terms of $c$. What about the unfixed BPS operators? For this example there are not that many options -- namely, there is the quarter BPS multiplet $\DD[0,4]$ and the $\BB$-series multiplets $\BB[0,2]_{\ell-1}$. The superconformal blocks for these multiplets are the limits of the blocks for the long multiplets $\LL[0,0]_{\Delta,\ell}$ at the unitarity bound. Thus the only unknown superconformal blocks appearing in the expansion of this four point function are those contributed by long multiplets $\LL[0,0]_{\Delta, \ell}$, possibly with $\Delta=\ell+6$. Later this will allow us to write the crossing symmetry equation for this four point function in terms of a single unknown function of conformal cross ratios.

Finally, let us make an auspicious observation about the quarter BPS multiplet $\DD[0,4]$ that is allowed in this correlation function. Since the chiral rings of the ADE $(2,0)$ theories are thought to be known, we may test for the presence of this multiplet in these theories. In general, there is a single such operator in the known $(2,0)$ theories. For example, in the $A_{N-1}$ theories the quarter BPS ring is given by
\begin{equation}
\RR_{\frac{1}{4}}[A_{N-1}]=\Cb[z_1,\ldots,z_{N};w_1,\ldots,w_{N}]^{S_{N}}\Bigg/ \left\{\sum z_i\sim\sum w_i\sim 0 \right\}~.
\end{equation}
Note that this is the same as the ring of $SU(N)$-invariant functions of two commuting, traceless $N\times N$ matrices $Z$ and $W$ (which is the same as the part of the chiral ring of $\NN=4$ super Yang-Mills in four dimensions that is generated by two of the three chiral superfields, say $Z$ and $W$). In these terms, a single quarter BPS operator is then generally present and can be written as 
\begin{equation}
\tr(Z^2)\tr(W^2)-\tr(ZW)^2~.
\end{equation}
However, for the special case of $N=2$, this operator is identically zero. What this tells us is that in precisely the $A_1$ theory, there will be no conformal block coming from a $\DD[0,4]$ multiplet in the four-point function of stress tensor multiplets. This observation will have major consequences in our interpretation of the bootstrap results later in this paper.

%% file: sections/3_correlator.tex

\section{The four-point function of stress tensor multiplets}
\label{sec:structure_of_correlator}

We are now in a position to describe the detailed structure of the four-point function of stress tensor multiplets. Maximal superconformal symmetry guarantees that (i) the stress tensor belongs to a multiplet whose superconformal primary is a Lorentz scalar operator; (ii) the four-point function of this supermultiplet admits a unique structure in superspace. Consequently we lose no information by restricting our attention to the four-point function of the scalar superconformal primary, which is a dramatically simpler object.\footnote{In fact, the technology to bootstrap four-point functions involving external tensorial operators, while conceptually straightforward, has not yet been fully developed. Rapid progress is being made in the area -- see \cite{Costa:2011mg,Dolan:2011dv,Costa:2011dw,SimmonsDuffin:2012uy,Siegel:2012di,Osborn:2012vt,Hogervorst:2013sma,Fitzpatrick:2013sya,Hogervorst:2013kva,Fitzpatrick:2014oza,Khandker:2014mpa,Elkhidir:2014woa,Costa:2014rya,Dymarsky:2013wla,Echeverri:2015rwa}.} This is a huge simplification in bootstrap studies, and has already been exploited for maximally superconformal field theories in four \cite{Beem:2013qxa} and three \cite{Chester:2014fya,Chester:2014mea} dimensions. The results described in this section rely heavily on the previous works \cite{Arutyunov:2002ff,Dolan:2004mu}.

The superconformal primary operator in the $\DD[2,0]$ multiplet is a half BPS scalar operators of dimension four that transforms in the $\mathbf{14}$ of the $\sof(5)_R$. We denote these scalar operators as $\Phi^{IJ}(x) = \Phi^{\{IJ\}}(x)$, where $I,J=1,\ldots,5$ are fundamental $\sof(5)_R$ indices, and the brackets denote symmetrization and tracelessness. A convenient way to deal with the $\sof(5)_R$ indices is to contract them with complex polarization vectors $Y^I$ and define
\begin{equation}
\Phi(x,Y) \colonequals \Phi^{IJ}(x) Y_I Y_J~.
\end{equation}
The polarization vectors can be taken to be commutative due to symmetrization of the two $\sof(5)_R$ indices, and tracelessness is encoded by the null condition:
\begin{equation}
Y^I Y_I = 0~.
\end{equation}
With these conventions, the two-point function of $\Phi(x,Y)$ is given by
\begin{equation}
\label{eq:2ptfn_indices}
\vev{\Phi(x_1,Y_1) \Phi(x_2,Y_2)} = \frac{4 (Y_1 \cdot Y_2)^2}{x_{12}^8}~.
\end{equation}
Homogeneity of correlators with respect to simultaneous rescalings of the $Y^I$ allows us to solve the null constraint as follows \cite{Dolan:2004mu},
\begin{equation}
Y^I = \left(y^i,\frac{1}{2}(1- y^i y_i), \frac{i}{2}(1+ y^i y_i)\right)~,
\label{eq:Yiny}
\end{equation}
with $y^i$ an arbitrary three-vector. The two-point function is now given by
\begin{equation}
\vev{\Phi(x_1,y_1) \Phi(x_2,y_2)} = \frac{y_{12}^4}{ x_{12}^8}~.
\label{eq:2ptfn}
\end{equation}
The normalization in \eqref{eq:2ptfn_indices} has led to unit normalization in these variables.

\subsection{Structure of the four-point function}

Conformal Ward identities and $R$-symmetry conservation dictate that the $\Phi(x,y)$ four-point function can be written as \cite{Dolan:2004mu}
\begin{equation}
\vev{\Phi(x_1,y_1) \Phi(x_2,y_2) \Phi(x_3,y_3) \Phi(x_4,y_4)} = \frac{y_{12}^4 y_{34}^4}{x_{12}^8 x_{34}^8} G(z,\bar z;\alpha, \bar \alpha)~,
\label{eq:fourptfn}
\end{equation}
where $z$ and $\bar z$ are related to the canonical conformally invariant cross-ratios,
\begin{equation}
u \colonequals \frac{x_{12}^2 x_{34}^2}{x_{13}^2 x_{24}^2} \equalscolon z \bar z~, \qquad \qquad v \colonequals \frac{x_{14}^2 x_{23}^2}{x_{13}^2 x_{24}^2} \equalscolon (1-z)(1-\bar z)~,
\end{equation}
and $\alpha$ and $\bar \alpha$ obey a similar relation with respect to ``cross-ratios'' of the polarization vectors,
\begin{equation}
\frac{1}{\a \bar \a} \colonequals \frac{y_{12}^2 y_{34}^2}{y_{13}^2 y_{24}^2}~, \qquad \qquad \frac{(\a -1)(\bar \a - 1)}{\a \bar \a} \colonequals \frac{y_{14}^2 y_{23}^2}{y_{13}^2 y_{24}^2}~.
\end{equation}
Although not manifest in this notation, the dependence of the full correlator on the $y^i$ is polynomial by construction.

The constraints of \emph{superconformal} invariance were investigated thoroughly in \cite{Dolan:2004mu}. Ultimately, the consequence of said constraints is that the four-point function must take the form
\begin{equation}
\label{eq:GinaH}
G(z,\bar z;\a,\bar \a) = u^4 \D_2   \left[ (z \a - 1) (z \bar \a - 1) (\bar z \a - 1)(\bar z \bar \a - 1) a(z,\bar z) \right] + z^2 \bar z^2 \HH_1^{(2)}(z,\bar z;\a,\bar \a)~.
\end{equation}
with
\begin{equation}
\label{eq:Hinh}
\HH_1^{(2)}(z,\bar z;\a,\bar \a) = D_2 \frac{(z \a - 1) (z \bar \a - 1) h(z) - (\bar z \a - 1)(\bar z \bar \a -1) h(\bar z)}{z - \bar z}~.
\end{equation}
Here $D_2$ and $\Delta_2$ are second order differential operators defined according to
\begin{equation}
\D_2 f(z,\bar z) \colonequals D_2 u f(z,\bar z) \colonequals \left( \frac{\del^2}{\del z \del \bar z} - \frac{2}{z - \bar z} \left( \frac{\del}{\del z} - \frac{\del}{\del \bar z} \right) \right) z \bar z f(z,\bar z)~.
\label{eq:diffops}
\end{equation}
The entire four-point function is determined in terms of a two-variable function $a(z,\bar z)$ and a single-variable function $h(\cdot)$. The superconformal Ward identities impose no further constraints on these functions.

As described in \cite{Beem:2014kka}, there exists a specific $R$-symmetry twist such that the correlation functions of $\Phi^{\{IJ\}}(x)$ devolve into those of a two-dimensional chiral algebra. For the two-point function \eqref{eq:2ptfn} this twist is implemented by taking $y_{12} = \bar z_{12}$, leading to 
\begin{equation}
\label{eq:twisted2pt}
\vev{\hat \Phi(z_1) \hat \Phi(z_2)} = \frac{1}{ z_{12}^4}~,
\end{equation}
where $\hat \Phi(z)$ denotes the twisted operator in the chiral algebra. We see that $\hat \Phi(z)$ behaves as a meromorphic operator of dimension two -- it is in fact the Virasoro stress tensor of the chiral algebra \cite{Beem:2014kka}. For the four-point function we set $\a = \bar \a = 1 / \bar z$ and obtain
\begin{equation}
\label{eq:twisted4pt}
\vev{\hat \Phi(z_1) \hat \Phi(z_2) \hat \Phi(z_3) \hat \Phi(z_4)} = \frac{- z^2 h'(z)}{z_{12}^4 z_{34}^4} ~.
\end{equation}
The dependence on the two-variable function $a(z,\bar z)$ completely drops out and the chiral correlator is determined by the derivative $h'(z)$ of the single-variable function introduced above.

\subsection{Constraints from crossing symmetry}
\label{subsec:constraints_from_crossing}

The correlation function \eqref{eq:fourptfn} must be invariant under permutations of the four operators. Interchanging the first and the second operators implies the constraint
\begin{equation}
\label{eq:Gaabcrossing}
G(z,\bar z; \a, \bar \a) = G\left( \frac{z}{z-1}, \frac{\bar z}{\bar z - 1}; 1- \a, 1 - \bar \a\right)~,
\end{equation}
whereas invariance under interchanging the first and the third operators requires
\begin{equation}
G(z,\bar z;\a, \bar \a) = \frac{z^4 \bar z^4 (\a - 1)^2 (\bar \a - 1)^2}{(1-z)^4 (1- \bar z)^4} G\left( 1-z, 1 - \bar z; \frac{\a}{\a - 1},\frac{\bar \a}{\bar \a - 1} \right)~.
\end{equation}
Additional permutations do not give rise to any additional constraints.

Using equations \eqref{eq:GinaH} and \eqref{eq:Hinh} we can express these constraints in terms of $a(z,\zb)$ and $h(z)$. Much as in \cite{Beem:2013qxa,Beem:2014zpa}, we find that from each of the above equations a constraint can be extracted that applies purely to the single variable function $h(z)$,
\begin{equation}
h'(z) = \frac{1}{(z-1)^2} h'\left( \frac{z}{z-1}\right) = \frac{z^2}{(z-1)^2} h'(1-z)~.
\end{equation}
Defining $g(z) \colonequals - z^2 h'(z)$, these constraints take the form
\begin{equation}
\label{eq:gcrossing}
g(z) = g\left(\frac{z}{z-1}\right) = \left( \frac{z}{z-1}\right)^4 g(1-z)~.
\end{equation}
This is precisely the crossing symmetry constraint for the four-point function of a chiral operator of dimension two. Of course none of this is a coincidence -- both the structure of this equation and the existence of a decoupled crossing relation for $h(z)$ are a direct consequence of the chiral algebra described in \cite{Beem:2014kka}. We will solve \eqref{eq:gcrossing} in the next subsection.

The remaining constraints from crossing symmetry amount to the following two relations for the two-variable function:
\begin{equation}
\begin{split}
a(z,\bar z) - \frac{1}{(z-1)^5 (\bar z-1)^5} &a\left(\frac{z}{z-1},\frac{\bar z}{\bar z-1}\right) = 0~, \\[10pt]
z \bar{z} a\left(z,\bar{z}\right)-(z-1) \left(\bar{z}-1\right) a\left(1-z,1-\bar{z}\right) &= \frac{1}{(z - \bar z)^3} \left( \frac{h\left(1-\bar{z}\right)-h(1-z)}{(z-1) \left(\bar{z}-1\right)}+\frac{h\left(\bar{z}\right)-h(z)}{z \bar{z}} \right)~.
\end{split}
\label{eq:crossingeq}
\end{equation}
When reformulated in terms of the conformal block expansion the first equation is easily solved, while the latter is the non-trivial crossing symmetry equation that we will be the subject of the numerical analysis.

\subsubsection{Solving for the meromorphic function}
\label{subsubsec:solving_meromorphic}
We will see in the next section that consistency with the six-dimensional OPE requires that $g(z)$ is meromorphic in $z$ and admits a regular Taylor series expansion around $z=0$ with integer powers. This leads to the ansatz
\begin{equation}
g(z) = \b_1 + \b_2 z + \b_3 z^2 + \b_4 z^3 + \ldots~.
\end{equation}
The crossing symmetry constraints \eqref{eq:gcrossing} imply that $\b_2 = 0$ and $\b_4 = \b_3$. The full form of $g(z)$ is then fixed in terms of $\b_{1,3}$ according to
\begin{equation}
g(z) = \b_1 \left(1 + z^4 + \frac{z^4}{(1-z)^4} \right) + \b_3 \left( z^2 + z^3 + \frac{z^4}{(1-z)^2} + \frac{z^4}{1-z}\right)~.
\label{eq:G_sol}
\end{equation}
which implies that
\begin{equation}
h(z) = - \b_1 \left(\frac{z^3}{3}-\frac{1}{z-1}-\frac{1}{(z-1)^2}-\frac{1}{3 (z-1)^3}-\frac{1}{z}\right)- \b_3 \left(z-\frac{1}{z-1}+\log (1-z)\right) + \b_5~,
\label{eq:h_sol}
\end{equation}
where $\b_5$ is an integration constant. From \eqref{eq:Hinh} we see that this constant does not affect $\HH_1^{(2)}(z,\bar z;\a,\bar \a)$ and therefore can be set to any convenient value.

This leaves us with the determination of the parameters $\b_1$ and $\b_3$. The former is determined from the normalization of the operator $\Phi(x,y)$. We fixed this normalization in \eqref{eq:2ptfn}, which led to a normalization of the twisted operator $\hat \Phi(z)$ as shown in \eqref{eq:twisted2pt}. Compatibility of \eqref{eq:twisted4pt} with this equation implies that $-z^2 h'(z) \to 1$ as $z \to 0$, and therefore
\begin{equation}
\beta_1 = 1~.
\end{equation}
The superconformal block decomposition described in the next section can be used to show that the parameter $\beta_3$ is a certain multiple of the squared OPE coefficient of the stress tensor. It would be relatively straightforward to work out the precise proportionality constant in this manner, but the chiral algebra provides an even more efficient way to find the same result. Indeed, the twisted correlator in \eqref{eq:twisted4pt} is proportional to the four-point function of Virasoro stress tensors in the chiral algebra \cite{Beem:2014kka}. The two-dimensional self-OPE of stress tensors takes the familiar form
\begin{equation}
T(z) T(0) \sim \frac{c/2}{z^4} + \frac{2 T(0)}{z^2} + \frac{\del T(0)}{z}~,
\end{equation}
with a two-dimensional central charge $c$ whose precise meaning will be discussed shortly. We chose to normalize the four-point function such that the unit operator appears with coefficient one, so in the chiral algebra $\hat \Phi(z) \leadsto T(z) \sqrt{2/c}$. Comparing the above OPE with \eqref{eq:G_sol} with $\b_1 = 1$ we find a match of the leading term, and at the first nontrivial subleading order we find
\begin{equation}
\b_3 = \frac{8}{c}~.
\end{equation}
Supersymmetry dictates that $c$ is related to the coefficient $C_T$ in the two-point function of the canonical six-dimensional stress tensor, which takes the form \cite{Osborn:1993cr}
\begin{eqnarray}
\label{eq:TT}
\langle T_{\mu \nu} (x) T_{\alpha \beta} (0) \rangle &=& \frac{C_T}{x^{2d}} \II_{\mu \nu \rho \sigma}(x)~,\nn \\
\II_{\mu \nu \rho \sigma}(x)&=& \frac{1}{2} \left( I_{\mu \rho}(x) I_{\nu \sigma}(x) + I_{\mu \sigma}(x) I_{\nu \rho}(x)-\frac{1}{6}\delta_{\mu \nu}\delta_{\rho \sigma}\right)~,\\
I_{\mu \nu} (x) &=& \delta_{\mu \nu}-2 \frac{x_\mu x_\nu}{x^2}~.\nn
\end{eqnarray}
The precise proportionality constant can be determined from the free tensor multiplet, for which $c=1$ \cite{Beem:2014kka} and $C_T=\frac{84}{\pi^6}$ \cite{Bastianelli:1999ab}. Therefore $C_T \equalscolon \frac{84}{\pi^6}c$. For the $(2,0)$ theories of type $\gf$ this central charge was determined in \cite{Beem:2014kka} to be
\begin{equation}
c = 4 d_\gf h^\vee_\gf + r_\gf~,
\end{equation}
with $d_\gf$, $h^\vee_\gf$, and $r_\gf$ the dimension, dual Coxeter number, and rank of $\gf$.

An important consequence of the above analysis is that the right-hand side of the second crossing symmetry equation in \eqref{eq:crossingeq} becomes manifestly dependent on $c$ through $h(z)$. The $c$ central charge thus becomes an input parameter for the numerical bootstrap analysis. This is reflected in the $c$-dependence of the numerical results shown below.

%% file: sections/4_block_decomposition_and_crossing.tex

\section{Superconformal block decomposition}
\label{Sec:blockdecomp}

The superconformal block decomposition of the four-point function of this decomposition \eqref{eq:fourptfn} was sketched in section \ref{sec:strategy} --  here we present the technical details. We will describe how the various multiplets that are allowed by the selection rules contribute to the functions $a(z,\zb)$ and $h(z)$. We will further show that the full contribution of many short multiplets can be determined from the known form of $h(z)$, allowing us to formulate a crossing symmetry equation that refers exclusively (and explicitly) to unknown SCFT data.

\subsection{Superconformal partial wave expansion }
\label{subsec:partial_waves}

We begin by considering the form of the regular (non-supersymmetric) conformal block decomposition of the four-point function. The operators $\Phi^{AB}$ transform in the $\bf{14}$, or the $[2,0]$, of $\sof(5)_R$. Consequently we may find operators in the following $\sof(5)_R$ representations in its self-OPE:
\begin{equation}
\begin{split}
([2,0]\otimes[2,0])_s &= [0,0] \oplus [2,0] \oplus [4,0] \oplus [0,4]~,\\
([2,0]\otimes[2,0])_a &= [0,2] \oplus [2,2]~.
\end{split}
\end{equation}
The first/second line contain the representations appearing in the symmetric/antisymmetric tensor product. The conformal block decomposition of $G(z,\bar z;\a,\bar \a)$ therefore takes the following form
\begin{equation} 
\label{eq:decompG}
G(z,\bar z; \a,\bar \a) = \sum_{r \in R} \left(Y^r(\a,\bar \a)  \sum_{k_r} \lambda_{k_r}^2 \GG_{\Delta_{k_r}}^{(\ell_{k_r})} (z,\bar z) \right)~,
\end{equation}
with $r \in R = \{ [0,0], [2,0], [4,0],[0,4],[0,2],[2,2]\}$ the set of $\sof(5)_R$ representations, $k_r$ labeling the different operators in representation $r$ appearing in the OPE, and $(\lambda_{k_r}, \Delta_{k_r}, \ell_{k_r})$ denoting the OPE coefficient, scaling dimension and spin of the operator, respectively. By Bose symmetry, only even/odd $\ell$ can appear for symmetric/anti-symmetric $r$. The $Y^r(\alpha,\alpha')$ are harmonic functions that encode the $\sof(5)$ invariant tensor structure associated with representation $r$. Their exact form is given in \eqref{eq:Yprojectors}. The functions $\GG_\Delta^{(\ell)}(z,\bar z)$ are the ordinary conformal blocks of six-dimensional CFT for a correlation function of identical scalars -- they are given by \eqref{eq:6dblock} with $\D_{12}=\D_{34}=0$. 

Supersymmetry introduces additional structure in the decomposition \eqref{eq:decompG}, since the conformal blocks corresponding to operators in the same supermultiplet can be grouped together into what we may call superblocks. The \emph{super}conformal block expansion can be written as
\begin{equation} \label{eq:superdecompG}
G(z,\bar z; \a,\bar \a) = \sum_{\XX} \lambda_{\XX}^2 \left( \sum_{r \in R} Y^r(\a,\bar \a) A_{r}^{\XX} (z,\bar z) \right)~,
\end{equation}
where the sum runs over superconformal multiplets $\XX$, with only one unknown squared OPE coefficient $\lambda_\XX^2$ per superconformal multiplet. The functions $A_{r}^{\XX} (z,\bar z)$ are finite sums, with known coefficients, of ordinary six-dimensional conformal blocks $\GG_{\Delta_{k_r}}^{(\ell_{k_r})}(z,\zb)$ corresponding to the different conformal primary operators $k_r$ that appear in the same $\XX$ multiplet.

Superconformal Ward identities dictate that each superconformal block -- that is, each expression of the form $\sum_{r \in R} Y^r(\a,\bar \a) A_r^\XX(z,\bar z)$ -- can also be written in the form given in \eqref{eq:GinaH} and \eqref{eq:Hinh}. To determine the superconformal blocks it therefore suffices to determine the contributions $a^\XX(z,\bar z)$ and $h^\XX(z)$ of a given multiplet $\XX$ to the two functions $a(z,\zb)$ and $h(z)$.

\subsection{Superconformal blocks}
\label{subsec:superconformal_blocks}

In Section \ref{sec:strategy} we introduced selection rules for which superconformal multiplets can make an appearance in the OPE of stress tensor multiplets, which we reproduce here for convenience:
\begin{equation} 
\label{eq:OPE}
\DD[2,0] \times \DD[2,0] \sim \mathbf{1} +  \DD[2,0] +  \DD[4,0]+ \DD[0,4] + \BB[2,0]_\ell +  \BB[0,2]_\ell +  \BB[0,0]_\ell + \LL[0,0]_{\D,\ell}~.
\end{equation}
In order to define the superconformal blocks corresponding to these multiplets we introduce two basic elements, which we call the \emph{atomic} contributions to $a(z,\bar z)$ or $h(z)$. These two functions take the form
\begin{equation}
\label{eq:ahatom}
\begin{split}
a^\text{at}_{\Delta,\ell}(z,\bar z) &= \frac{4}{z^{6} \bar z^{6}(\Delta-\ell-2)(\Delta+\ell+2)} \GG_{\Delta+4}^{(\ell)} (0,-2;z,\bar z)~, \\
h^{\text{at}}_\beta(z) &=  \frac{z^{\b -1}}{1 - \b} {}_2F_1[\b -1,\b;2\b,z]~,
\end{split}
\end{equation}
where $\GG_{\D}^{(\ell)}(\D_1 - \D_2,\D_3-\D_4;z,\bar z)$ is the ordinary conformal block for a correlation function of four operators with unequal scaling dimensions $\D_i$ given in \eqref{eq:6dblock}.\footnote{In \cite{Dolan:2004mu} the function that we call $a^\text{at}_{\Delta,\ell}(z,\bar z)$ was  given as an infinite sum of Jack polynomials. Our more transparent expression was derived by pulling the conformal Casimir operator through $\Delta_2$ given in \eqref{eq:diffops}.} We claim that the superconformal block for any representation in \ref{eq:OPE} can be recovered from these atomic contributions in the manner specified in Table \ref{tab:superblocks}.

\begin{table}[h!t]
\centering
\begin{tabular}{>$l<$ >$c<$ >$l<$ >$l<$ l}
\XX & \D &  a^\XX(z,\bar z) &  h^\XX(z) & \text{comments} \\
\hline \hline
\LL[0,0]_{\D,\ell}  & \D  & a^{\text{at}}_{\D,\ell}(z,\bar z) & 0 &  generic long multiplet, $\D > \ell + 6$\\
\BB[0,2]_{\ell -1} & \ell + 7  & a^{\text{at}}_{\ell + 6, \ell}(z,\bar z) & 0 &  $\ell > 0$\\
\DD[0,4] & 8  & a^{\text{at}}_{6,0}(z,\bar z) & 0 &   \\
\BB[2,0]_{\ell - 2} & \ell + 6 &   a^{\text{at}}_{\ell + 4, \ell}(z,\bar z) & 2^{-\ell} h^{\text{at}}_{\ell + 4}(z) & $\ell > 0$\\
\DD[4,0] & 8 &  a^{\text{at}}_{4,0}(z,\bar z) & h^{\text{at}}_{4}(z) & \\
\BB[0,0]_{\ell} & \ell + 4 & 0 &  h^{\text{at}}_{\ell + 4}(z) & higher spin currents, $\ell \geqslant 0$\\
\DD[2,0] & 4 & 0 &  h^{\text{at}}_2(z) & stress tensor multiplet\\
{\bf 1} & 0 & 0 & h^{\text{at}}_0(z) &identity
\end{tabular}
\caption{\label{tab:superblocks} Superconformal blocks contribution from all superconformal multiplets appearing in the OPE of two stress tensor multiplets. The contributions are determined from the atomic building blocks. Bose symmetry requires that $\ell$ is an even integer. Here $\D$ is the dimension of the superconformal primary.}
\end{table}

Let us first momentarily take for granted that there are no other atomic contributions besides $a^\text{at}_{\Delta,\ell}(z,\bar z)$ and $h^\text{at}_\beta (z)$. The particular combinations shown in Table \ref{tab:superblocks} are then uniquely determined by decomposing the superconformal blocks into ordinary conformal blocks and requiring that the only conformal multiplets appearing are actually included in the superconformal multiplet under consideration. This decomposition can be done with the help of \eqref{eq:Ainah}, which for each $a^{\text{at}}_{\Delta,\ell}(z,\bar z)$ or $h_\beta^\text{at}(z)$ leads to six functions $A_{r}^\text{at} (z,\bar z)$ that describe the exchange of operators in the representation $r$ of $\sof(5)_R$. Each of these $A_r^\text{at}(z,\bar z)$ in turn admits a decomposition in a finite number of conformal blocks, and by enumerating these blocks we arrive at the conformal primary operator content of the superconformal multiplet. As an example we can consider $a^\text{at}_{\D,\ell}(z,\bar z)$ with $\ell \geqslant 4$ and $\Delta > \ell + 6$, where this procedure leads to the following primary operator content:
\begin{equation}
\label{tab:longcontent}
\begin{array}{llllll}
 \text{[0 0]} & \text{[0 2]} & \text{[2 0]} & \text{[0 4]} & \text{[2 2]} & \text{[4 0]} \\
 \hline
 \text{($\Delta $)}_{\ell } & \text{($\Delta $ + 1)}_{\ell -1} & \text{($\Delta $ + 2)}_{\ell -2} & \text{($\Delta $ + 2)}_{\ell } & \text{($\Delta $ + 3)}_{\ell -1} & \text{($\Delta $ + 4)}_{\ell } \\
 \text{($\Delta $ + 2)}_{\ell -2} & \text{($\Delta $ + 1)}_{\ell +1} & \text{($\Delta $ + 2)}_{\ell } & \text{($\Delta $ + 4)}_{\ell -2} & \text{($\Delta $ + 3)}_{\ell +1} & \text{} \\
 \text{($\Delta $ + 2)}_{\ell }   & \text{($\Delta $ + 3)}_{\ell -3} & \text{($\Delta $ + 2)}_{\ell +2} & \text{($\Delta $ + 4)}_{\ell } & \text{($\Delta $ + 5)}_{\ell -1} & \text{} \\
 \text{($\Delta $ + 2)}_{\ell +2} & \text{($\Delta $ + 3)}_{\ell -1} & \text{($\Delta $ + 4)}_{\ell -2} & \text{($\Delta $ + 4)}_{\ell +2} & \text{($\Delta $ + 5)}_{\ell +1} & \text{} \\
 \text{($\Delta $ + 4)}_{\ell -4} & \text{($\Delta $ + 3)}_{\ell +1} & \text{($\Delta $ + 4)}_{\ell } & \text{($\Delta $ + 6)}_{\ell } & \text{} & \text{} \\
 \text{($\Delta $ + 4)}_{\ell -2} & \text{($\Delta $ + 3)}_{\ell +3} & \text{($\Delta $ + 4)}_{\ell +2} & \text{} & \text{} & \text{} \\
 \text{($\Delta $ + 4)}_{\ell }   & \text{($\Delta $ + 5)}_{\ell -3} & \text{($\Delta $ + 6)}_{\ell -2} & \text{} & \text{} & \text{} \\
 \text{($\Delta $ + 4)}_{\ell +2} & \text{($\Delta $ + 5)}_{\ell -1} & \text{($\Delta $ + 6)}_{\ell } & \text{} & \text{} & \text{} \\
 \text{($\Delta $ + 4)}_{\ell +4} & \text{($\Delta $ + 5)}_{\ell +1} & \text{($\Delta $ + 6)}_{\ell +2} & \text{} & \text{} & \text{} \\
 \text{($\Delta $ + 6)}_{\ell -2} & \text{($\Delta $ + 5)}_{\ell +3} & \text{} & \text{} & \text{} & \text{} \\
 \text{($\Delta $ + 6)}_{\ell }   & \text{($\Delta $ + 7)}_{\ell -1} & \text{} & \text{} & \text{} & \text{} \\
 \text{($\Delta $ + 6)}_{\ell +2} & \text{($\Delta $ + 7)}_{\ell +1} & \text{} & \text{} & \text{} & \text{} \\
 \text{($\Delta $ + 8)}_{\ell }   & \text{} & \text{} & \text{} & \text{} & \text{} \\
\end{array}
\end{equation}
This list contains precisely the conformal primaries in the superconformal multiplet $\LL[0,0]_{\Delta > \ell + 6,\,\ell \geqslant 4}$ that could appear in the self-OPE of $\Phi^{\{AB\}}(x)$. It is easy to check that spurious operators would be introduced by adding any additional $a^\text{at}_{\Delta,\ell}(z,\bar z)$ or $h_{\beta}^\text{at}(z)$, and so we arrive at the first entry of Table~\ref{tab:superblocks}. For $h^{\text{at}}_\beta(z)$ with $\beta = \ell + 4$ and $\ell \geqslant 0$ the same analysis yields:
\begin{equation}
\begin{array}{lll}
	\text{[0 0]} & \text{[0 2]} & \text{[2 0]}  \\
	\hline
	\text{($\ell $ + 4)}_{\ell } & \text{($\ell $ + 5)}_{\ell +1} & \text{($\ell $ + 6)}_{\ell +2} \\
	\text{($\ell $ + 6)}_{\ell +2} & \text{($\ell $ + 7)}_{\ell +3} & \text{} \\
	\text{($\ell $ + 8)}_{\ell +4} & \text{} & \text{}\\
\end{array}
\label{tab:shortcontent}
\end{equation}
Using the same logic we conclude that this has to be the superconformal block for a $\BB[0,0]_\ell$ multiplet. For the other entries in Table \ref{tab:superblocks} the analysis is analogous, though computationally more subtle: certain conformal multiplets vanish from \eqref{tab:longcontent} for $\Delta = \ell + 6$ and $\Delta = \ell + 4$, and in some cases cancellations occur between $a^\text{at}_{\ell+4,\ell}(z,\bar z)$ and $h_{\ell+4}^\text{at}(z)$.

It remains to justify our claim that the atomic functions given in \eqref{eq:ahatom} are unique. For the single-variable part this follows immediately from the structure of the chiral algebra \cite{Beem:2014kka}, since $h^{\text{at}}_\beta(z)$ is just the standard $\slf_2$ conformal block.
For the two-variable part we prove the claim by contradiction. Suppose there exists a hypothetical alternative atomic function $\tilde a^\text{at}(z,\bar z)$. This function is, by assumption, a building block for a superconformal block and so must admit a decomposition into a finite number of conformal blocks in each $R$-symmetry channel. We can assume in this decomposition there is no block in the $R$-symmetry channel with Dynkin labels $[4,0]$, because according to \eqref{eq:adecomp} we can always remove those by adding $a^{\text{at}}_{\Delta,\ell}(z,\bar z)$. (Although equation \eqref{tab:longcontent} is modified for low values of $\Delta$ and $\ell$, the function $a^{\text{at}}_{\Delta,\ell}(z,\bar z)$ continues to contribute exactly one block in this channel.) From the first equation in \eqref{eq:Ainah} we then find that
\begin{equation}
\Delta_\epsilon \left[ u^2 \tilde a^\text{at}(z,\bar z) \right] = 0~.
\end{equation}
The solution space to this equation can be parameterized in terms of Jack polynomials. Substituting these solutions in the second equation in \eqref{eq:Ainah} produces the operator content in the $[2,2]$ channel. For any nonzero $\tilde a^\text{at}(z,\bar z)$ this operator content will always  contain operators of twist zero or twist four with nonzero spin. The former are not allowed by unitarity and the latter do not appear in any of the superconformal multiplets allowed by the selection rules. We therefore conclude that $\tilde a^\text{at}(z,\bar z)$ cannot exist, and $a^\text{at}_{\Delta,\ell}(z,\bar z)$ and $h^\text{at}_\beta (z)$ are the only allowed building blocks.

The second and third entries in Table \ref{tab:superblocks} can be understood in terms of the decomposition of a long multiplet at the unitarity bound (see the second and last equations in \eqref{eq:decomposition}). The selection rules forbid superconformal multiplets of type $\AA$ to appear in the OPE, and this is corroborated by the vanishing of the OPE coefficient of the superconformal primary (\ie, the singlet operator $(\Delta)_\ell$ in \eqref{tab:longcontent}) precisely when $\Delta \to 6 + \ell$ in $a^{\text{at}}_{\Delta,\ell}(z,\bar z)$. The superconformal block for the long multiplet then smoothly becomes the superconformal block for the $\BB[0,2]_\ell$ (for $\ell > 0$) or $\DD[0,4]$ (for $\ell = 0$) multiplets at the unitarity bound.

The full list of superconformal blocks in Table \ref{tab:superblocks} allows us to trivially solve the first crossing equation in \eqref{eq:crossingeq} by restricting $\ell$ to be an even integer in both $a^{\text{at}}_{\Delta, \ell}(z, \bar z)$ and $h^\text{at}_\ell(z)$. This is just the manifestation of Bose symmetry at the level of these functions. We can also justify the assumption made in deriving \eqref{eq:G_sol}, namely that $h(z)$ admits a regular Taylor series expansion around $z=0$ with integer powers: this follows from the behavior of $h^{\text{at}}_\beta(z)$ near $z=0$.

\subsection{Solving for the short multiplet contributions}
\label{subsec:solving_short_multiplet_contribution}

From the solution for $h(z)$ given in \eqref{eq:h_sol} and the expression for superconformal blocks in terms of atomic blocks in Table \ref{tab:superblocks} we will be able to determine the full contribution to the four-point function of all short multiplets that contribute to $h(z)$. In principle there could be an ambiguity in this procedure since the contributions of some supermultiplets to $h(z)$ may cancel. However, multiplets of type $\BB[0,0]_{\ell}$ contain higher spin currents. These are the hallmark of a free theory and we will therefore impose their absence from the spectrum of operators. With those out of the way, the superconformal block interpretation of $h(z)$ is unambiguous.

The decomposition of $h(z)$ into atomic blocks takes the following form
\begin{eqnarray}
h(z)&=& h^{at}_{0}(z) + \sum_{\ell=-2\,, \ell~\mathrm{even}}^\infty b_\ell \; h^{at}_{\ell+4}(z)~,  \\
b_\ell&=&\frac{(\ell +1) (\ell +3) (\ell +2)^2 \frac{\ell }{2}! \left(\frac{\ell }{2}+2\right)!! \left(\frac{\ell }{2}+3\right)!! (\ell +5)!!}{18 (\ell +2)!! (2 \ell +5)!!} \nn \\
&& + \frac{8}{c}\frac{  \left(2^{-\frac{\ell }{2}-1} (\ell  (\ell +7)+11) (\ell +3)!! \Gamma \left(\frac{\ell }{2}+2\right)\right)}{(2 \ell +5)!!}~,\nn
\end{eqnarray}
where $b_{-2}$ should be thought of as the limit of the above expression as $\ell\rightarrow2$, which gives $b_{-2}= 8/c$. In this decomposition we have set the unphysical integration constant in \eqref{eq:h_sol} to $\b_5 = -1/6 + 8/c$.

The coefficients $2^{\ell} b_\ell$ with $\ell \geqslant 0$ are now the squared OPE coefficients of $\BB[2,0]_{\ell-2}$ and $\DD[4,0]$ multiplets. We can therefore split the two-variable function into two parts,
\begin{equation} 
\label{eq:adecomp}
a(z, \bar z) =a^{\chi}(z,\bar z) + a^{\mathrm{u}}(z,\bar z)~,
\end{equation}
where we now have
\begin{equation} 
a^{\chi}(z,\bar z)= \sum_{\ell=0,~\ell~\mathrm{even}}^\infty 2^{\ell} b_{\ell} \, a^\text{at}_{\ell+4,\ell}(z,\bar z)~.
\label{eq:ashort}
\end{equation}
and the unknown function $a^{\mathrm{u}}(z,\bar z)$ encodes the contribution of all the blocks on the first three lines of table \ref{tab:superblocks},
\begin{equation} 
a^{\mathrm{u}}(z,\bar z)= \sum_{\substack{\Delta \geqslant \ell+6,\\ \ell \geqslant 0,~\ell~\mathrm{even}}} \lambda_{\Delta, \ell}^2 \, a^\text{at}_{\Delta,\ell}(z,\bar z)~.
\label{eq:aunfixed}
\end{equation}
We note that $a^{\mathrm{u}}(z,\bar z)$ includes both the long multiplets and the $\BB[0,2]$ and $\DD[0,4]$ short multiplets, whose dimensions are protected but OPE coefficients unknown.

The crossing symmetry problem can now be put into its final form by substituting the decomposition \eqref{eq:adecomp} into \eqref{eq:crossingeq}:
\begin{equation}
\boxed{
\begin{aligned}
&z \bar z a^\mathrm{u}\left(z,\bar{z}\right)+ (z-1) \left(\bar{z}-1\right) a^\mathrm{u}\left(1-z,1-\bar{z}\right) = \\ &  \frac{1}{(z - \bar z)^3} \left( \frac{h\left(1-\bar{z}\right)-h(1-z)}{(z-1) \left(\bar{z}-1\right)}+\frac{h\left(\bar{z}\right)-h(z)}{z \bar{z}} \right) + (z-1) \left(\bar{z}-1\right) a^\chi\left(1-z,1-\bar{z}\right) - z \bar{z} a^\chi\left(z,\bar{z}\right)~.
\end{aligned}
}
\label{eq:finalcrossingeq}
\end{equation}
The right-hand side of this equation is known from \eqref{eq:h_sol} and \eqref{eq:ashort}, and depends on a single parameter $c$. The left-hand side is given by \eqref{eq:aunfixed} and features an infinite set of undetermined OPE coefficients and scaling dimensions. Equation \eqref{eq:finalcrossingeq} can now be exploited to constrain these parameters. This analysis can proceed either numerically or (using a particular limit) analytically. The aim of the next two sections is to present the details and the results of an extensive numerical analysis. Analytic results can be found in Appendix~\ref{App:lightcone}.

%% file: sections/5_numerics.tex

\section{Numerical methods}
\label{sec:numerics}

We will numerically analyze the crossing relation \eqref{eq:finalcrossingeq} to obtain bounds on the allowed CFT data. In this section we briefly review the now-standard approach to deriving such bounds. We first repackage the terms in the crossing equation as a sum rule:
\begin{equation} 
\label{eq:shortcrossingeq}
\sum_{\Delta, \ell} \lambda_{\Delta, \ell}^2 \AA^\text{at}_{\Delta, \ell} (z, \bar z) = \AA^\chi(z, \bar z; c)~.
\end{equation}
Here the sum runs over the unfixed spectrum of the theory, for which the OPE coefficients are also unknown, and we have defined
\begin{equation}
\label{eq:crossed_blocks}
\AA^\text{at}_{\Delta, \ell} (z, \bar z)= (z- \bar z)^3 z \bar{z}\; a^\text{at}_{\Delta,\ell}(z,\bar z) + (\bar z - z)^3 (1-z) \left(1-\bar{z}\right)a^\text{at}_{\Delta,\ell}(1-z,1-\bar z)~.
\end{equation}
$\AA^\chi(z, \bar z; c)$ denotes the right-hand side of \eqref{eq:finalcrossingeq}, which is completely fixed in terms of the central charge $c$.

Following \cite{Rattazzi:2008pe} we can put constraints on the allowed spectrum of operators by acting on the crossing equation with $\Rb$-valued linear functionals. The rough idea is to start with an \emph{ansatz} about the spectrum of operators, and then to try to rule out this ansatz by producing a nonzero linear functional such that
\begin{eqnarray}
\label{eq:func_dimbound}
\phi \left[ \AA^\chi(z,\bar z;c) \right] &\leqslant& 0~,\nn\\
\phi \left[ \AA^\text{at}_{\Delta, \ell} (z, \bar z) \right] & \geqslant & 0 \,, \quad  \forall\,(\Delta,\,\ell) \in \text{trial spectrum}~, 
\end{eqnarray}
A typical example of such an ansatz would be to pick a large number $\Delta_0 > 6$ and to assume that all multiplets of type $\LL[0,0]_{\Delta,0}$ have $\Delta > \Delta_0$. If we can produce a functional satisfying \eqref{eq:func_dimbound} then this ansatz is inconsistent with \eqref{eq:shortcrossingeq} and we conclude that the theory must  have a multiplet of type $\LL[0,0]_{\Delta,0}$ with $\Delta \leqslant \Delta_0$. Lowering $\Delta_0$ and repeating the process leads to a lowest possible $\Delta_0$ for which such a functional exists; this $\Delta_0$ is then the best upper bound. An analogous procedure can be used to produce upper bounds on the scaling dimensions of the lowest unprotected operators of higher spins.

In a similar vein, we can bound the squared OPE coefficients of a particular operator contributing as $\AA^\text{at}_{\Delta^\star, \ell^\star} (z, \bar z)$ to the crossing symmetry equation. Such bounds are obtained by performing the following optimization problem
\begin{align}
\text{Minimize } & \phi \left[ \AA^\chi(z,\bar z; c) \right] ~,\nn\\
& \phi \left[ \AA^\text{at}_{\Delta^\star, \ell^\star} (z, \bar z) \right]  = \,  1 ~, \label{eq:func_opebound} \\
& \phi \left[\AA^\text{at}_{\Delta, \ell} (z, \bar z) \right]  \geqslant\, 0 \,, \quad  \forall \{\Delta, \ell\}\neq \{\Delta^\star,\ell^\star\}  \in \text{trial spectrum}~. \nn
\end{align}
Denoting the minimum of the optimization by $\phi_{\rm min} \left[\AA^\chi (z,\bar z; c) \right]=M$, there is an upper bound on the squared OPE coefficient
\begin{equation} 
\lambda_{\Delta^\star, \ell^\star}^2 \leqslant M~.
\end{equation}
If the minimum is negative then this bound rules out the spectrum that went into the minimization problem entirely, because a negative upper bound is inconsistent with the unitarity requirement $\lambda_{\Delta^\star, \ell^\star}^2 \geqslant 0$. In other words, the functional in this case will satisfy the conditions of \eqref{eq:func_dimbound}. 

If it happens that the block  $\AA^\text{at}_{\Delta^\star, \ell^\star} (z, \bar z)$ is isolated, in the sense that it is not continuously connected to the set of blocks for which the functional is required to be positive in \eqref{eq:func_opebound}, lower bounds on the squared OPE coefficient of the corresponding operator can also be obtained. This is accomplished by instead fixing $\phi \left[ \AA^\text{at}_{\Delta^\star, \ell^\star} (z, \bar z) \right]  = - 1$. The same minimization problem then gives 
\begin{equation} 
\lambda_{\Delta^\star, \ell^\star}^2  \geqslant - M~.
\end{equation}
If  $M \geqslant 0$ this constraint is redundant, as unitarity already required the squared OPE coefficient to be non-negative.

Bounds on the central charge follow a similar recipe. The right hand side of \eqref{eq:shortcrossingeq} depends in a simple manner on the central charge:
\begin{equation}
\AA^\chi(z, \bar z; c) = \AA_1^\chi(z, \bar z) + \frac{1}{c}\AA_2^\chi(z, \bar z)~.
\end{equation}
We require the functional to be one on $\AA_1^\chi(z,\zb)$ and maximize the action of the functional on $-\AA_2^\chi(z,\zb)$. This produces a lower bound
\begin{equation}
c\geqslant\phi_{\rm max}[-\AA_2^\chi(z,\zb)]~.
\end{equation}

As has become the standard in the numerical bootstrap literature, we pick a basis of functionals consisting of derivatives evaluated at the crossing symmetric point $z=\bar{z}=\hf$. The functionals in our setup can therefore be written as
\begin{equation} 
\phi[f(z,\bar z)] = \sum_{m,n=0}^{\Lambda} \a_{mn} \del_z^m \del_{\bar z}^n f(z,\bar z) \big|_{z = \bar z = \hf}~.
\end{equation}
Within such a family we then search for real coefficients $a_{mn}$ that produce the best possible bound. We truncate the sum by only considering derivative combinations such that $m,n \leqslant \Lambda$. Another way of thinking about this truncation in the space of functionals is that the bounds for fixed $\Lambda$ only probe the \emph{truncated crossing equation} that arises by expanding \eqref{eq:crossingeq} in a Taylor series of order $\Lambda$ in $z$ and $\zb$ about $z=\zb=\frac12$. These bounds for each $\Lambda$ must be obeyed by physical theories, but as we increase $\Lambda$ we become sensitive to more and more of the structure of the crossing equation, so the bounds will improve. The symmetries of the functions appearing in \eqref{eq:shortcrossingeq} allow us to only consider $m \leqslant n$ with $m+n$ even. 

To find functionals we use linear and semi-definite programming techniques, which were pioneered in \cite{Rattazzi:2008pe} and \cite{Poland:2011ey}, respectively. For most of the results shown in this paper we have used the \texttt{IBM ILOG CPLEX} linear programming optimizer, interfaced with \texttt{Mathematica}.\footnote{As explained in, \eg, \cite{Rattazzi:2008pe}, the use of these linear programming techniques requires the discretization of the scaling dimensions appearing in the trial spectrum and truncation in spins. We have checked numerically that our results are not sensitive to refinements of these approximations.} This optimizer works with machine precision and is consequently quite fast, but in practice we find that it is limited to $\Lambda \lesssim 22$ before serious precision issues arise. Higher values of $\Lambda$ were needed for reliable extrapolations in Figs.~\ref{Fig:cbound} and \ref{Fig:extrapol_l0}. Those results were obtained using the semi-definite programming approach with the arbitrary precision solver \texttt{SDPB} \cite{Simmons-Duffin:2015qma}. Readers interested in the technical details of our numerical implementation -- such as the discretization used for linear programming, the degree of polynomial approximations used for the semi-definite approach, or \texttt{SDPB} parameter files -- should feel free to e-mail the authors.

Finally, let us note that the numerical bounds derived using $\texttt{CPLEX}$ in the following section are generally not entirely smooth for $\Lambda > 18$. Instead, we often find sparsely distributed outlier points. This is because machine precision is barely sufficient to obtain bounds with these values of $\Lambda$. Because these ``failed searches'' occur rather infrequently and the tendency of the bounds as a function of $c$ is still clearly distinguishable, we have included all the values up to $\Lambda = 22$ in our plots.\footnote{In several cases we have verified that our results do not significantly change when we repeat the analysis at arbitrary precision with \texttt{SDPB} \cite{Simmons-Duffin:2015qma}.}

%% file: sections/6_results.tex
\section{Results}
\label{sec:results}

The undetermined CFT data that appear in our crossing symmetry relation amount to:
\begin{itemize}
	\item OPE coefficients $\lambda^2_{\DD[0,4]}$ and $\lambda^2_{\BB[0,2]_{\ell}}$ of the short multiplets that are unfixed by the chiral algebra.
	\item Scaling dimensions $\Delta_\ell$, $\Delta_\ell'$, \ldots~and OPE coefficients $\lambda_\ell$, $\lambda_\ell'$, \ldots~of all long ($\LL[0,0]_{\Delta,\ell}$) multiplets.
\end{itemize}
In the following subsections we present numerical results which constrain a subset of these parameters. These constraints will in all cases depend on the $c$ central charge of the theory, which (as we explained previously) enters the crossing symmetry equation \eqref{eq:finalcrossingeq} via the coefficients of the predetermined multiplets. However, we begin by investigating a more elementary question:
\begin{itemize}
	\item[] \emph{Are all positive values of $c$ consistent with crossing symmetry, unitarity, and the absence of higher spin currents?}
\end{itemize}
As we explain below, the existence of a lower bound for $c$ has profound implications for the $A_1$ theory.

\input{sections/6_1_cbound}
\input{sections/6_2_opecoeffbounds}
\input{sections/6_3_dimbounds}
\input{sections/6_4_a1bounds}

%% file: sections/6_1_cbound.tex

\subsection{Central charge bounds}
\label{subsec:cbounds}

Let us start by asking which values of the $c$ central charge are allowed in unitary theories. The numerical methods presented in the preceding section allowed us to obtain a lower bound for $c$. Our best numerical result can be summarized as the following
\begin{result*}
Every unitary and local six-dimensional $(2,0)$ superconformal theory without higher spin currents must have $c > 21.45$.
\end{result*}
This bound was obtained with $\Lambda = 59$.\footnote{Recall from Section \ref{sec:numerics} that the dimension of the search space increases $\propto \Lambda^2$. Consequently, for larger values of $\Lambda$ we obtain better bounds but at a greater computational cost. This best result with $\Lambda=59$ had a search space of dimension $870$.} While this result is a welcome discovery, the bound is still a ways off from the lowest central charge of any of the known theories -- namely $c=25$ for the $A_1$ theory.

\begin{figure}[t!]
             \begin{center}           
              \includegraphics[scale=0.4]{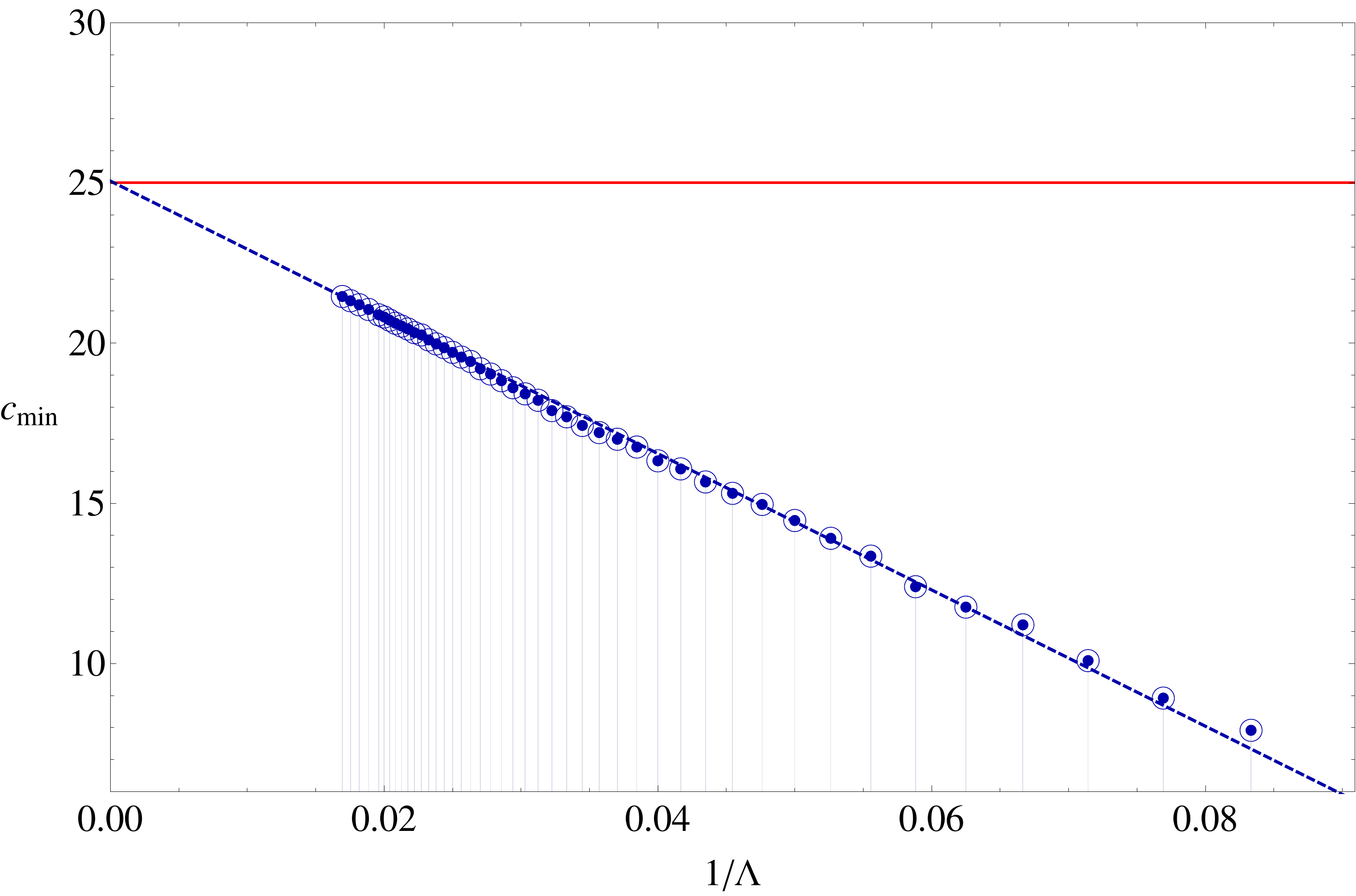}
              \caption{Bound on the central charge $c$ as a function of $1/\Lambda$, which is a good proxy for the numerical cost of the result. Central charges below the data points are excluded. The dotted line shows a linear extrapolation, which indicates that with infinite numerical power the lower bound  converges to $c\simeq25$. This is precisely the value for the $A_1$ theory as indicated by the horizontal line.}
              \label{Fig:cbound}
            \end{center}
\end{figure}

However, we have found that the behavior of this lower bound as a function of $\Lambda$ is incredibly regular. This is shown in Fig.~\ref{Fig:cbound}, which contains -- in addition to the value quoted above -- also a large number of data points corresponding to bounds for lower values of $\Lambda$. From the figure it is clear that our best lower bound has not yet converged as a function of $\Lambda$, and is likely to improve substantially if $\Lambda$ is increased significantly. However the most obvious feature of Fig.~\ref{Fig:cbound} is that the data points display an extraordinarily linear dependence on $1/\Lambda$. We can use a linear fit extrapolate to infinite cutoff, leading to the prediction that the bound converges to $c \simeq 25$ as $\Lambda \to \infty$.\footnote{As an example, we find $c_{\rm min} \simeq 25.06 - 212.7 \Lambda^{-1}$ from an ordinary least-squares regression through the last ten data points.} This is our first indication that our numerical study of the crossing symmetry constraints is more than a mathematical exercise: the value $c=25$ is precisely the value corresponding to the $A_1$ theory!

We should note that the approximately linear behavior for large $\Lambda$ in Fig.~\ref{Fig:cbound} is a little surprising, because there is currently no precise theory that parametrizes the $\Lambda$ dependence of the bounds even in the asymptotic regime. Nevertheless it seems entirely plausible to state the following
\begin{conj*}
The lower bound on $c$ converges exactly to $25$ as $\Lambda \to \infty$.
\end{conj*}
The validity of this conjecture would have important physical consequences. The numerical problem of finding a lower bound on $c$ has a \emph{dual} formulation where one finds a solution to the truncated crossing symmetry equations rather than a functional. Solving this dual problem is equivalent to proving that a functional does not exist and vice versa. It was pointed out in \cite{Poland:2010wg,ElShowk:2012hu} and used extensively in \cite{El-Showk:2014dwa} that at the lowest possible value of $c$ this dual solution is \emph{unique}, and in all known cases it appears to converge to a complete crossing symmetric four-point function. In our case, this uniqueness implies the following corollary to our conjecture:

\begin{cor*}
For a unitary $(2,0)$ superconformal theory with $c = 25$ and without higher spin currents there is a \emph{unique} crossing symmetric four-point function of the stress tensor multiplet.
\end{cor*}

Therefore, at the level of this correlation function, the $A_1$ theory can be completely bootstrapped. We emphasize that the determination of a single correlation function is no small feat: it contains information about infinitely many scaling dimension of unprotected operators (in this case the $R$-symmetry singlet operators of even spin) and their OPE coefficients. There would then be little room for the other crossing symmetry equations to exhibit any freedom whatsoever, and in this scenario the full $A_1$ theory is likely to be nothing more than the unique solution of the crossing symmetry equations at $c = 25$! In Subsection \ref{subsec:a1bounds} we will investigate the possibilities for bootstrapping the $A_1$ theory in more detail, and discuss what we can learn about this theory with finite numerical precision.

%% file: sections/6_2_opecoeffbounds.tex

\subsection{Bounds on OPE coefficients}
\label{subsec:opebounds}

In this section we present bounds on the OPE coefficients of the short multiplets ($\DD[0,4]$ and $\BB[0,2]_\ell$, with $\ell=1,3,\ldots$) that are not determined by the chiral algebra. In Appendix \ref{App:lightcone} we derive from crossing symmetry that the $\BB[0,2]_\ell$ multiplets must be present for all sufficiently large $\ell$. To study the low-spin cases we must use numerical methods. Here we focus on the $\DD[0,4]$, $\BB[0,2]_{\ell=1}$, and $\BB[0,2]_{\ell=3}$ multiplets.

\subsubsection{$\DD[0,4]$ OPE coefficient bounds}
\label{subsubsec:D04_OPE_bounds}

\begin{figure}[t!]
             \begin{center}           
              \includegraphics[scale=0.35]{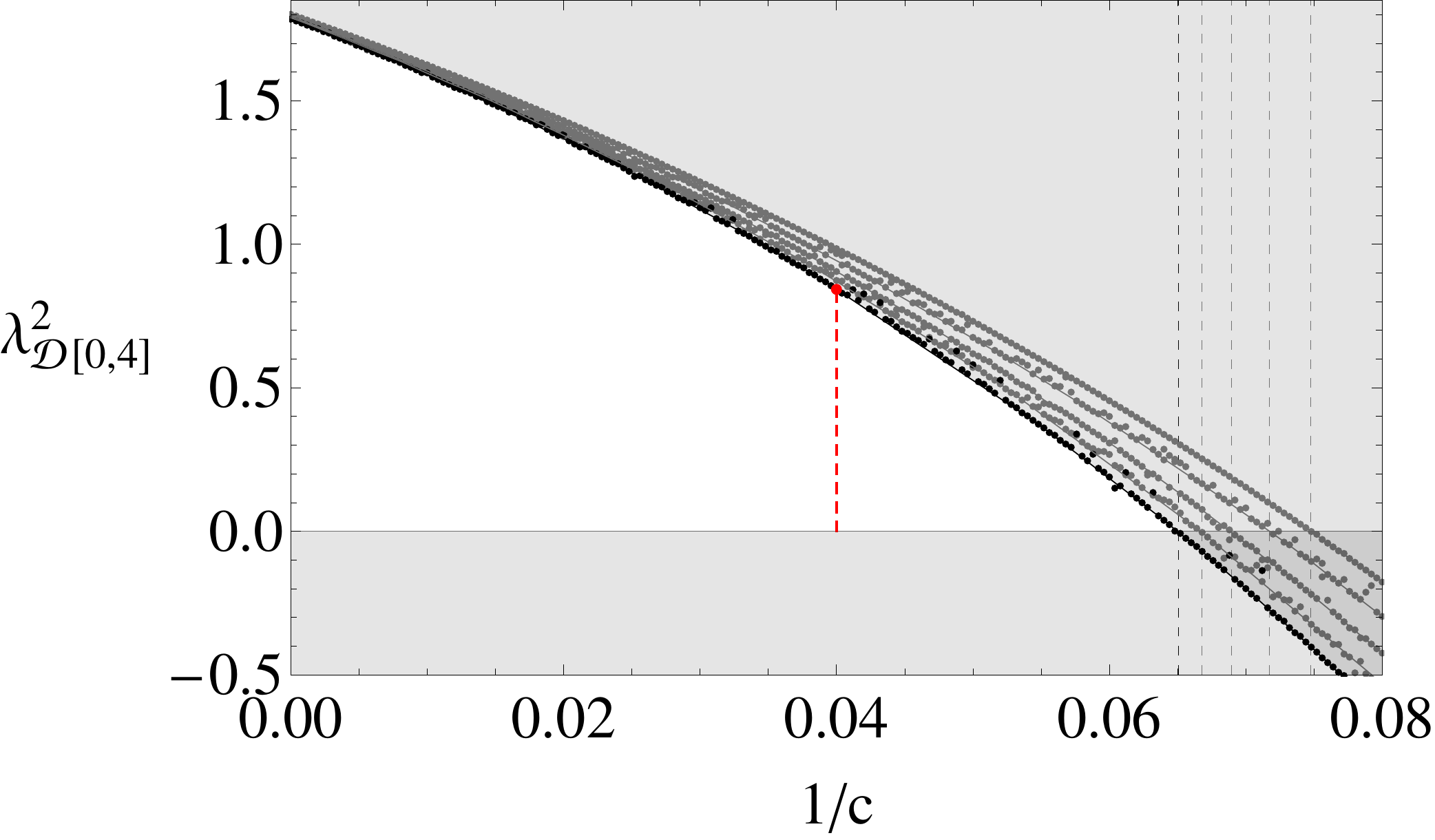} \includegraphics[scale=0.35]{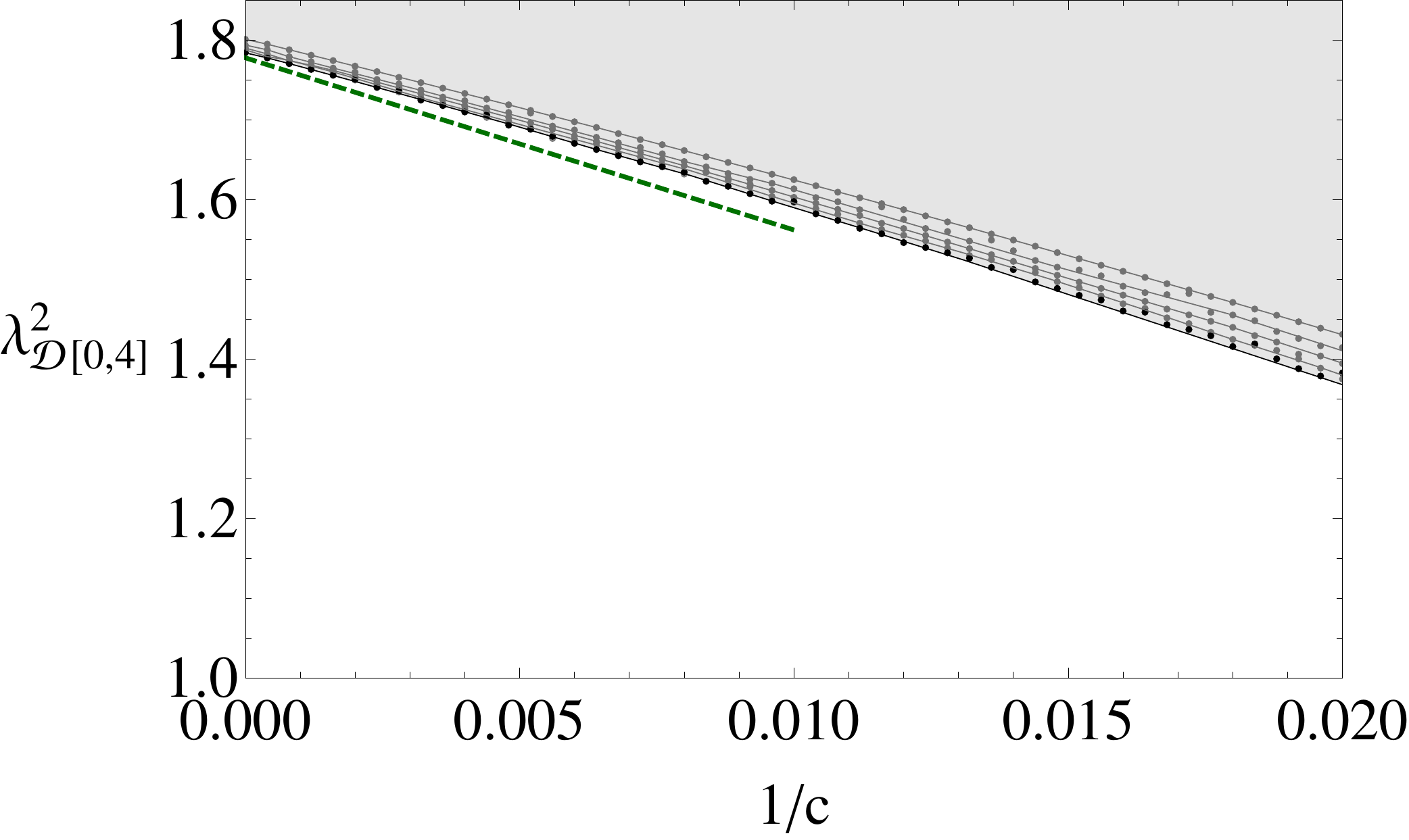}
              \caption{Upper bound on the OPE coefficient squared of the $\DD[0,4]$ multiplet as a function of the inverse central charge $c$ for $\Lambda=18,\ldots,22$, with the strongest bound shown in black. The shaded region is excluded by the numerics and unitarity ($\lambda_{\DD[0,4]}^2 \geqslant 0$). The red vertical line corresponds to $c=25$, the central charge of the $A_1$ theory. The vertical dashed lines denote the minimum allowed central charge $c_{\rm min}(\Lambda)$ from Fig.~\ref{Fig:cbound} for the same values of $\Lambda$. The right plot is a magnification of the large central charge region. The dashed green line is the prediction from supergravity including the first $1/c$ correction \eqref{eq:OPE_large_k_small_l}.
              }
            	\label{Fig:OPE_D_bound}
            \end{center}
\end{figure}

In Fig.~\ref{Fig:OPE_D_bound} we show upper bounds for the (squared) OPE coefficient of the $\DD[0,4]$ multiplet as a function of the central charge. Unitarity requires $\lambda_{\DD[0,4]}^2 \geqslant 0$, so the squared OPE coefficient is restricted to the unshaded region of Fig.~\ref{Fig:OPE_D_bound}. Of particular interest are the regions at small and large central charge. Let us first discuss the small central charge regime. The upper bound crosses zero for small values of the central charge, and to the right of these crossing points the upper bound is negative so any consistent solution to the crossing symmetry equations is forbidden. These crossing points thus translate into a lower bound for $c$, which are precisely the lower bounds from Fig.~\ref{Fig:cbound}. We then see that the vanishing of $\lambda^2_{\DD[0,4]}$ is in some sense responsible for the lower bound on $c$, and a solution to crossing at $c_{\rm min}$ cannot have a $\DD[0,4]$ multiplet appearing in this OPE. We saw in Section \ref{sec:strategy} that the $\DD[0,4]$ multiplet is absent only from the $A_1$ theory, so there is a nice consistency between this result for $\lambda^2_{\DD[0,4]}$ and our conjecture above regarding the asymptotic value of $c_{\rm min}(\Lambda)$. For good measure, we also report that without extrapolation to $\Lambda \to \infty$ these results give a rigorous bound $0 \leqslant \lambda^2_{\DD[0,4]} \leqslant 0.843$ for the $A_1$ theory -- this is the red interval in Fig.~\ref{Fig:OPE_D_bound}.

The large central charge limit is shown in detail on the right plot of Fig.~\ref{Fig:OPE_D_bound}. The supergravity solution lies below our upper bound, which is an important consistency test of our numerics. What is more striking is that the two results are so close. For infinite $\Lambda$ the numerical result may well coincide with the supergravity result at very large $c$. This provides another strong indication that the numerical analysis can indeed ``mine'' the crossing symmetry constraints and recover the physics of the true $(2,0)$ theories, as opposed to simply producing bounds.\footnote{A similar match between supergravity results and numerical bounds, including $1/c$ corrections, was observed in \cite{Beem:2013qxa} for $\NN = 4$ super Yang-Mills.}

For intermediate values of $c$ the $\DD[0,4]$ multiplets should be present for all $c > 25$. This is consistent with our bounds, which we expect to remain strictly positive in this region. In the limit $\Lambda \to \infty$ one would hope that the bounds will again be saturated by the other known $(2,0)$ theories, simply because these are the only known solutions to crossing symmetry that can prevent the bounds from decreasing even further. This would imply that the deviations of our bound from the straight-line behavior at large central charges should correspond quantum M-theoretic corrections to eleven-dimensional supergravity.

\subsubsection{$\BB[0,2]$ OPE coefficient bounds}
\label{subsubsec:B02_OPE_bounds}

\begin{figure}[t!]
             \begin{center}
             \includegraphics[scale=0.36]{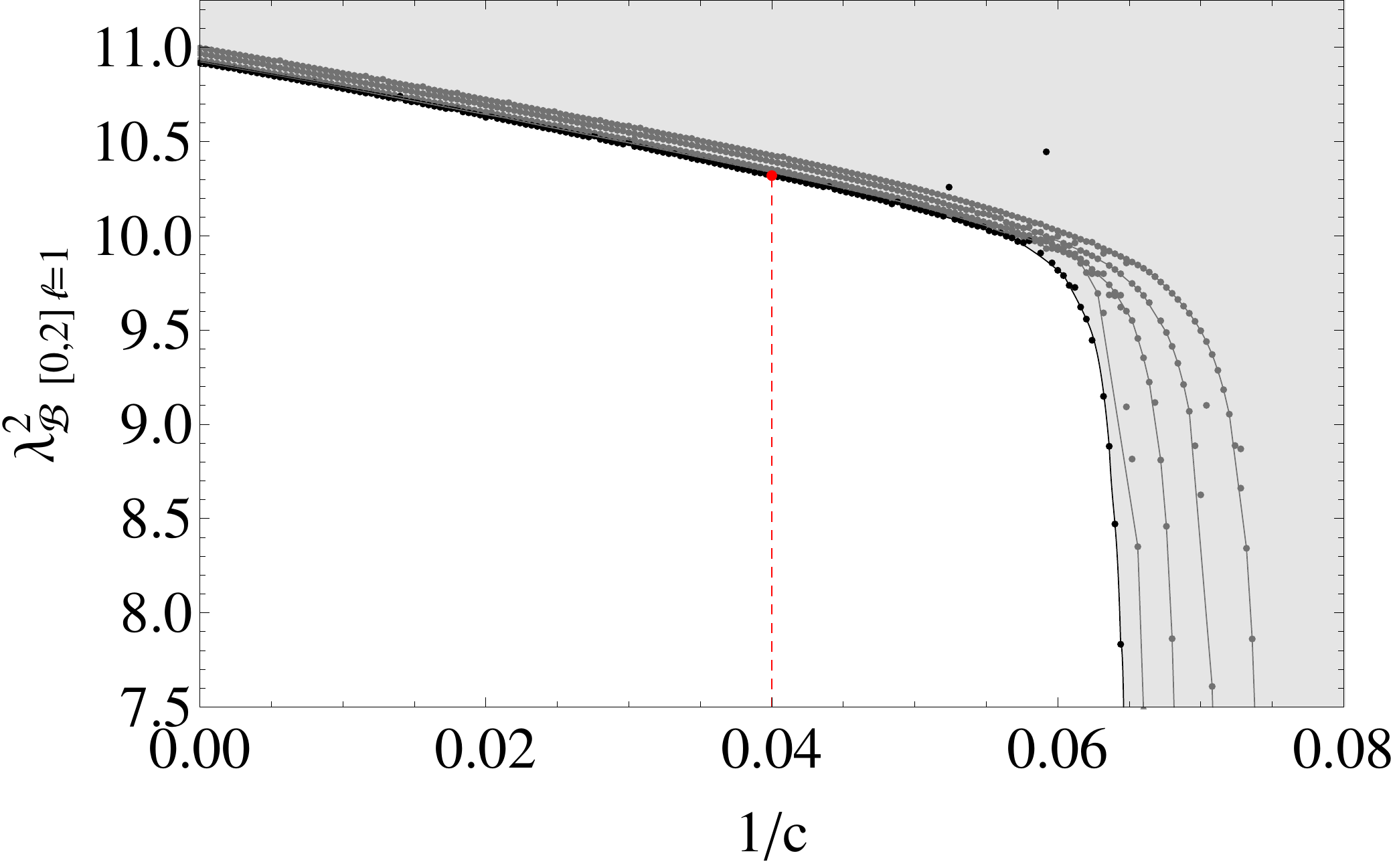}
             \includegraphics[scale=0.36]{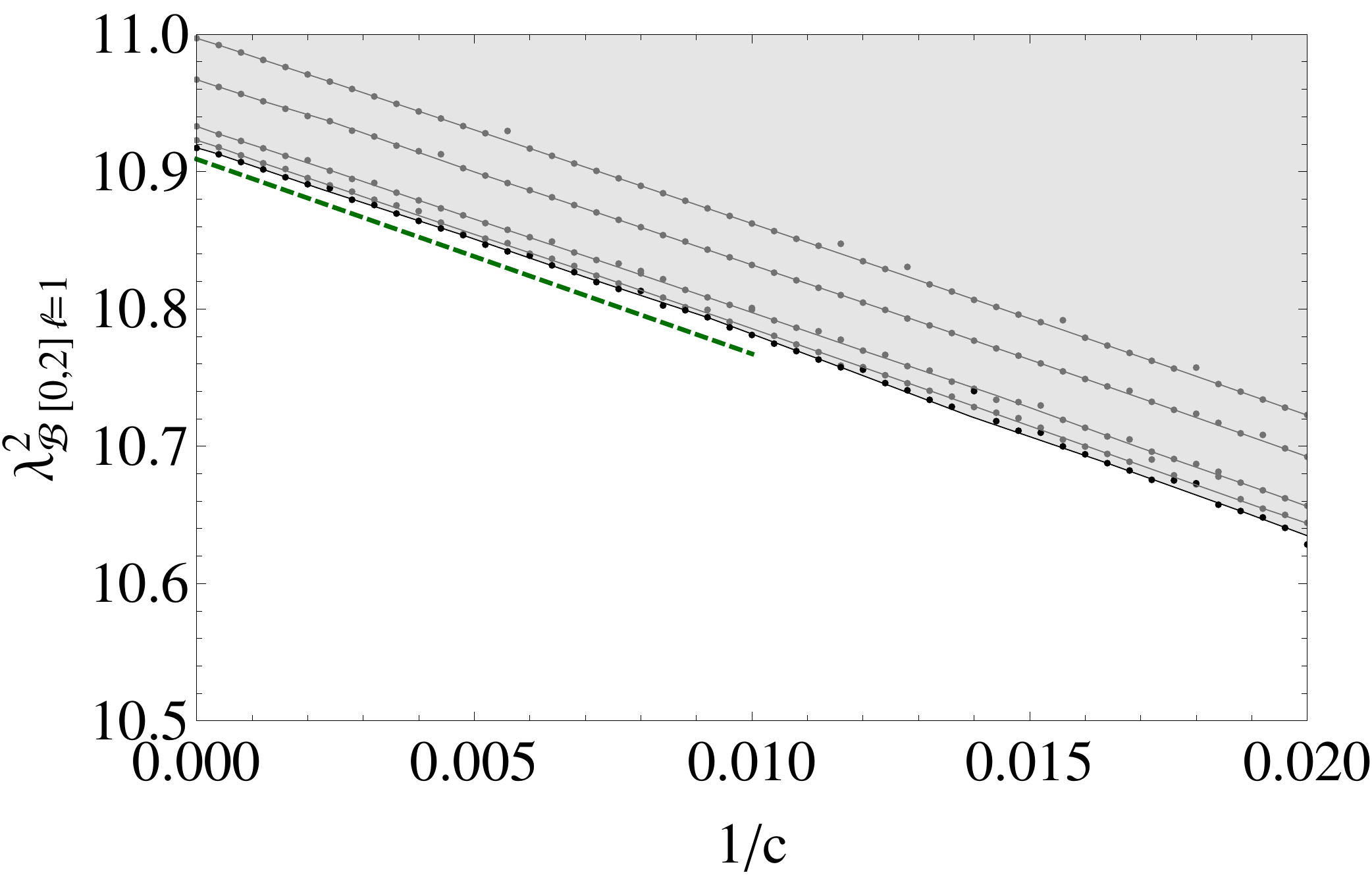}
             \includegraphics[scale=0.36]{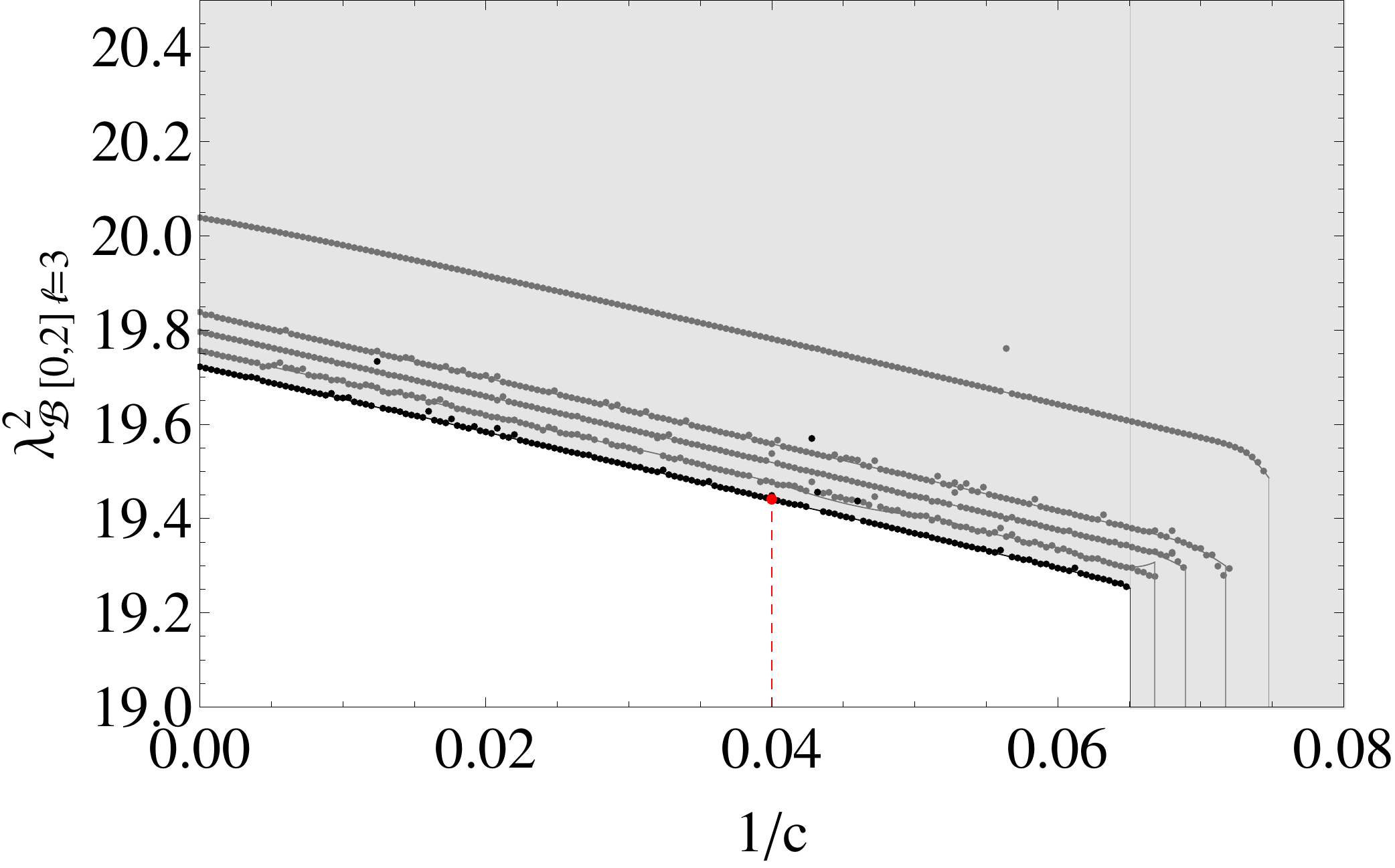}
             \includegraphics[scale=0.36]{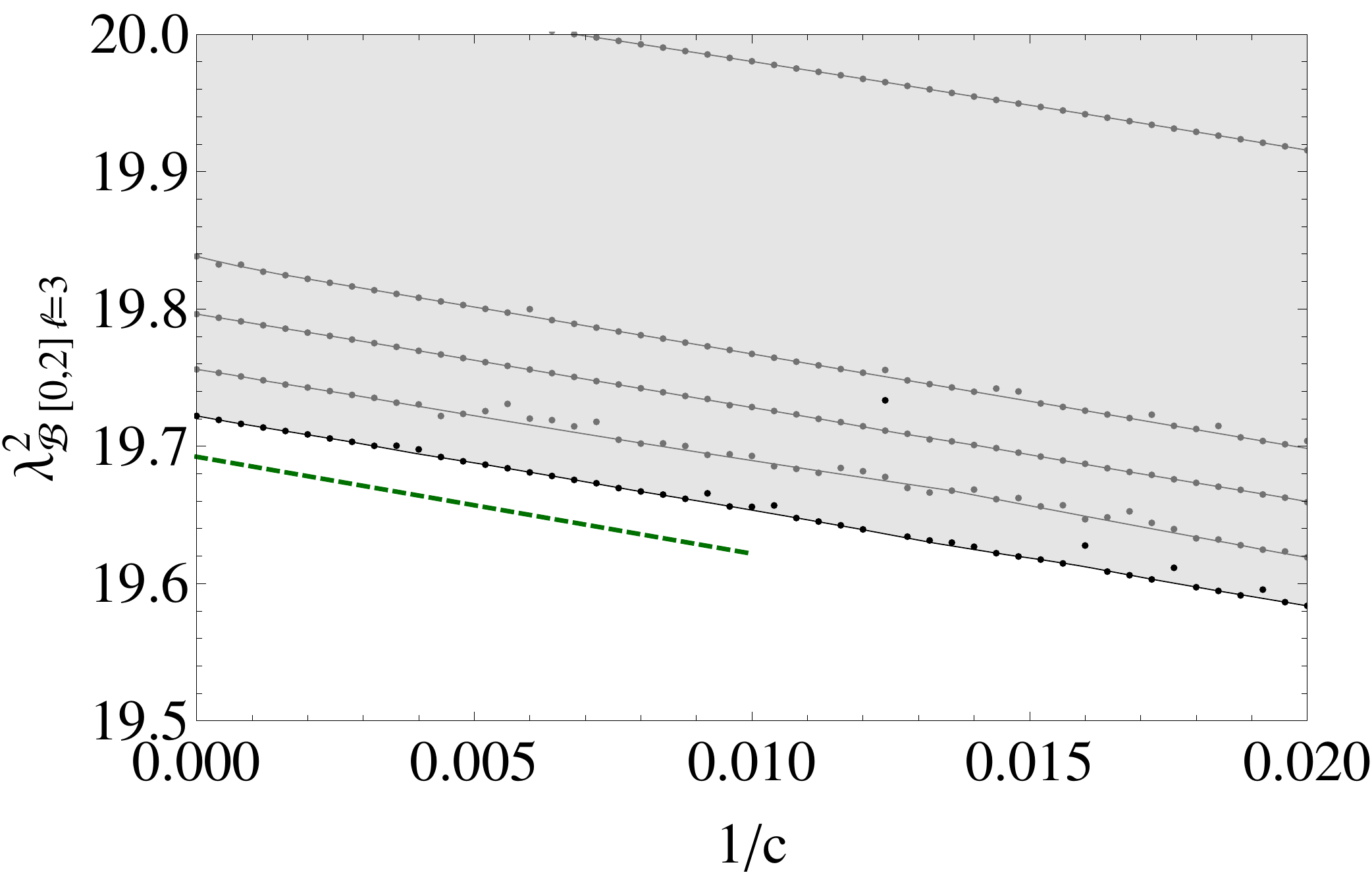}
             \caption{Upper bounds on the squared OPE coefficients of $\BB[0,2]_{\ell}$ multiplets with $\ell=1$ (top) and $\ell = 3$ (bottom) as a function of the inverse central charge $c$. The different curves correspond to different values of $\Lambda=18,\ldots,22$, with the black curve representing the strongest bound. The red vertical line represents the $A_1$ theory. The right plots are magnified at very large central charge. The dashed green line corresponds to the  supergravity answer quoted in \eqref{eq:OPE_large_k_small_l}.}
             \label{Fig:OPE_B_bounds}
             \end{center}
\end{figure}

In Fig.~\ref{Fig:OPE_B_bounds} we show upper bounds for the (squared) OPE coefficients of the $\BB[0,2]_1$ and $\BB[0,2]_3$ multiplets. The behavior of the bounds near $c_{\rm min}$ can be seen in the left two plots. General consistency of the bounds implies that they have to be zero when $c < c_{\rm min}$, but the way they approach zero is rather different both from the $\DD[0,4]$ case and from one another. The $\ell = 1$ bound tends to zero sharply but relatively smoothly, whereas the $\ell = 3$ bound displays genuine step function behavior. This brings up a certain subtlety concerning the solution to crossing symmetry at $c_{\rm min}$. As has been investigated in, \eg, \cite{ElShowk:2012hu}, the extremal solution at finite $\Lambda$ is unique but only an approximate solution of the full crossing equation. For the case $\ell = 3$, the step function behavior indicates that the corresponding multiplet is present in the approximate solution with a coefficient that is given by its value at the kink -- so approximately $19.25$ for $\Lambda = 22$. This number should decrease somewhat as $\Lambda$ increases, but will probably stay finite. The $\ell = 1$ bound, on the other hand, is strictly speaking equal to zero at $c_{\rm min}$ and the corresponding multiplet is absent from the extremal approximate solution. But the bound increases sharply as we move away from $c_{\rm min}$ up to a value of $~10$. It may well be the case that this bound will ultimately develop the same step function behavior as observed for $\ell = 3$ as $\Lambda$ is increased. In that case the absence of the $\ell = 1$ multiplet in the approximate solution would be a numerical artifact, and the true $\Lambda\to\infty$ extremal solution would include such a multiplet with a coefficient of~$\sim 10$.

The large central charge behavior is shown in the plots on the right of Fig.~\ref{Fig:OPE_B_bounds} with dashed lines indicating the supergravity results \cite{Heslop:2004du}. The numerical bounds converge very well towards the supergravity results, confirming once more that these bounds are sensitive to the physics of the actual $(2,0)$ theories.

It follows from the general analysis in appendix \ref{App:lightcone} that for $\ell \to \infty$ the OPE coefficients $\lambda^2_{\BB[0,2]_\ell}$ will converge to a linear function of $1/c$ for all the allowed values of $c$, with coefficients that can be extracted from the supergravity solution. The bottom left plot in Fig.~\ref{Fig:OPE_B_bounds} corresponds to $\ell = 3$ but is already strikingly linear. For higher $\ell$ we expect an even better match with the lightcone prediction.

%% file: sections/6_3_dimbounds.tex

\subsection{Bounds on scaling dimensions}
\label{subsec:deltabounds}

In this subsection we turn our attention to the long multiplets. For the four-point function under consideration these multiplets are necessarily of type $\LL[0,0]_{\Delta,\ell}$ with $\ell$ even. We will be solely concerned with the quantum number $\Delta$ -- the scaling dimensions of these multiplets. An investigation of the OPE coefficients for long multiplets is left for future work.

\subsubsection{Scalar operators}
\label{subsubsec:scalar_long_bounds}

\begin{figure}[t!]
             \begin{center}           
              \includegraphics[scale=0.35]{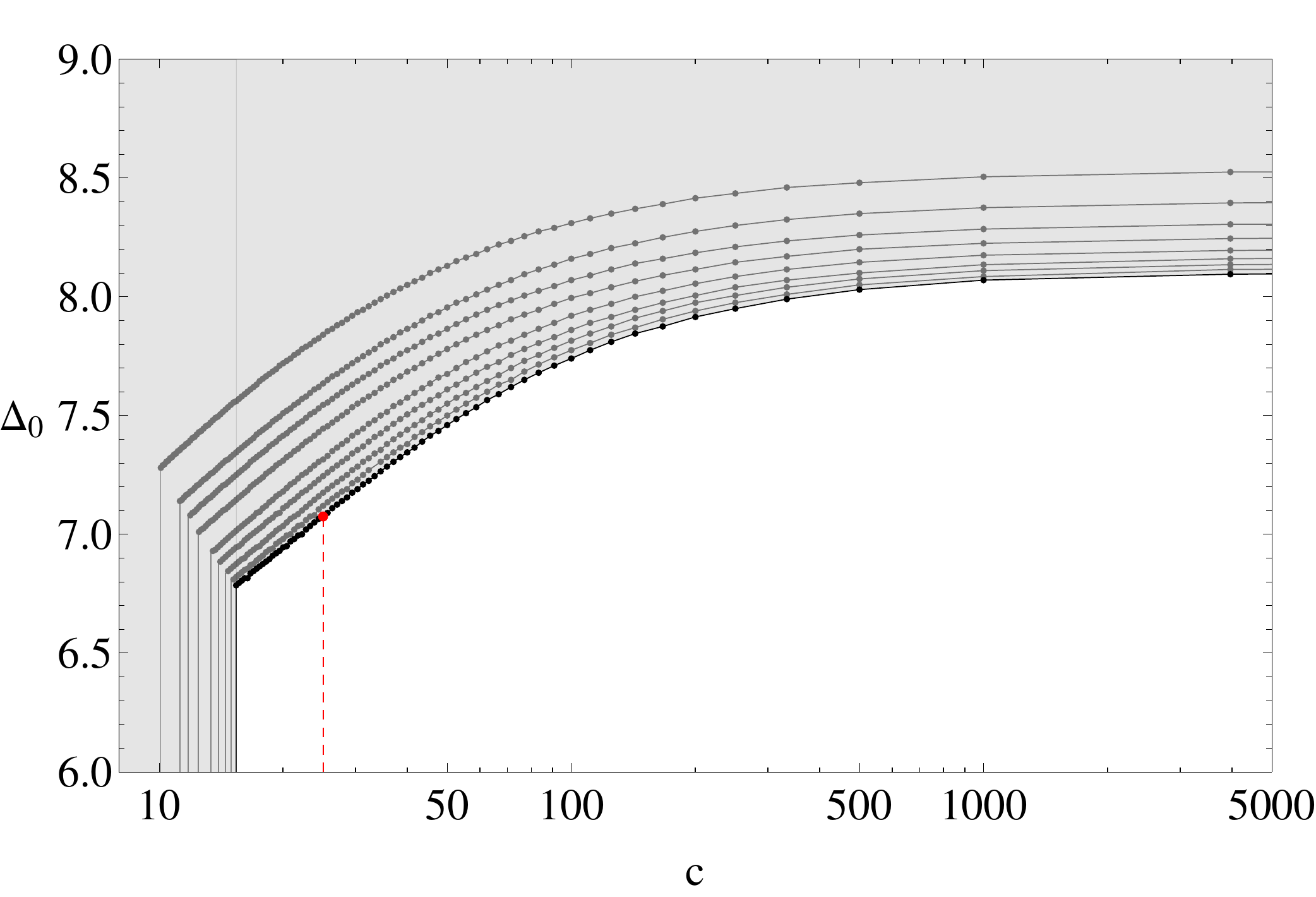} \includegraphics[scale=0.35]{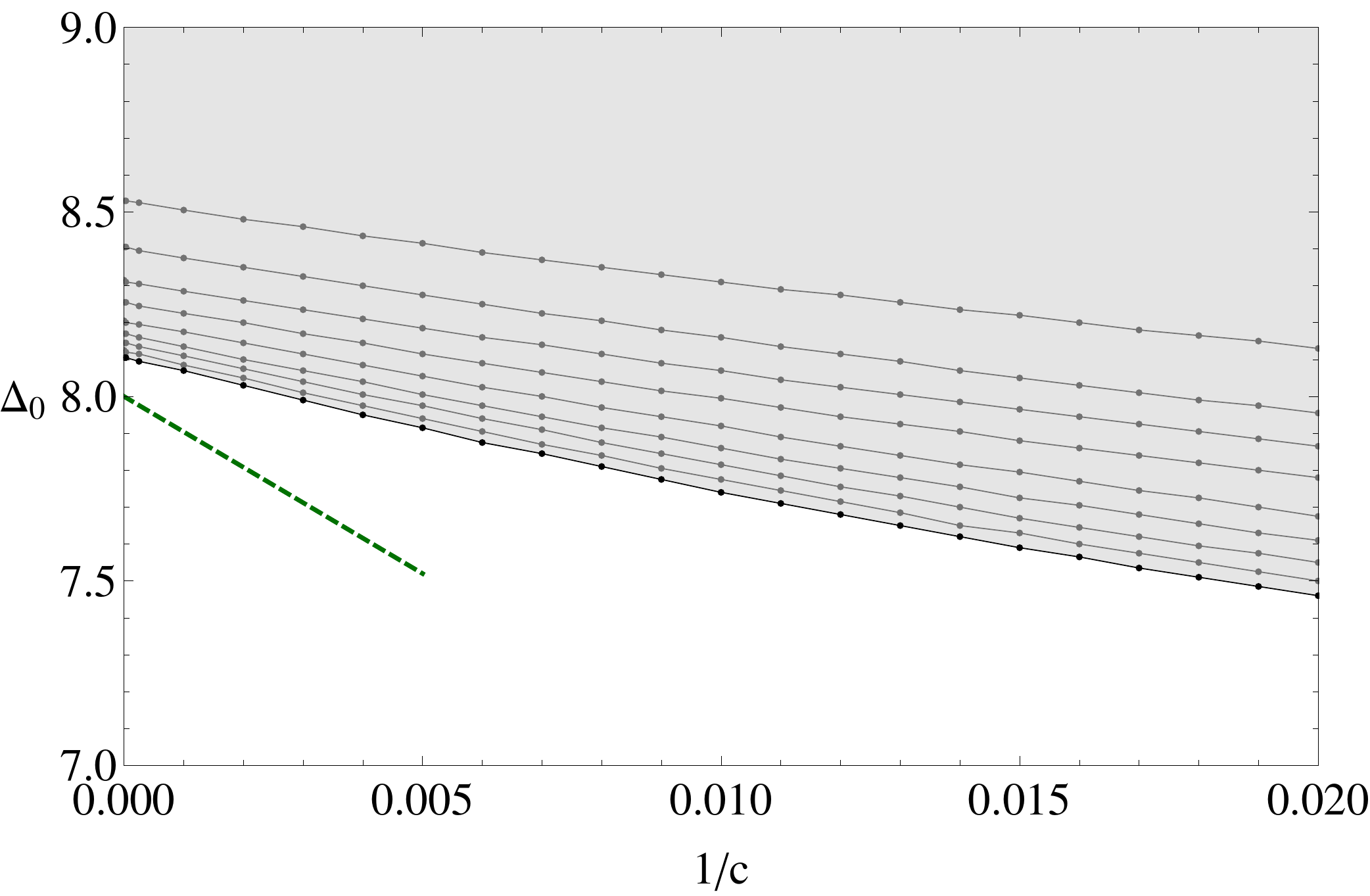}
              \caption{Upper bound for the dimension of the first long scalar multiplet. The different curves correspond to $\Lambda=18,\ldots,22$, with the black curve representing the strongest bound. The shaded region is excluded by the numerics. The vertical red line is located at the central charge of the $A_1$ theory. On the right we display the bound for very large $c$, with the green dashed line corresponding to the supergravity result \eqref{DTsugra}.}
              \label{Fig:long_scalar_bound}
            \end{center}
\end{figure}

In Fig.~\ref{Fig:long_scalar_bound} we present upper bounds on the dimension $\Delta_0$ of the first long scalar multiplet. We recall that unitarity of the corresponding representation of the superconformal algebra requires that $\Delta_0 \geqslant 6$. Below the value $c_{\rm min}(\Lambda)$ there can be no solution to crossing symmetry, so we have a sharp cutoff at that value for each $\Lambda$.

With $\Lambda=22$ we find an upper bound of approximately $7.08$ for the $A_1$ theory at $c=25$ that increases monotonically with $c$ until reaching a value of approximately $8.11$ at infinite central charge.\footnote{As discussed in \cite{Beem:2014zpa} this monotonicity is a generic property of the kind of bounds studied here.} The latter value is quite close to the generalized free-field solution at $\Delta_0 = 8$, to which it presumably would converge at higher $\Lambda$. The leading $1/c$ behavior obtained from supergravity, while consistent with the bound, does not appear to follow it very closely. We believe that this is simply an artifact of slow convergence in the scalar sector.\footnote{A similar pattern was observed for $4d$ $\NN=4$ SCFTs in \cite{Beem:2013qxa}: the bounds are saturated with much better accuracy for spin greater than zero.} Indeed, the large $c$ behavior shown in Figs.~\ref{Fig:OPE_D_bound} and \ref{Fig:OPE_B_bounds}, and also in Figs.~\ref{Fig:long_spin2_bound} and \ref{Fig:long_spin4_bound} below, suggests that as $\Lambda \to \infty$ the bounds will be exactly saturated by the supergravity result. This is the most natural option from a physical perspective because we do not expect any other theories to exist at very large central charge.

For intermediate values of $c$ we have upper bounds for $\Delta_0$ that are valid for all the physical $(2,0)$ theories. It is again natural to suspect that these bounds will be saturated by the actual theories and in this way the bounds actually offer a (very rough) estimate of the actual scaling dimensions. For example, we see that the $(A_2, A_3, A_4)$ theories should have unprotected $\LL[0,0]_{\Delta,0}$ scalar multiplets with primaries of dimensions $\Delta_0 \lesssim (7.7, 7.9, 8.0)$, respectively. (For the $A_1$ theory we provide a more refined estimate below.) We emphasize that these are the first estimates of unprotected operator dimensions in the $(2,0)$ theories. It would be very interesting if they could be verified through other means.

\begin{figure}[ht!]
             \begin{center}           
              \includegraphics[scale=0.4]{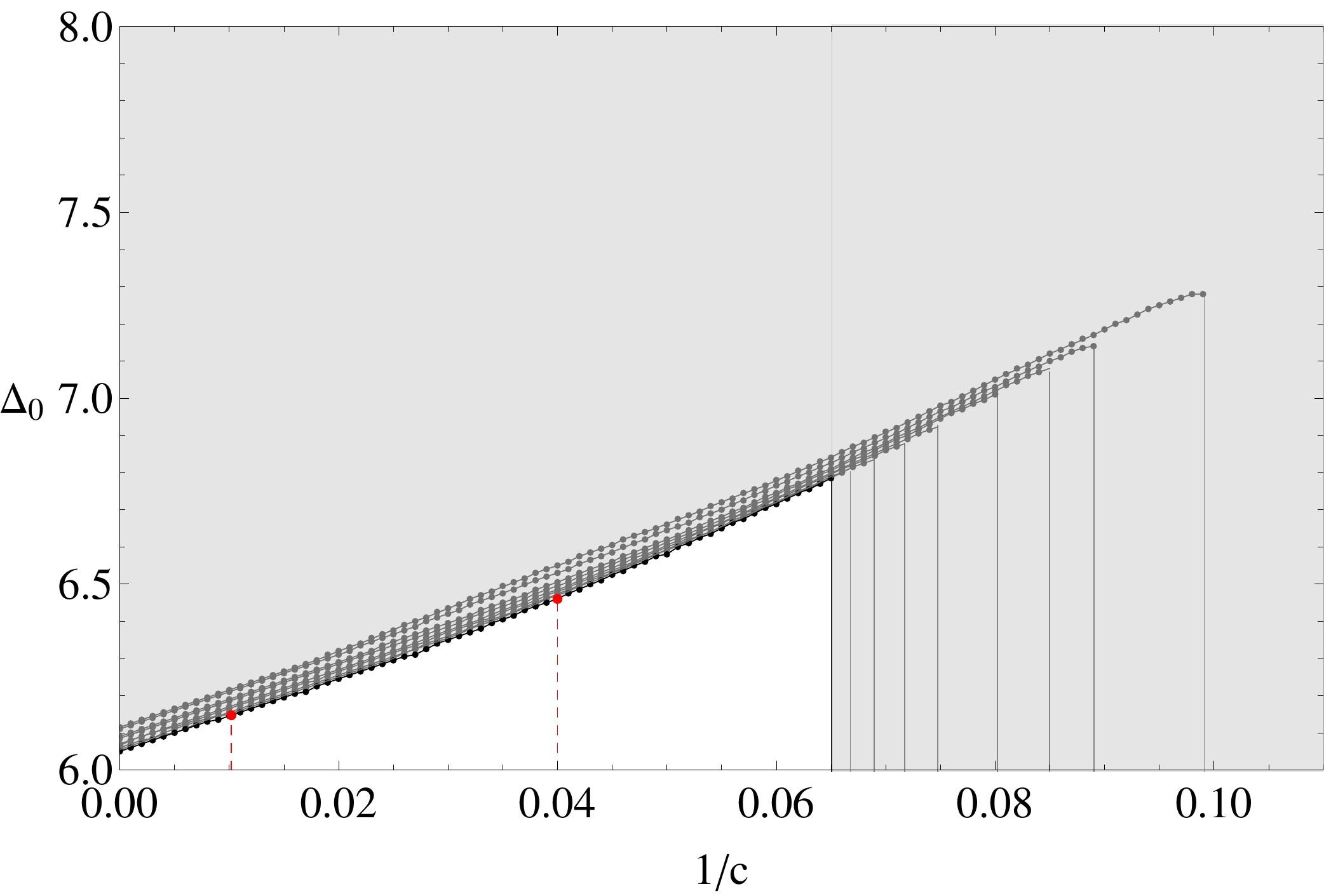}
              \caption{Bound on the dimension of the first spin zero long multiplet as a function of the inverse central charge $c$, with the $\DD[0,4]$ short multiplet excluded from the spectrum. The different bounds correspond to $\Lambda=14,15,\ldots22$. The vertical red lines marks the central charges of the $A_1$ and $A_2$ theories.}
              	\label{Fig:long_scalar_bound_noD}
            \end{center}
\end{figure}

\subsubsection*{Removing the \texorpdfstring{$\DD[0,4]$}{D[0,4]} multiplet}

The bounds in Fig.~\ref{Fig:long_scalar_bound} were obtained without imposing any constraints (besides non-negativity) on the OPE coefficient $\lambda^2_{\DD[0,4]}$. However, since the $\DD[0,4]$ short multiplet is absent from the $A_1$ theory, we can give our bounds for the $A_1$ theory a boost by imposing by hand that $\lambda^2_{\DD[0,4]}=0$ and re-computing the upper bound $\Delta_0$. This also gives us some insight into the possibility of additional theories beyond the known $A_1$ theory that may have no $\DD[0,4]$ in the spectrum. 

The resulting plot is shown in Fig.~\ref{Fig:long_scalar_bound_noD}. For small $c$ we see that we can get by without the $\DD[0,4]$ multiplet -- crossing symmetry can easily be satisfied as long as the theory has an unprotected operator in the unshaded region. Precisely at $c=c_{\rm min}$ the bounds in Figs.~\ref{Fig:long_scalar_bound_noD} and \ref{Fig:long_scalar_bound} coincide, since we already know that $\lambda^2_{\DD[0,4]} = 0$ at $c_{\rm min}$. For larger $c$ the bound falls off quickly and approaches the unitarity bound $\Delta_0 = 6$. An extrapolation of the bounds at $c = 98$ (corresponding to the $A_2$ theory) suggests that for $\Lambda \to \infty$ the bound will end up at $\Delta \approx 6$. For higher $c$ the rate of convergence is even better. Since the contribution from the $\DD[0,4]$ multiplet is exactly the same as that of a long multiplet at $\Delta=6$, we may then conclude that one \emph{must} re-introduce the $\DD[0,4]$ multiplet for $c \geqslant 98$ in order to satisfy crossing symmetry. In this sense crossing symmetry dictates the presence of these multiplets for theories with sufficiently large $c$.

We also note that the bounds in Fig.~\ref{Fig:long_scalar_bound_noD} seem to be converge much better than those in Fig.~\ref{Fig:long_scalar_bound}. We will take advantage of this in Subsection \ref{subsec:a1bounds} where we will focus more on the $A_1$ theory.

\subsubsection*{Adding a lower bound}

\begin{figure}[t!]
             \begin{center}           
              \includegraphics[scale=0.4]{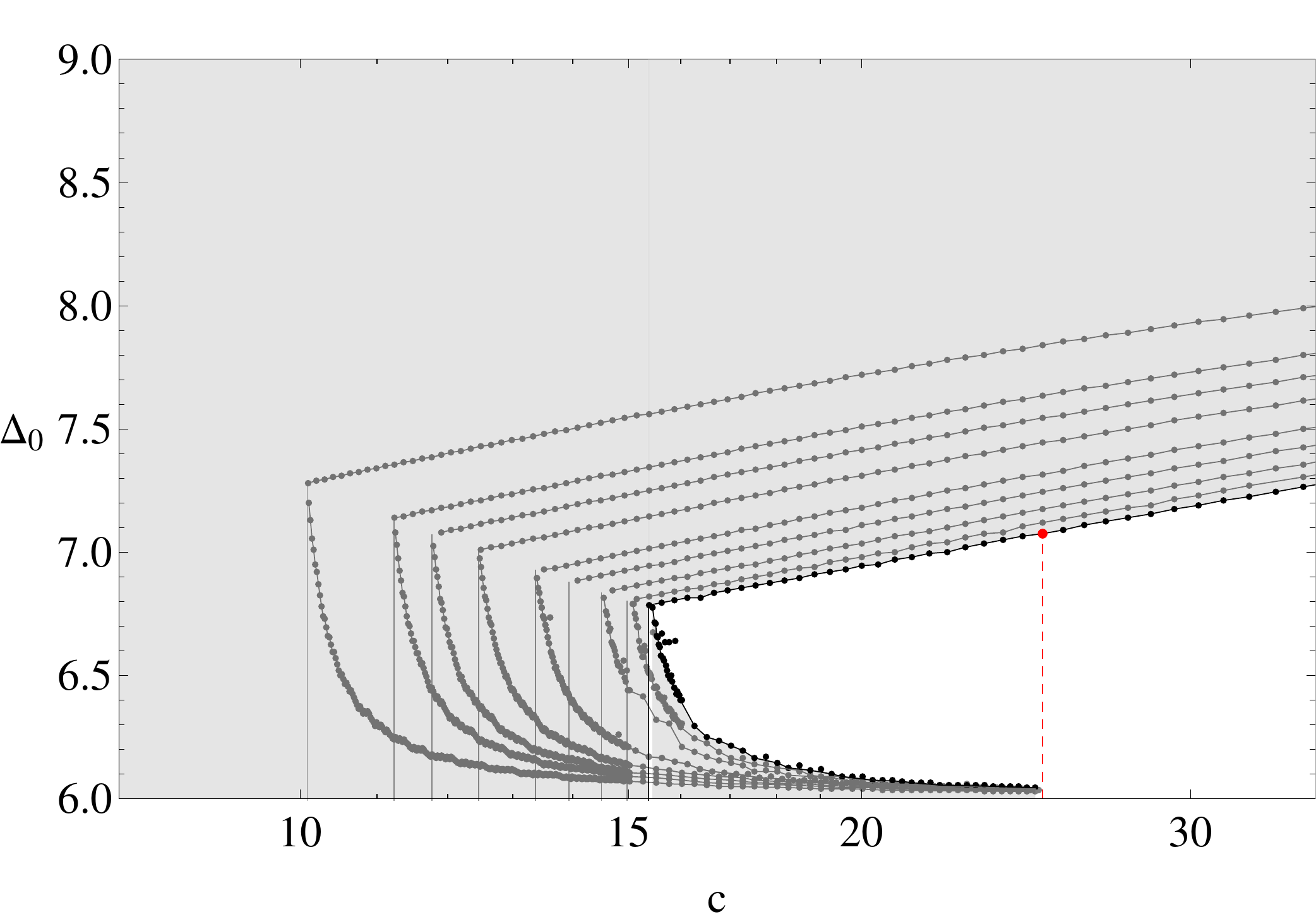}
              \caption{Upper and lower bound for the dimension of an unprotected scalar operator. Every consistent theory must have an operator in the unshaded region. This region collapses to a point precisely at $c_{\rm min}$, which demonstrates the uniqueness of the corresponding truncated solution. The different curves correspond to $\Lambda=18,\ldots,22$, with the black curve representing the strongest bound. The vertical red line marks the central charge corresponding to the $A_1$ theory. Following the extrapolation in Fig.~\ref{Fig:cbound}, we expect the cusp to converge to a point on this line and thereby determine the scaling dimension of the first unprotected scalar operator in the $A_1$theory.}
              \label{Fig:long_scalar_bound_lower}
            \end{center}
\end{figure}

The solution to the crossing symmetry equation at $c_{\rm min}$ is expected to be unique. Fig.~\ref{Fig:long_scalar_bound} does nothing to display this uniqueness, because it merely shows that the theory at $c_{\rm min}$ needs to contain an unprotected scalar operator anywhere between the unitarity bound and the best upper bound of approximately $6.8$. We can improve the situation by adding a lower bound as shown in Fig.~\ref{Fig:long_scalar_bound_lower}. In general the lower bound is rather weak and for sufficiently large $c$ it hits the unitarity bound where it becomes meaningless. However, close to $c_{\rm min}$ the lower bound is strong, and at $c_{\rm min}$ it is practically coincident with the upper bound. At this point there is no freedom left, and for that value of $c_{\rm min}$ there has to be an operator precisely at the cusp in order to satisfy the truncated crossing symmetry equations. In this way Fig.~\ref{Fig:long_scalar_bound_lower} more accurately reflects the uniqueness of the  solution to the truncated crossing equations.

The lower bound was found by searching for a functional that is positive everywhere except in an interval ending at the upper bound.\footnote{Because of numerical subtleties the endpoint of the interval has to be chosen to lie slightly higher than the upper bound. We have taken it to be $\Delta_0^{\text{upper bound}}+0.05$.} The lower bound is then obtained by making this interval as small as possible. A small caveat is in order: the existence of such a functional implies that there \emph{must} be an operator whose dimension is contained in the interval, but in principle there \emph{could} be additional operators also below the lower bound. Although we do not expect these operators to be present on physical grounds, we can never completely rule out their existence because we can always take their OPE coefficients to be infinitesimally small.

\subsubsection{Spinning operators}
\label{subsubsec:spin_long_bounds}

\begin{figure}[h!tb]
             \begin{center}
              \includegraphics[scale=0.36]{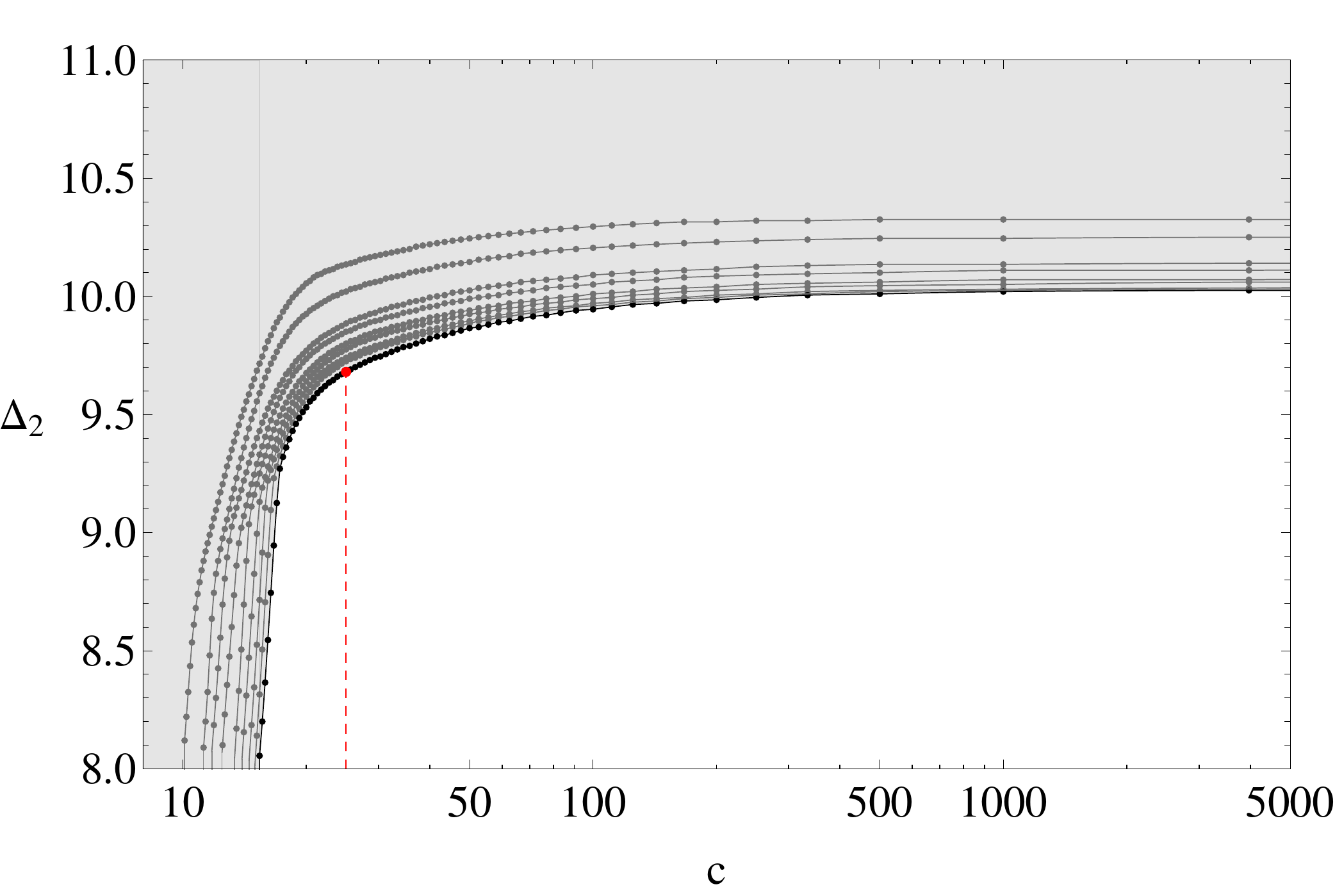} \includegraphics[scale=0.36]{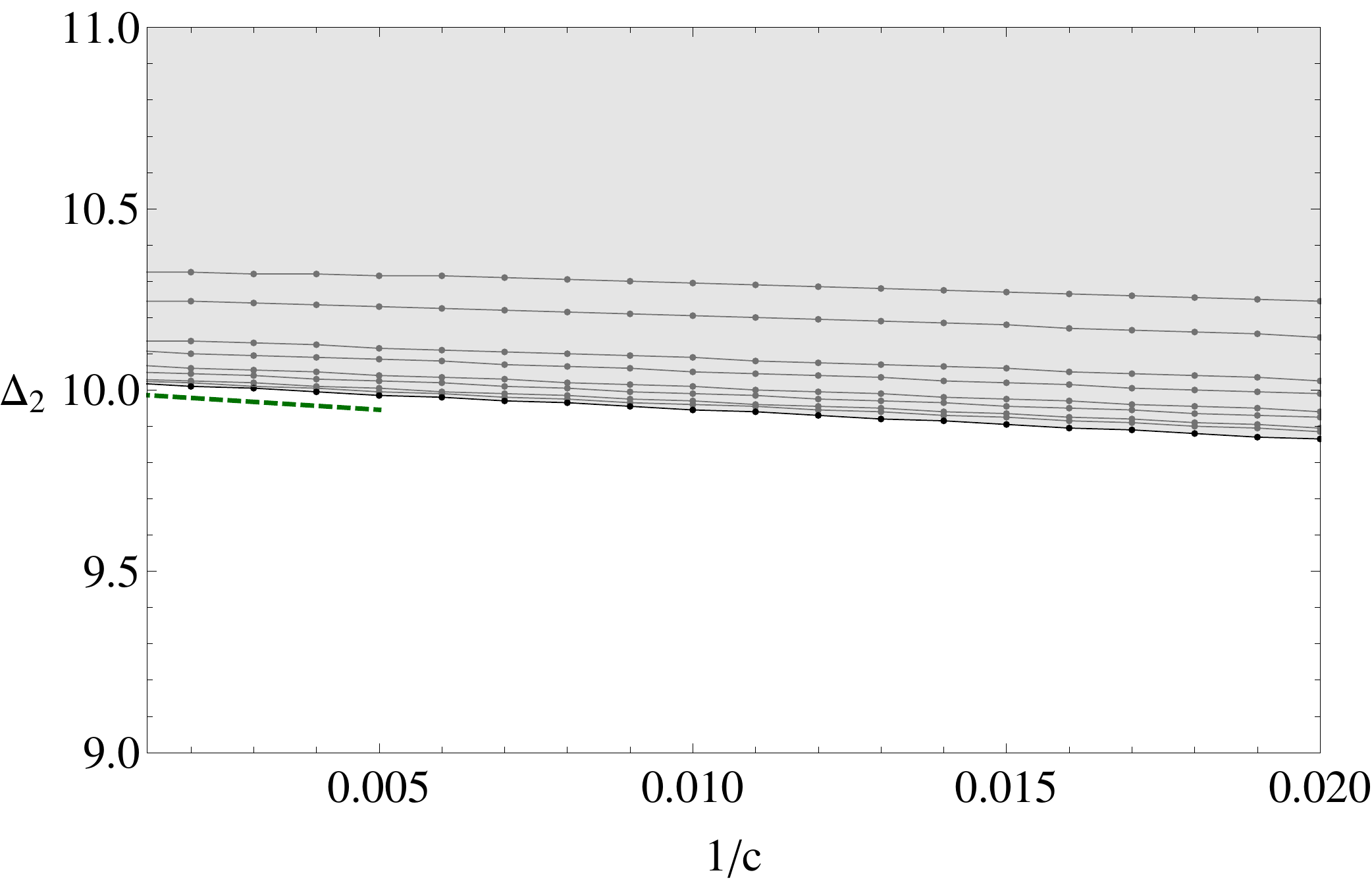}\\
              \end{center}
              $\,\,$\includegraphics[scale=0.36]{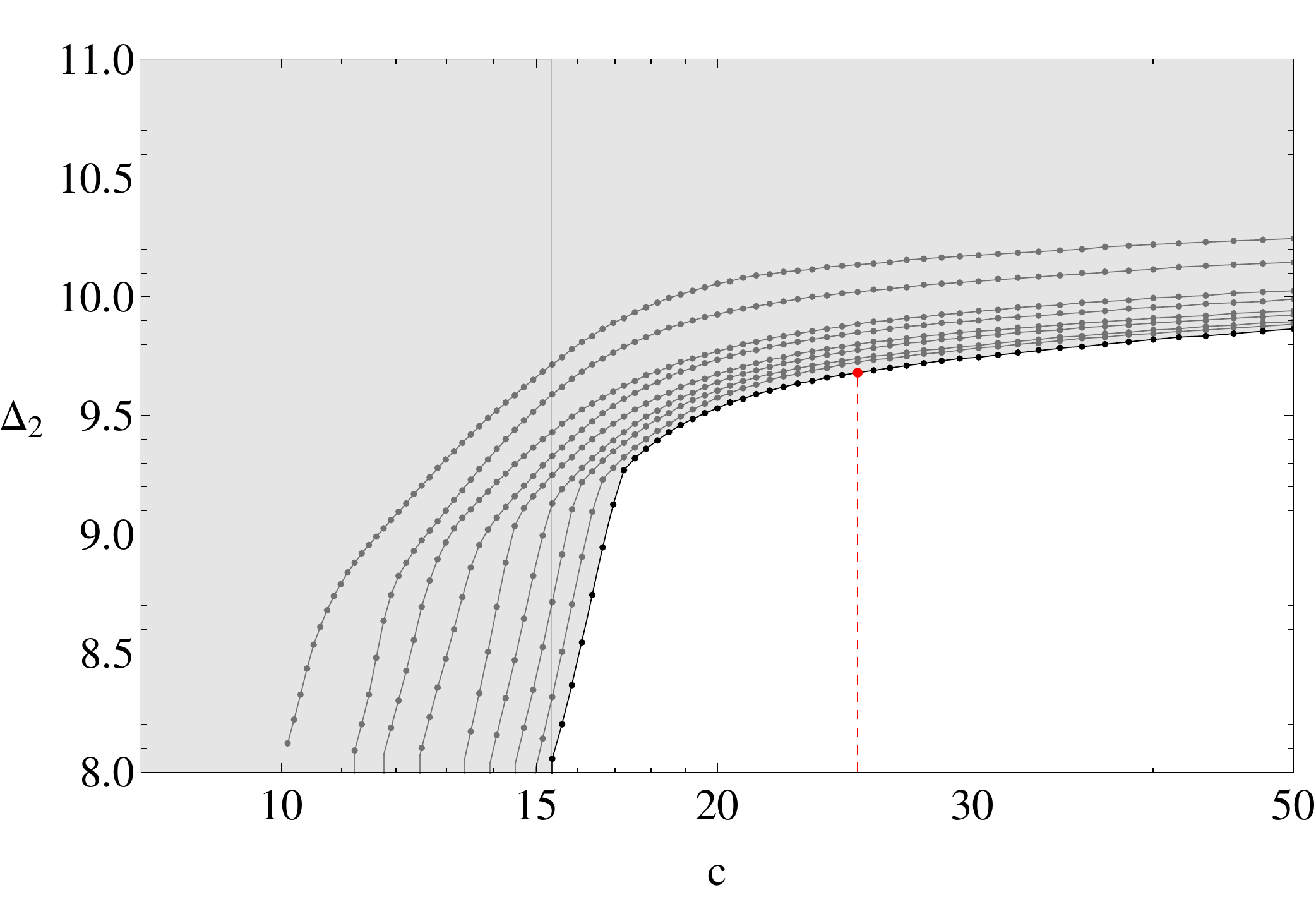}
              \begin{center}
              \caption{Upper bounds for the dimension of the first unprotected spin two operator. The different curves correspond to $\Lambda=18,\ldots,22$, with the black curve representing the strongest bound, and the shaded region is excluded by the numerics. The vertical red line on the left plots corresponds to the central charge of the $A_1$ theory. The plot on the right is a zoomed in result for very large $c$, with the green dashed line corresponding to the known supergravity answer given in \eqref{DTsugra}. The third plot is a magnification of the small central charge region.}
              \label{Fig:long_spin2_bound}
            \end{center}
\end{figure}

\begin{figure}[h!tb]
             \begin{center}
              \includegraphics[scale=0.36]{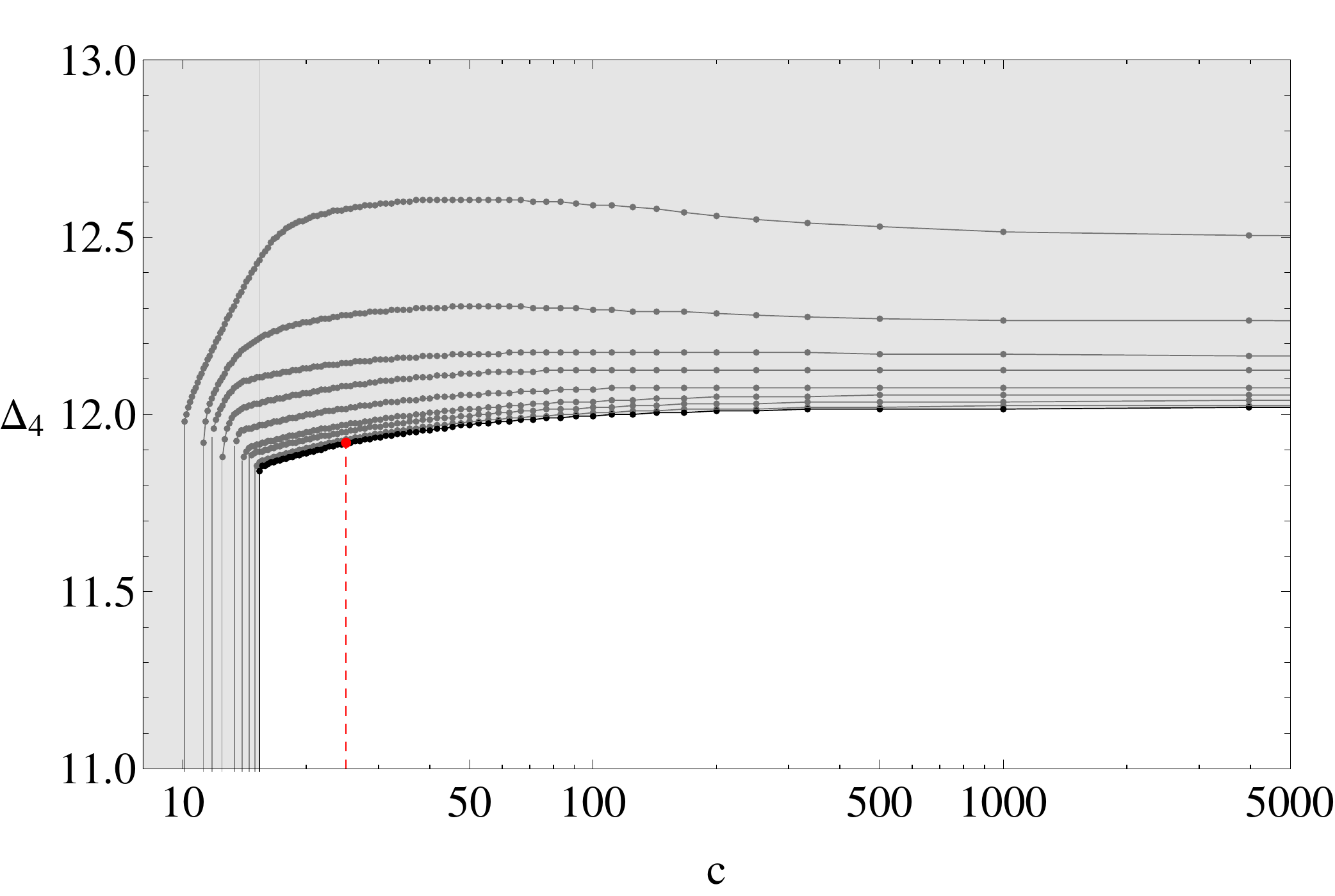} \includegraphics[scale=0.36]{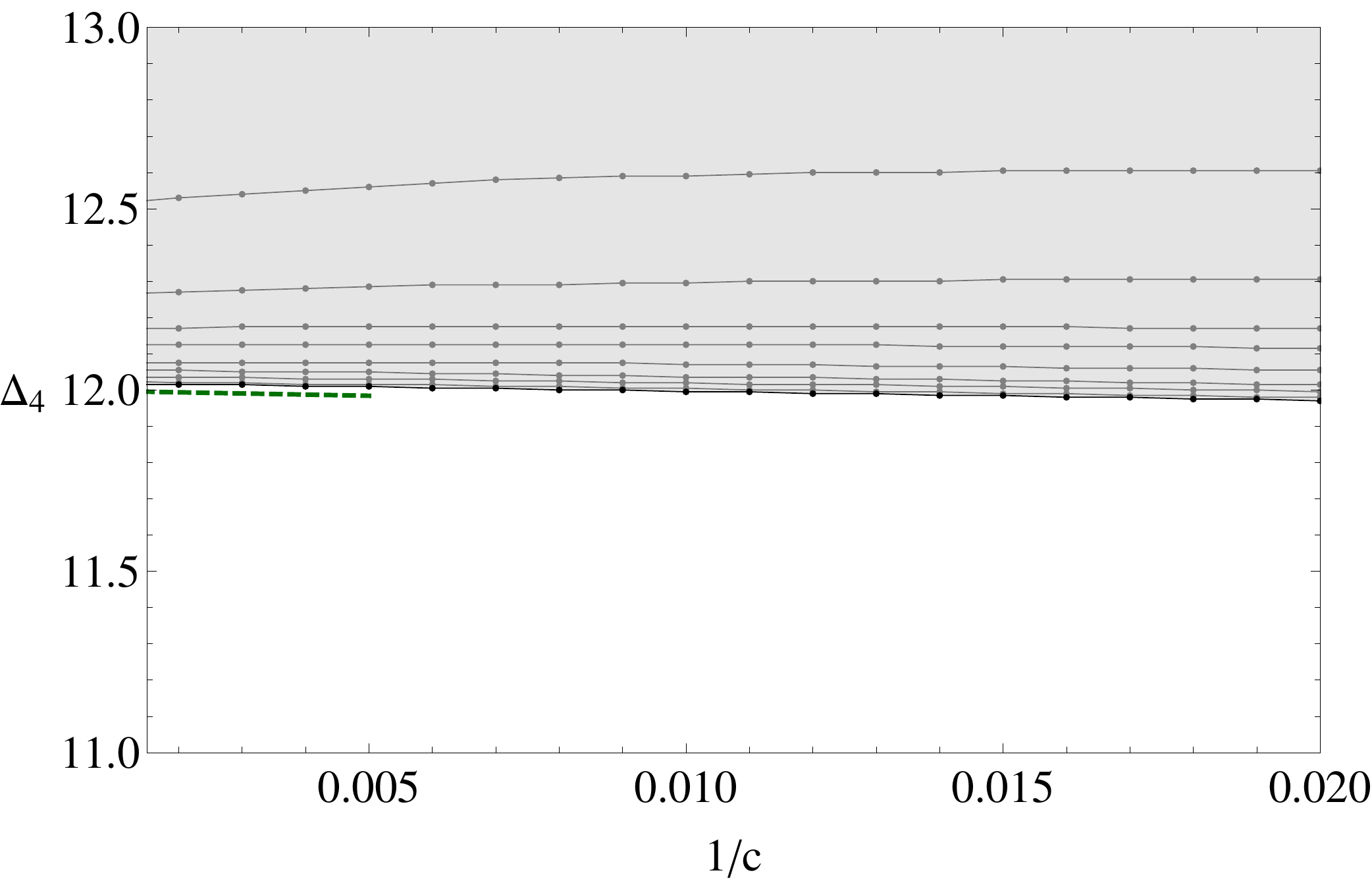}
              \caption{Upper bound for the dimension of the first unprotected spin four operator.  The different curves correspond to different values of $\Lambda=18,\ldots,22$, with the black curve representing the strongest bound, and the shaded region is excluded by the numerics. The vertical red lines on the left correspond to the central charges of known $(2,0)$ theories. The plot on the right is a zoomed in result for very large $c$, with the green dashed line corresponding to the known supergravity solution given in \eqref{DTsugra}.}
                \label{Fig:long_spin4_bound}
            \end{center}
\end{figure}

Figs.~\ref{Fig:long_spin2_bound} and \ref{Fig:long_spin4_bound} present upper bounds on the first unprotected spin 2 and spin 4 operators (operators of type $\LL[0,0]_{\Delta, \ell}$ for  $\ell =0,2$). The structure of these plots is the same as before, and we again would expect these bounds to be saturated by physical theories. This is exemplified at very large $c$ where the bounds agree very well with mean field theory and the $1/c$ correction obtained from supergravity \eqref{DTsugra}. No gap is assumed in the spectrum of scalar operators when obtaining these bounds, so the presence or absence of the $\DD[0,4]$ short multiplet is irrelevant.

In contrast to the scalar and the spin 4 bounds (below), we do not observe step function behavior at $c_{\rm min}$, but rather a more gradual decrease of the bound towards the unitarity bound. We recall that the $\BB[0,2]_1$ block masquerades as an $\LL[0,0]_{\Delta_2, 2}$ block at the unitarity bound $\Delta_2 = 8$, so the non-step function behavior in Fig.~\ref{Fig:long_spin2_bound} is presumably related to the same phenomenon the top left plot of Fig.~\ref{Fig:OPE_B_bounds}.

Although we have not performed a more detailed investigation, the following provides a likely explanation of the behavior in the spin 2 channel.\footnote{This paragraph is rather technical. The uninitiated reader may wish to skip to its last sentence.} Suppose that the approximate solution to crossing symmetry obtained at $c_{\rm min}$ with finite $\Lambda$ has a small bias: instead of a $\BB[0,2]_1$ block it has an $\LL[0,0]_{\Delta_2,2}$ block which sits just above the unitarity bound. As in the scalar and spin 4 channel, the presence of such a block would technically imply step function behavior of the bound at $c_{\rm min}$, but since the block appears only slightly above the unitarity bound the step can be quite small and we would not observe it in Fig.~\ref{Fig:long_spin2_bound}. This long block is very similar to the $\BB[0,2]_1$ short block, and therefore effectively replaces it in the approximate solution to crossing symmetry. In this way the upper bound on the $\BB[0,2]_1$ OPE coefficient at $c_{\rm min}$ can consistently be zero, which is precisely what is observed in the top left figure of Fig.~\ref{Fig:OPE_B_bounds}. Of course we expect the bias to disappear in the limit where $\Lambda \to \infty$. In the current scenario this happens  through a decrease of the dimension $\Delta_2$ of the $\LL[0,0]_{\Delta_2, 2}$ block towards the unitarity bound, where it degenerates into a $\BB[0,2]_1$ block. At this point we would find a step function at $c_{\rm min}$ in \ref{Fig:long_spin2_bound}, and indeed the transition already appears to become sharper for higher $\Lambda$. Similarly, the bound in Fig.~\ref{Fig:OPE_B_bounds} at $c_{\rm min}$ will have to transition towards the dimension of the first unprotected operator, and therefore also become infinitely sharp in the limit of large $\Lambda$. In summary, then, the relative smoothness of these particular transitions at $c_{\min}$ is plausibly a numerical artifact and we expect to recover genuine step function behavior as $\Lambda \to \infty$.

\subsubsection{Combining spins}
\label{subsubsec:combined_spin_bounds}

\begin{figure}[t!]
             \begin{center}           
              \includegraphics[scale=0.35]{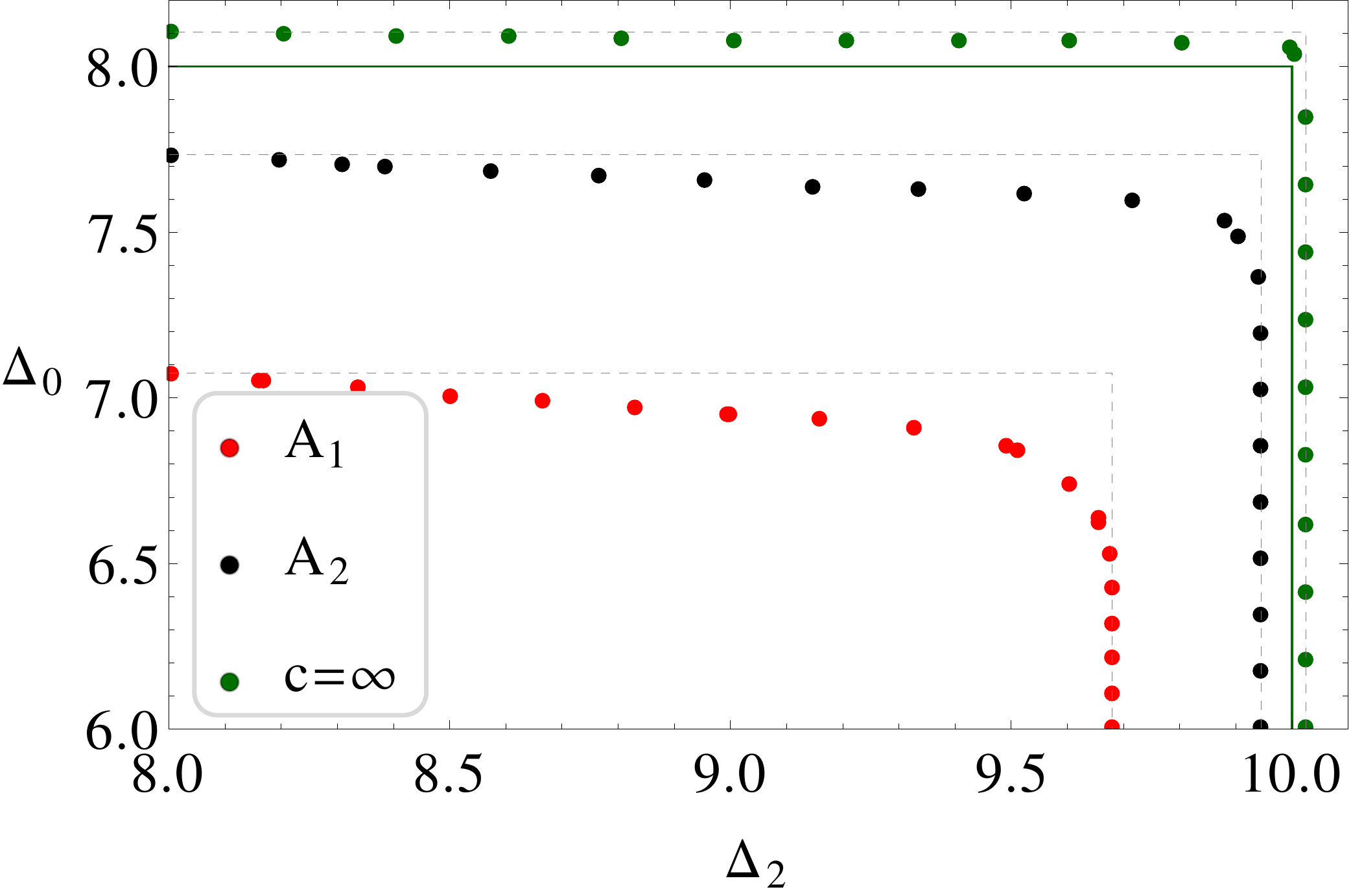}
              \includegraphics[scale=0.35]{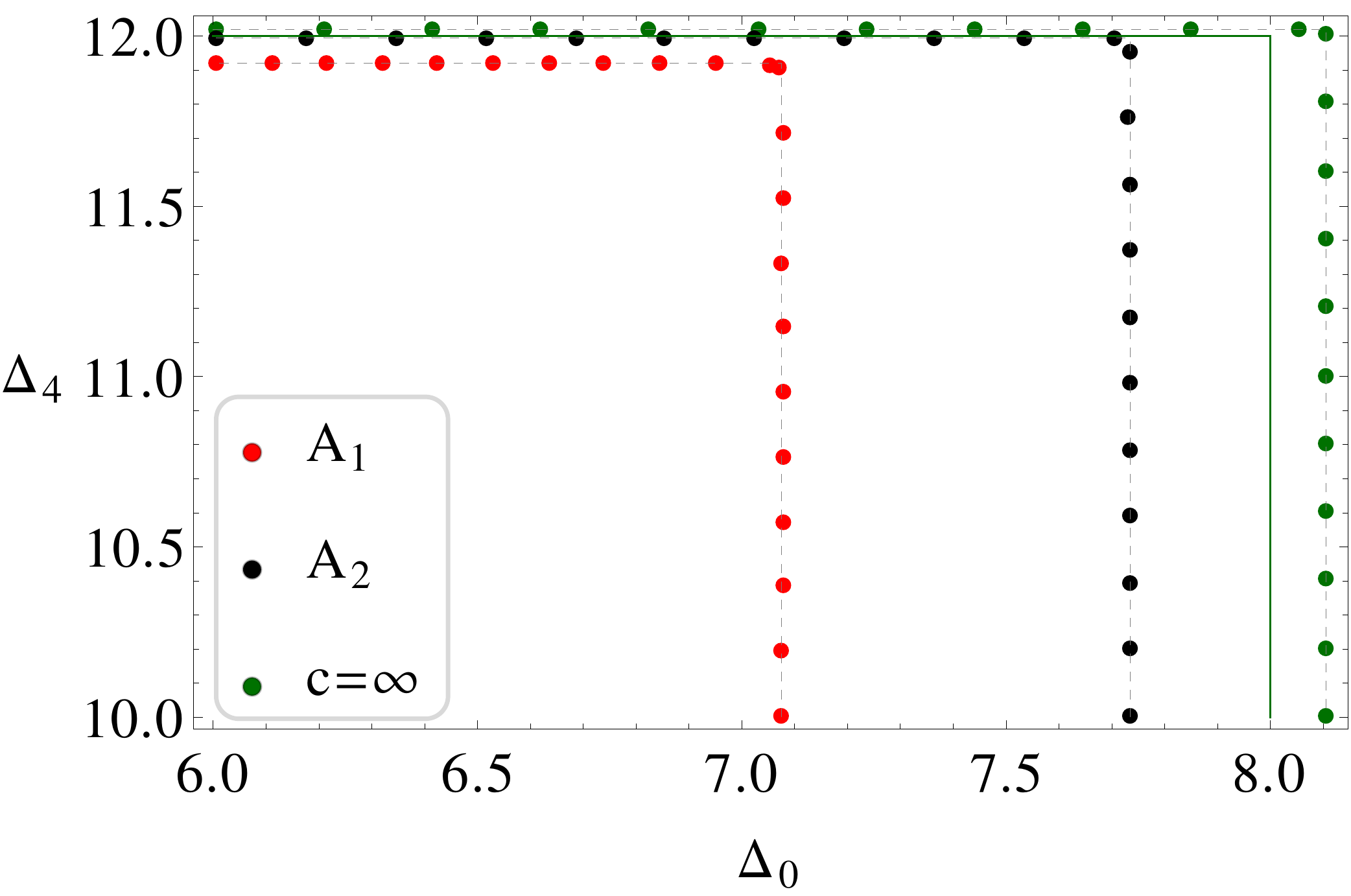}
              \includegraphics[scale=0.35]{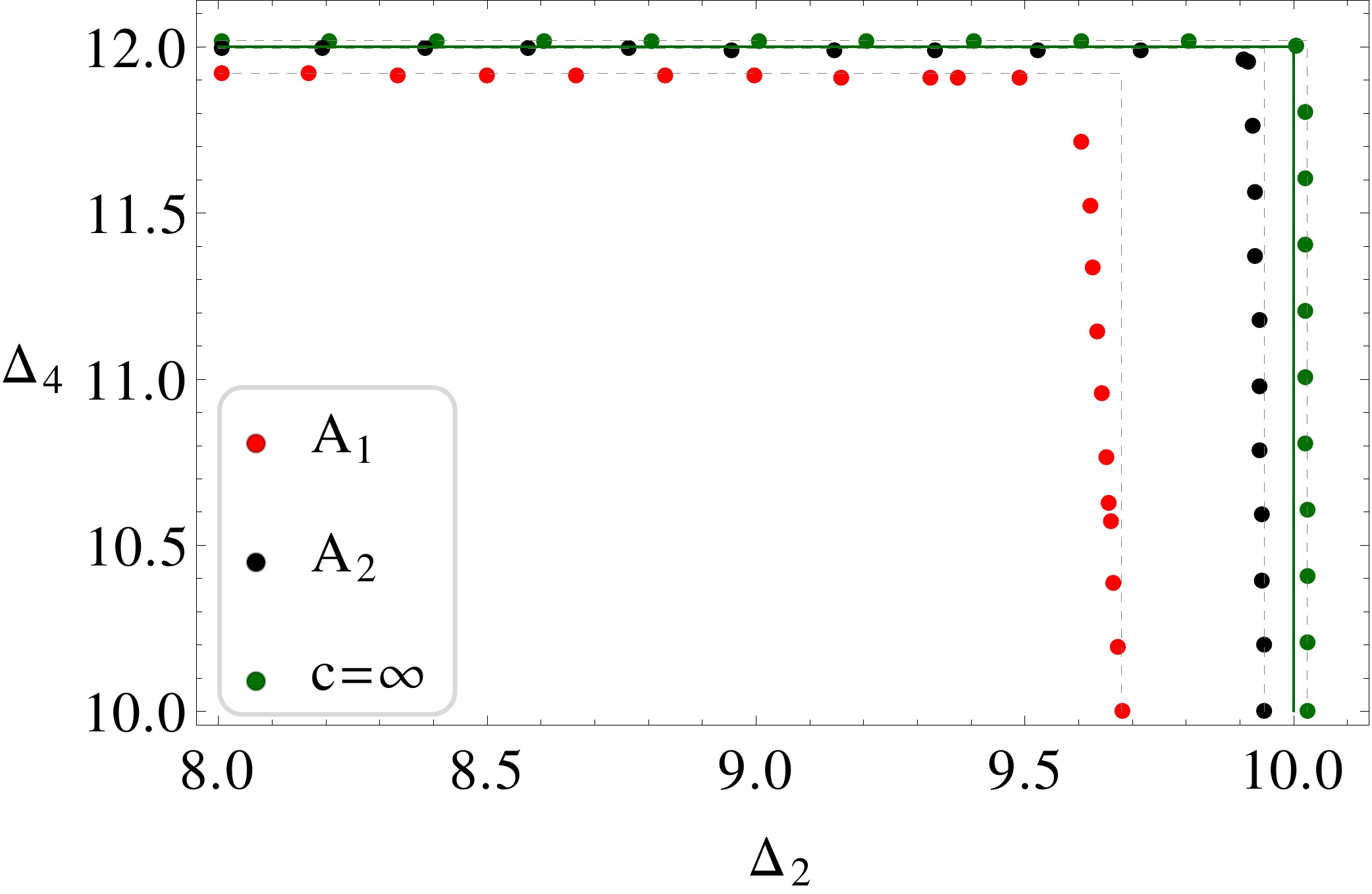}
              \caption{Bounds on the spin $0,2,4$ superconformal primary dimensions when a gap is imposed in one of the other channels for a cutoff of $\Lambda=22$. These bounds are for the central charges corresponding to the $A_1$ and $A_2$ theories and to the generalized free field theory limit $c=\infty$, and are obtained with the addition of the short multiplet $\DD[0,4]$. The dashed lines show the bounds on $(\Delta_0, \Delta_2, \Delta_4)$ from imposing gaps in a single channel, and the full green lines denote the dimensions obtained from generalized free field theory. The allowed region corresponds to the inside of the ``rectangles'' delimited by the dots.}
              	\label{Fig:square}
            \end{center}
\end{figure}

So far our upper bounds for the scaling dimensions have been for just a single spin channel. We can also combine spins and obtain (for fixed $c$) exclusion plots in the higher-dimensional space spanned by $(\Delta_0, \Delta_2, \Delta_4, \ldots)$. In Fig.~\ref{Fig:square} we show such exclusion plots in the two-dimensional subspaces spanned by the pairs $(\Delta_0,\Delta_2)$, $(\Delta_0,\Delta_4)$ and $(\Delta_2,\Delta_4)$. We have the fixed value of $c$ to correspond to either the $A_1$ theory, the $A_2$ theory, or infinity. The bounds obtained in the preceding subsections already dictate that the allowed dimensions are inside the squares delimited by the dashed lines. As we impose gaps simultaneously in two channels we numerically carve out a smaller part of this square, and the dimensions must now be below the dots shown in Fig.~\ref{Fig:square}.

We have claimed above that both at large and at small central charge the bounds are saturated by physical theories. In particular, we claim that the spin 0 and the spin 2 bounds should converge to the \emph{same} solution of the crossing symmetry equations. If this is the case then the combined bounds plotted in Fig.~\ref{Fig:square} should converge to perfect rectangles, and deviations from this shape may indicate that something is amiss. 

For $c=\infty$ we in addition know that the vertex of this square should be localized at the known values $(\Delta_0,\Delta_2,\Delta_4) = (8,10,12)$. We indicated this with the green lines in Fig.~\ref{Fig:square}, and observe that the numerical bounds indeed nicely follow the outline of a square. Again, for $\Lambda \to \infty$ we expect these points to converge precisely onto these squares.

For finite $c$, the absence of noteworthy features in Fig.~\ref{Fig:square} is also reassuring. The fact that we do not yet find sharp rectangles can be ascribed to the relatively poor convergence of the bounds, and we expect improvement for larger values of $\Lambda$. Turning the logic around, given that we are forced to work with finite $\Lambda$ we can obtain somewhat improved estimates of scaling dimensions by estimating the location of the corner points.\footnote{In \cite{Beem:2013qxa} similar methods were used to improve the estimate of the $1/c$ corrections in four-dimensional $\NN = 4$ theories.}

%% file: sections/6_4_a1bounds.tex

\subsection{Bootstrapping the \texorpdfstring{$A_1$}{A1} theory}
\label{subsec:a1bounds}

We now focus our attention on the $A_1$ theory, for which we have argued that the correlator under consideration is be completely fixed by crossing symmetry. Its full determination using these numerical methods would, however, require infinite computational resources. The aim of this section is to show a few examples of results that can be obtained at a finite numerical cost.

\subsubsection{Estimates for the lowest-dimensional scalar operator}
\begin{figure}[t!]
	\begin{center}           
		\includegraphics[scale=0.5]{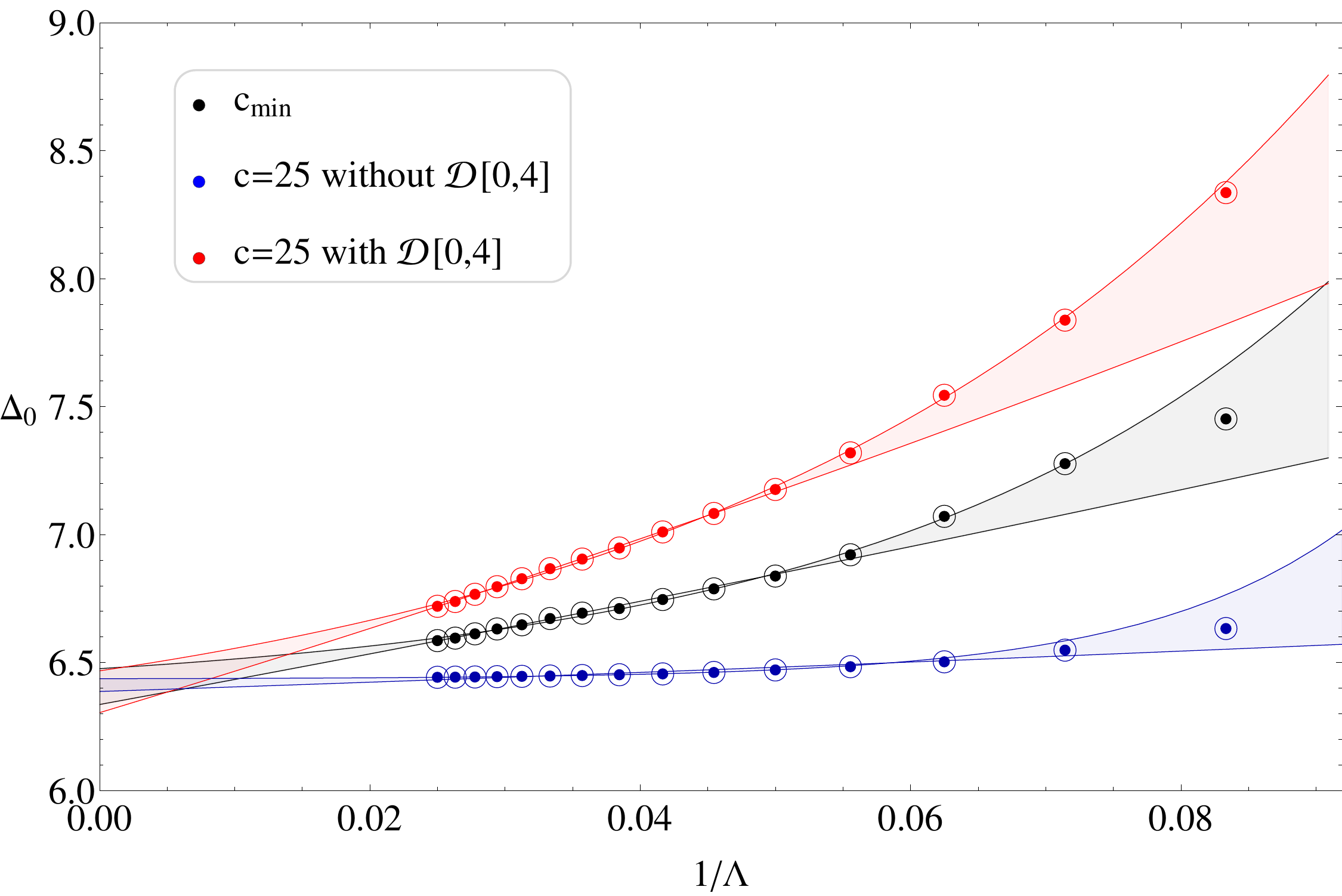}
			\caption{Upper bounds on the dimension of the first long spin $0$ multiplet as a function of the inverse of the cutoff $\Lambda$ for the minimum central charge $c_{\rm min}(\Lambda)$ (black) and for $c=25$ with (red) and without (blue) the  $\DD[0,4]$ short multiplet.}
			\label{Fig:extrapol_l0}
	\end{center}
\end{figure}
Let us first estimate the dimension of the first long scalar multiplet. We have three ways of doing so: we can extrapolate from the cusp at $c_{\rm min}$ in Fig.~\ref{Fig:long_scalar_bound_lower}, vertically down from the bound at $c = 25$ in the same figure, and finally vertically down from the bound at $c=25$ obtained without the $\DD[0,4]$ multiplet in Fig.~\ref{Fig:long_scalar_bound_noD}. These three estimates should converge to the same scaling dimension and this offers a good cross-check of the extrapolations.

These bounds are shown in Fig.~\ref{Fig:extrapol_l0}. The data points in this plot were obtained using the semidefinite approach with \texttt{SDPB} -- consequently we could extend to higher values of $\Lambda$ than in Figs.~\ref{Fig:long_scalar_bound_noD} and~\ref{Fig:long_scalar_bound_lower}. We have added extrapolations which were obtained by fitting the last $n$ data points for various values of $n$ with an exponential function and showing the ones that give the lowest and highest extrapolated values.

Fig.~\ref{Fig:extrapol_l0} gives us enough confidence to claim that all three approaches will indeed ultimately converge to the same point. The most reliable way to accurately estimate this point is from the bottom curve, which corresponds to the bound without the $\DD[0,4]$ multiplet and has almost converged. From its best value at $\Lambda = 40$ we extract the following bound:

\begin{result*}
If the $A_1$ theory does not have a $\DD[0,4]$ multiplet of operators, then it must have an unprotected scalar operator of dimension $\Delta_0 < 6.443$.
\end{result*}

Besides the strict upper bound, the three extrapolations in Fig.~\ref{Fig:extrapol_l0} together with the aforementioned uniqueness of the theory at $c_{\rm min}$ encourages us to put forward an additional

\begin{conj*}
The $A_1$ theory has an unprotected scalar operator of dimension $6.387 < \Delta_0 < 6.443$.
\end{conj*}

Proving this conjecture would require a rigorous estimate of the lower bound. This might be possible with the use of more sophisticated computational techniques, for example by studying multiple correlators following the blueprint of \cite{Kos:2014bka}.

\subsubsection{Estimates for the second lowest-dimensional scalar operator}
\label{subsubsec:second_scalar_estimates}

\begin{figure}[t!]
             \begin{center}           
              \includegraphics[scale=0.45]{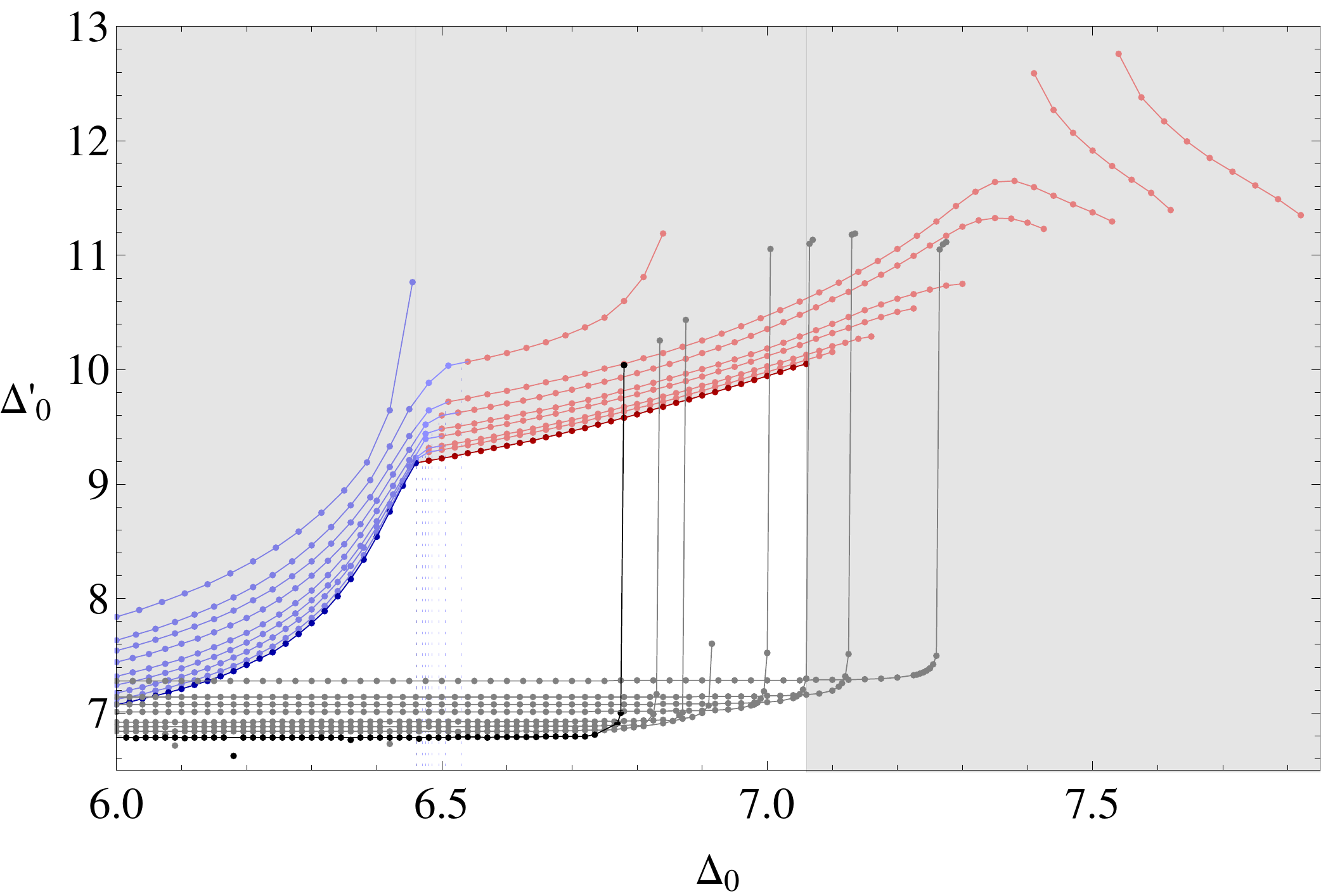}
              \caption{Bound on the dimension of the second scalar superconformal primary dimension $\Delta_0^\prime$, as a function of the dimension of the dimension of the first scalar superconformal primary $\Delta_0$. See the main text for a detailed explanation. The cutoff is increased from $\Lambda=14$ to $22$, with the darker lines corresponding to the strongest bound $\Lambda=22$. The region shaded in gray is the one excluded by the strongest bound.
              }
              \label{Fig:kinkl0}
            \end{center}
\end{figure}

We can also constrain the dimension $\Delta^\prime_0$ of the second unprotected scalar operator in the spectrum of the $A_1$ theory. We can bound $\Delta^\prime_0$ from above if we are willing to commit to a value $\Delta_0$ of the first unprotected scalar operator. Since we do not exactly know $\Delta_0$, we plot in Fig.~\ref{Fig:kinkl0} the upper bound on $\Delta^\prime_0$ as a function of $\Delta_0$. The figure again contains three sets of curves corresponding to the three different methods discussed above, with a color-coding that matches Fig.~\ref{Fig:extrapol_l0}. We will explain each set of curves in turn.

The black curves were computed\footnote{The bounds shown are computed at $c=c_{\rm min}+0.01$, since at $c_{\rm min}$ a functional is always found, even if no bound is imposed, whose zeros are the dimensions of the operators in the unique solution to the truncated crossing equation. This makes it hard to obtain bounds on operator dimensions.} at $c_{\rm min}(\Lambda)$. The uniqueness of the solution to crossing for these values of $c$ can be seen in that when $\Delta_0<\Delta_*$ -- the first scalar dimension in the unique solution -- we always rediscover that operator and have $\Delta_0^\prime=\Delta_*$. Hence the horizontal plateaus in the black curves for low $\Delta_0$. Once $\Delta_0$ reaches $\Delta_*$ then $\Delta^\prime_0$ can rise to the dimension of the next operator, and this is leads to the sharp jump precisely along the diagonal $\Delta_0 = \Delta^\prime_0$. If we increase $\Delta_0$ further then we have no solution to crossing symmetry. The top of each peak is then the best estimate for the next scalar operator.

In the red and blue curves we have set $c = 25$ with and without the $\DD[0,4]$ multiplet. Both curves end when $\Delta_0$ reaches the upper bound given in Fig.~\ref{Fig:extrapol_l0} as a function of $\Lambda$. The red curve coincides with the blue curve for small $\Delta_0$ - in that case the presence of the $\DD[0,4]$ multiplet apparently does not change the bound\footnote{This is corroborated by an analysis of the dual problem, which shows that along this segment the $\DD[0,4]$ multiplet is absent from the truncated crossing symmetry equations, even if we in principle allow for its presence.} on $\Delta^\prime_0$. On the other hand, for higher values of $\Delta_0$ we can only satisfy crossing symmetry if the $\DD[0,4]$ is also present, and the kink marks the transition between these two regimes. The location of the kink is then our best estimate for the pair $(\Delta_0, \Delta^\prime_0)$ in an $A_1$ theory without a $\DD[0,4]$ multiplet.

We can speculate about the shape of these curves in the limit where $\Lambda \to \infty$. In that case we expect all three bounds to coincide. The red segment past the kink will therefore shrink to zero size, and the shape of the blue curve below the kink will be increasingly convex, with a limiting shape similar to the sharp peak that we already observe in the black curve. The highest point of the black curve in turn will move further left. Its position will have to coincide with the eventual location of the kink, and this is then where we can read off $\Delta^\prime_0$. Notice that all these tendencies are already visible by extrapolating from the curves obtained for lower $\Lambda$.

Let us finally provide a few numbers. If the $A_1$ theory does not have a $\DD[0,4]$ multiplet then we can use the bounds given by the blue curve. We conclude that the second scalar in the $A_1$ theory must have $\Delta^\prime_0 < 9.19$. Furthermore, from a cautious extrapolation from the location of the kink we conjecture that in addition $\Delta^\prime_0 \gtrsim 8.3$. 

\subsubsection{Bounds for OPE coefficients}
\label{subsubsec:A1_OPE_bounds}

\begin{figure}[t!]
             \begin{center}           
              \includegraphics[scale=0.36]{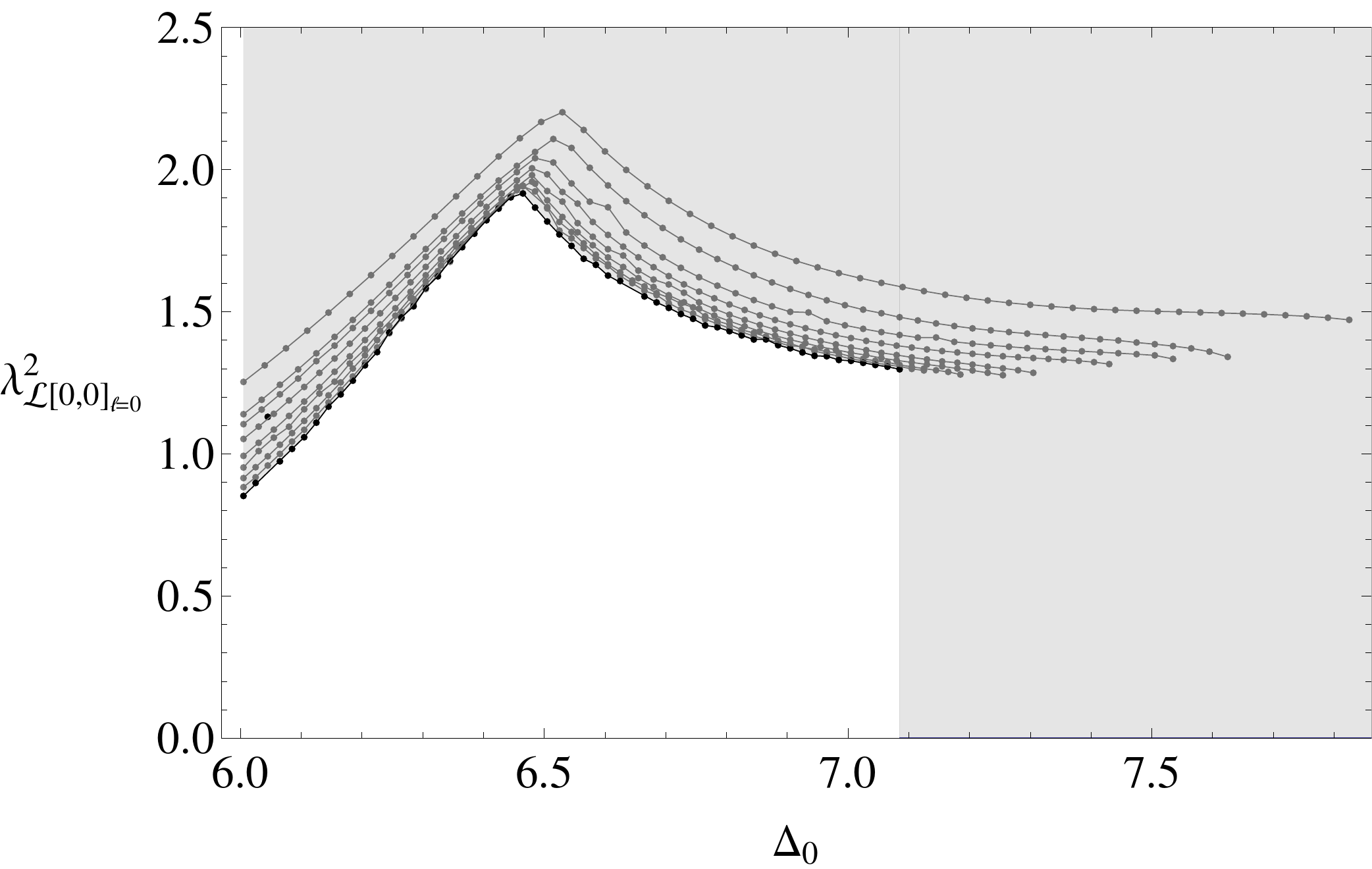}
              \includegraphics[scale=0.36]{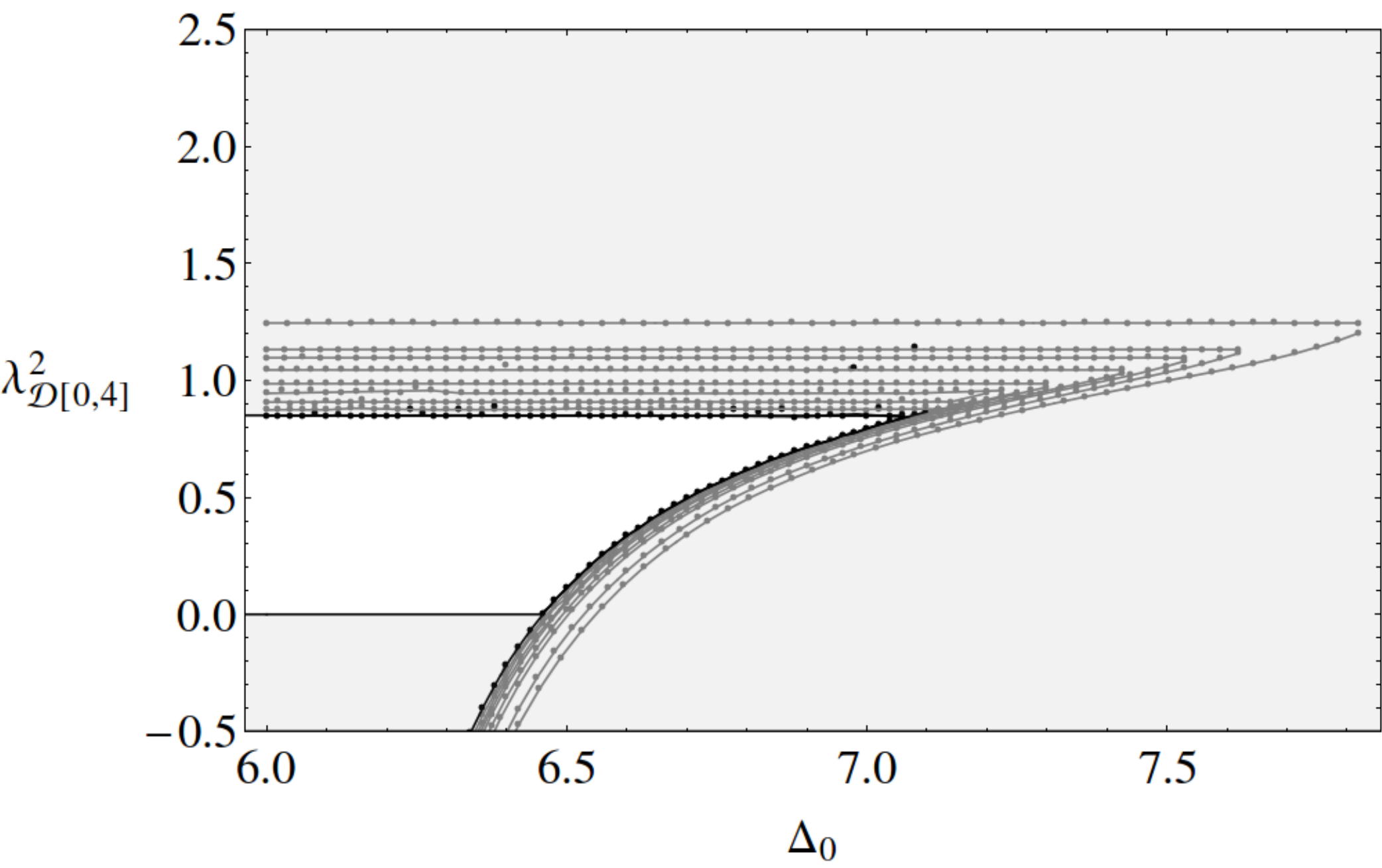}
              \caption{Left: Bound on the OPE coefficient (squared) of a scalar operator of dimension $\Delta_0$ in the $A_1$ theory. We vary $\Delta_0$ from the unitarity bound to the dimension bound obtained in Fig.~\ref{Fig:long_scalar_bound}. For this plot we allow for the presence of the short operator $\DD[0,4]$. Right: Lower and upper bound on the OPE coefficient squared of a possible $\DD[0,4]$ multiplet in the $A_1$ theory, as a function of the dimension of $\Delta_0$. In both plots the cutoff is increased from $\Lambda=14$ to $22$. The excluded values of the OPE coefficient correspond to the shaded region.}
              	\label{Fig:opekink}
            \end{center}
\end{figure}

Next we present bounds on various OPE coefficients, again as a function of the dimension $\Delta_0$ of the first unprotected scalar operator. In Fig.~\ref{Fig:opekink} we show an upper bound for the $A_1$ theory on $\lambda^2_{\LL[0,0]}$, the coefficient of the unprotected operator of dimension $\Delta_0$, and an upper and a lower bound for $\lambda^2_{\DD[0,4]}$. As explained in Section \ref{sec:numerics}, a lower bound on $\lambda^2_{\DD[0,4]}$ is possible because for $\Delta_0 > 6$ this multiplet is isolated. The numerical lower bound however becomes useless for small $\Delta_0$ since $\lambda^2_{\DD[0,4]} \geqslant 0$ by unitarity.

The kink in the left plot of Fig.~\ref{Fig:opekink} coincides with the kink in Fig.~\ref{Fig:kinkl0} and also agrees with the point where the lower bound on $\lambda^2_{\DD[0,4]}$ crosses zero in the right plot of Fig.~\ref{Fig:opekink}. In this way we observe again that a $\DD[0,4]$ multiplet would be required if we try to increase $\Delta_0$ past the kink. Although we cannot rigorously prove the absence of the $\DD[0,4]$ multiplet, the right plot of Fig.~\ref{Fig:opekink} does provide a rigorous upper bound on the coefficient of the corresponding conformal block.

As illustrated by Fig.~\ref{Fig:OPE_D_bound}, for infinite $\Lambda$ we expect the upper bound in the right plot to decrease to zero. We also expect the upper bound on $\lambda^2_{\LL[0,0]}$ to exhibit a sharp peak in this limit, for the same reason as in the previous discussion concerning the large $\Lambda$ behavior of the blue and red curves in Fig.~\ref{Fig:kinkl0}. The eventual location of the kink is therefore again our best estimate for the pair $(\Delta_0, \lambda^2_{\LL[0,0]})$. From the left plot in Fig.~\ref{Fig:opekink} we conclude that the $A_1$ theory, with or without a $\DD[0,4]$ multiplet, must have $\lambda^2_{\LL[0,0]} < 1.92$ for the first unprotected scalar operator. We expect it to not be smaller than $1.8$ based on extrapolation of our current results. 

\subsubsection{Bounds on the second lowest-dimensional spinning operators}
\label{subsubsec:second_spin_bounds}

\begin{figure}[t!]
             \begin{center}           
              \includegraphics[scale=0.36]{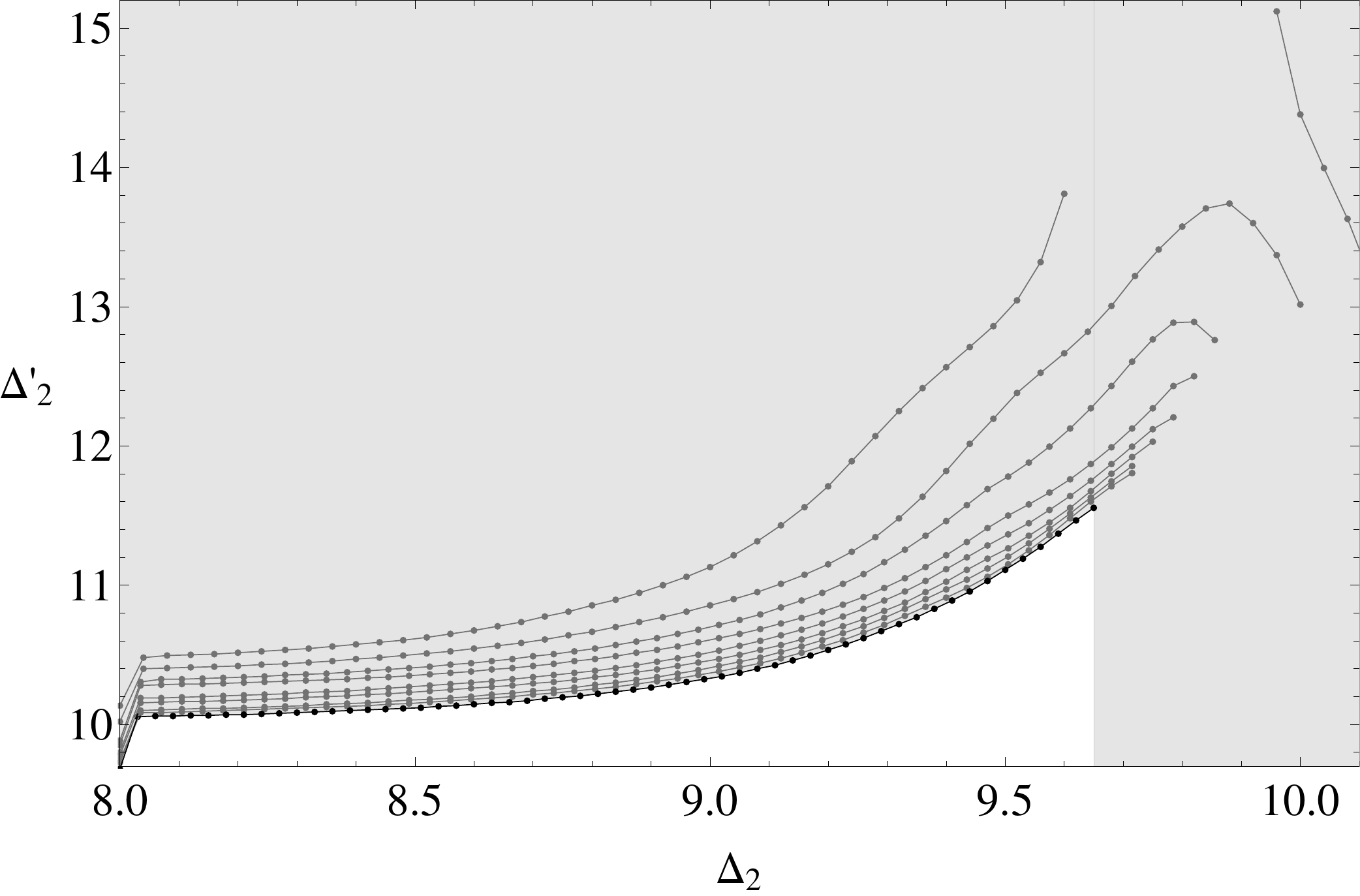}
              \includegraphics[scale=0.36]{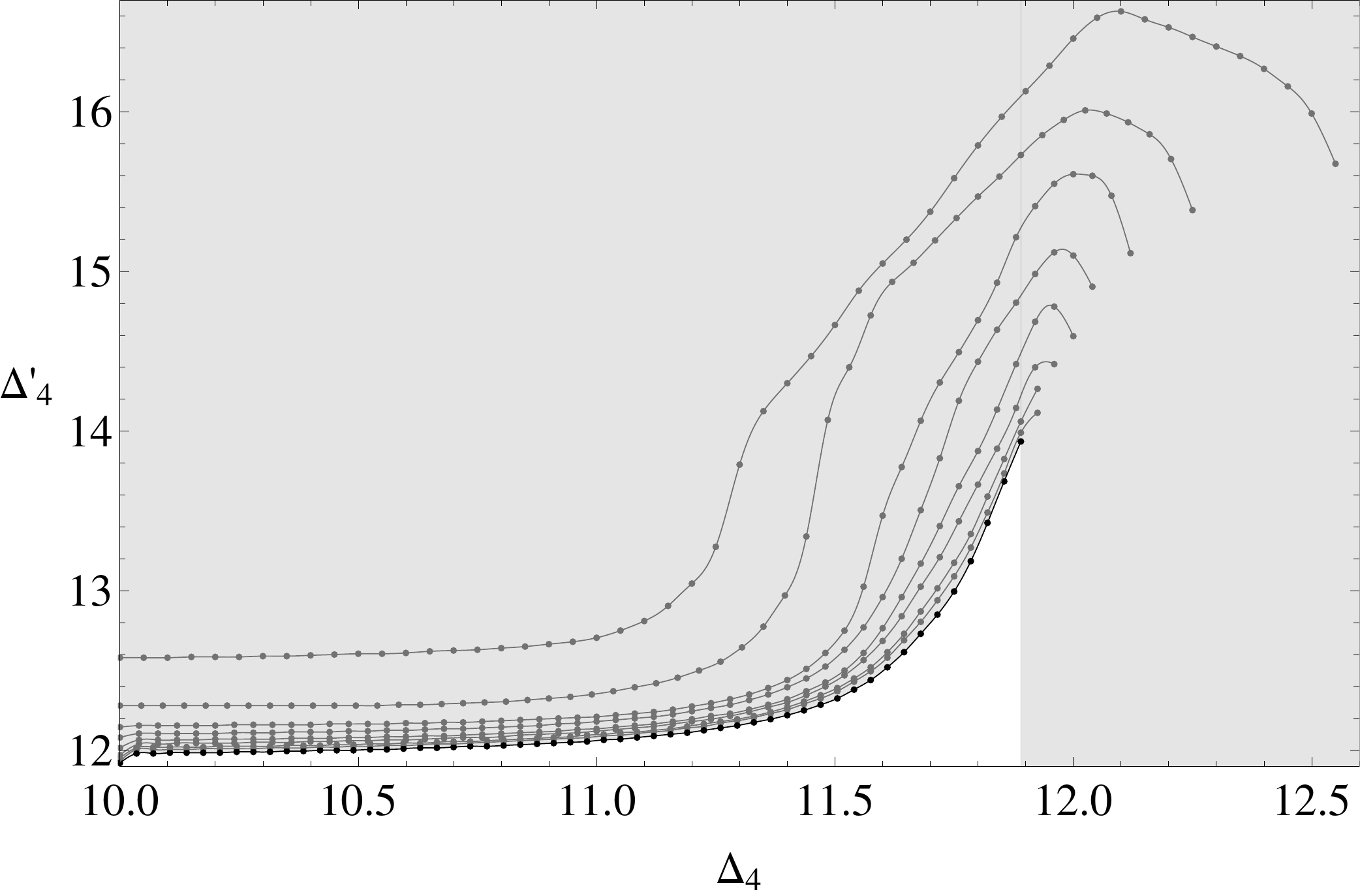}
              \caption{Bound on the dimension of the second spin $\ell=2,4$ superconformal primary dimension $\Delta_{2,4}^\prime$, as a function of the dimension of the dimension of the first spin $\ell=2,4$ superconformal primary $\Delta_{2,4}$ for $c=25$. The cutoff is increased from $\Lambda=14$ to $22$.}
              \label{Fig:kinkl2l4}
            \end{center}
\end{figure}

We have also investigated the dimensions of the second operators of spin two and four. The results are shown in Fig.~\ref{Fig:kinkl2l4} where we assume there to exist an operator of dimension $\Delta_{2,4}$ in the allowed range, and bound the dimension of the second operator $\Delta_{2,4}^\prime$.

For both spins we observe a small step-like feature for a small value of $\Delta_2$ and $\Delta_4$ which is most likely an artifact of working at finite $\Lambda$, similarly to what was observed in Fig.~\ref{Fig:long_spin2_bound}. Increasing the value of $\Delta_2$ and $\Delta_4$ we find an upward sloping upper bound for the next operator, and again a cutoff at the maximal allowed value of $\Delta_{2,4}$. These cutoff values can also be read off from Figs.~\ref{Fig:long_spin2_bound} and \ref{Fig:long_spin4_bound} at $c = 25$. These plots are the spin 2 and 4 analogues of Fig.~\ref{Fig:kinkl0}, and in the large $\Lambda$ limit we again expect them to converge towards a sharp peak. This behavior is already very much apparent in the spin 4 plot.\footnote{For these plots we do not impose any gap in the scalar sector, and the in- or exclusion of the $\DD[0,4]$ multiplet therefore does not affect the numerics.} Again, the location of the peak in principle provides us with reasonable estimates for the pairs $(\Delta_{2,4}, \Delta^\prime_{2,4})$.

\subsubsection{Combining spins}
\label{subsubsec:A1_combined_spins}

\begin{figure}[t!]
             \begin{center}           
              \includegraphics[scale=0.35]{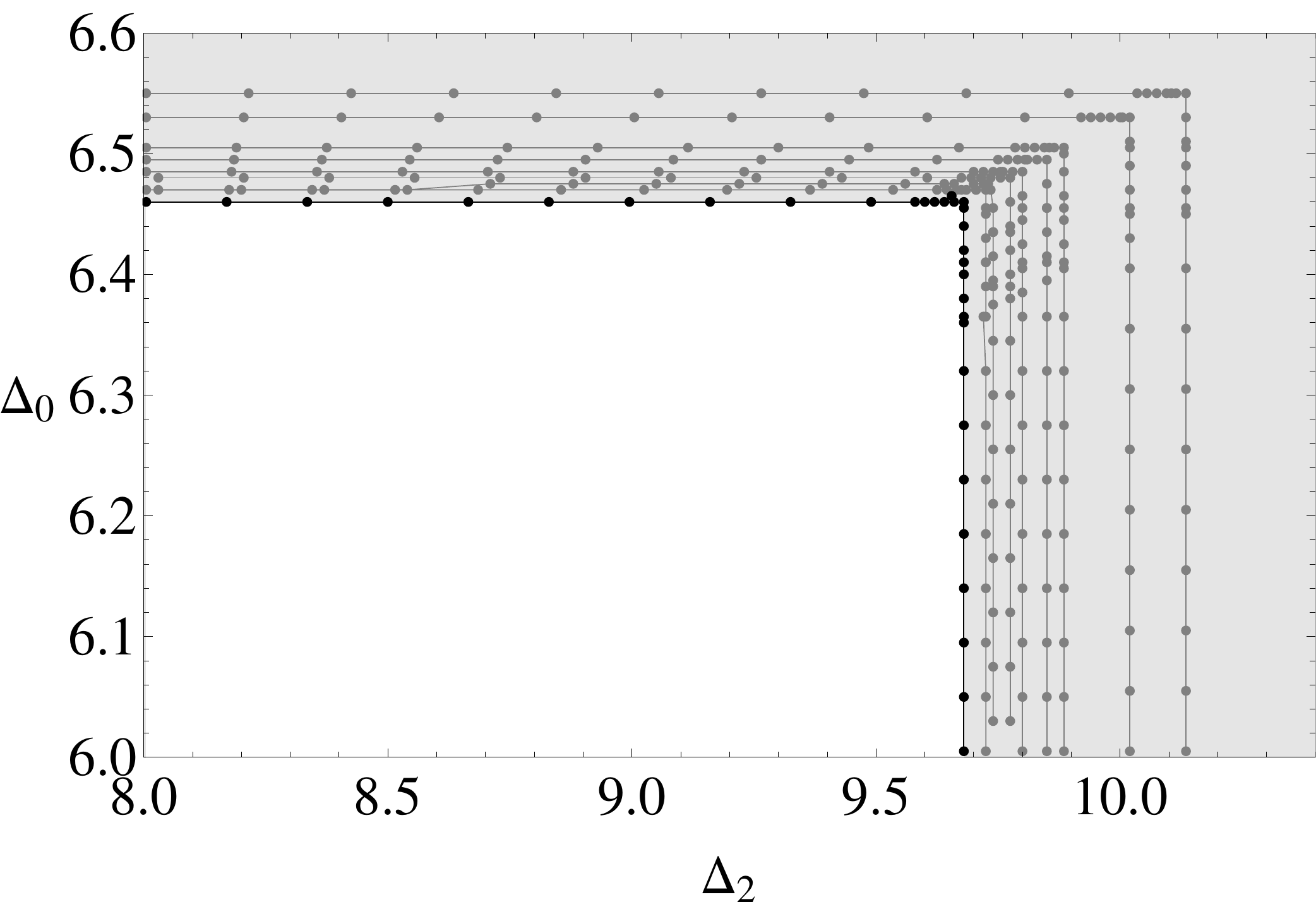}
              \includegraphics[scale=0.35]{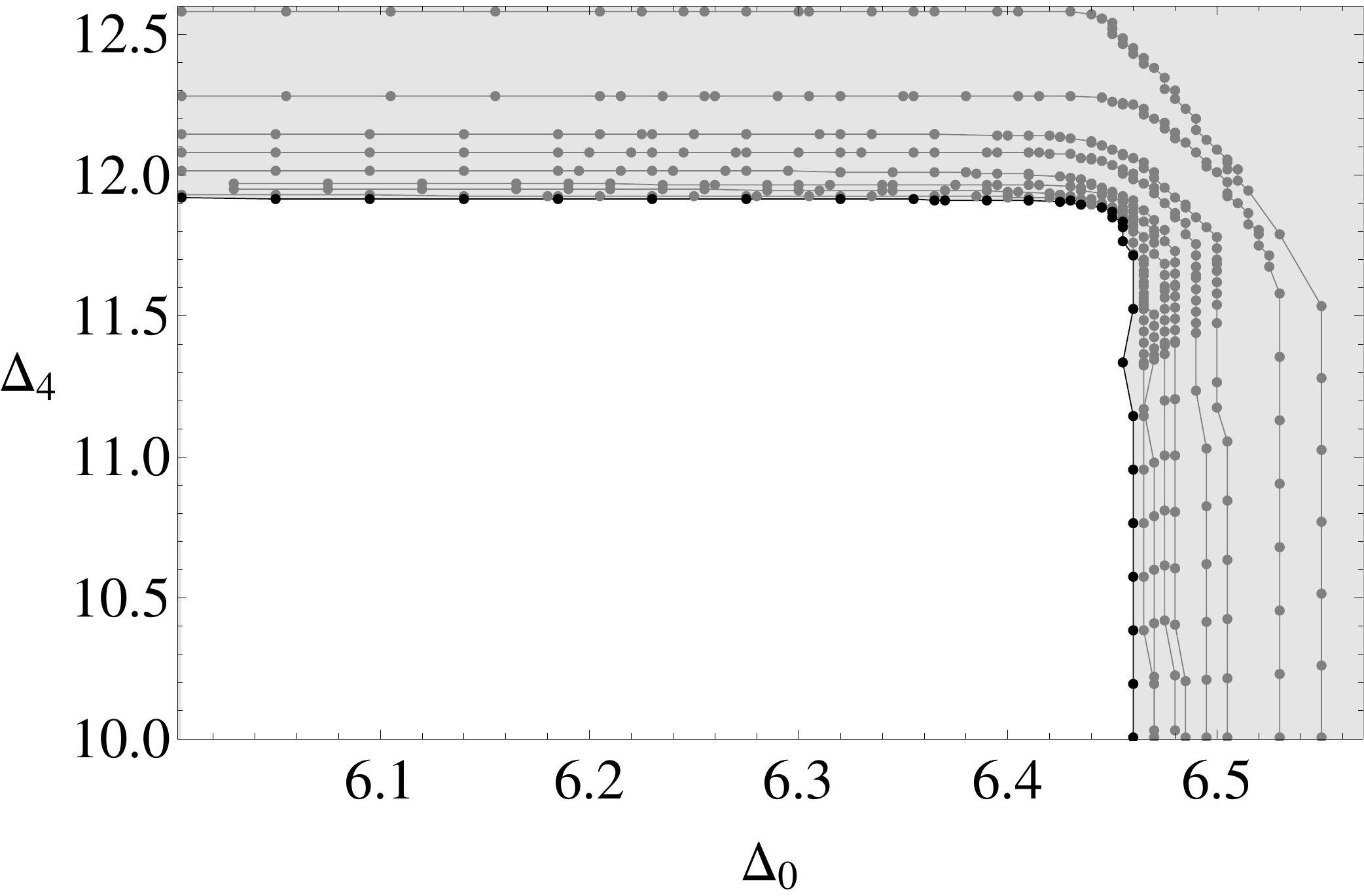}
              \caption{Combined exclusion plots for the lowest dimension operators of spin $0$, $2$ and $4$ in the $A_1$ theory with a cutoff $\Lambda=14,\ldots,22$. These bounds improve on Fig.~\ref{Fig:square} because we enforced the absence of the $\DD[0,4]$ multiplet.}
                \label{Fig:a1square_noD}
            \end{center}
\end{figure}

Finally, we can repeat the analysis leading to Fig.~\ref{Fig:square} while disallowing the $\DD[0,4]$ multiplet. The resulting bounds for the combinations $(\Delta_0,\Delta_2)$ and $(\Delta_0,\Delta_4)$ are shown in \ref{Fig:a1square_noD}. Notice that the $(\Delta_2,\Delta_4)$ plot would be the same as in Fig.~\ref{Fig:square}, because in that case no gap is imposed in the scalar sector and a scalar long multiplet approaching the unitarity bound mimics precisely the $\DD[0,4]$ short operator. Fig.~\ref{Fig:a1square_noD} shows significant improvement over Fig.~\ref{Fig:square}: the shape is more rectangular and the numerical values of the bounds are also significantly lower. We in addition observe a better rate of convergence (not shown in Fig.~\ref{Fig:square}).

\begin{figure}[t!]
             \begin{center}   
              \includegraphics[scale=0.35]{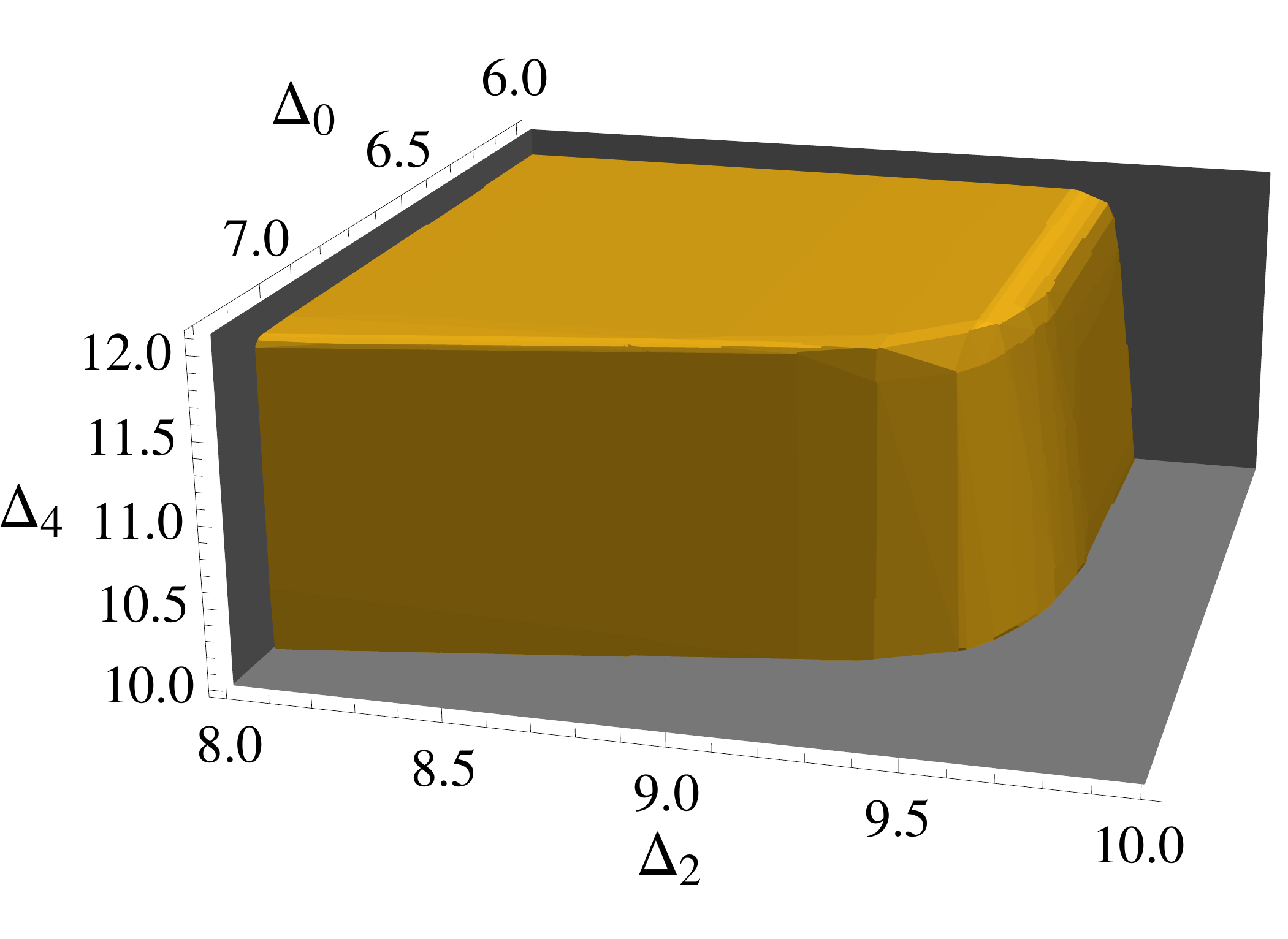}\;     
              \includegraphics[scale=0.35]{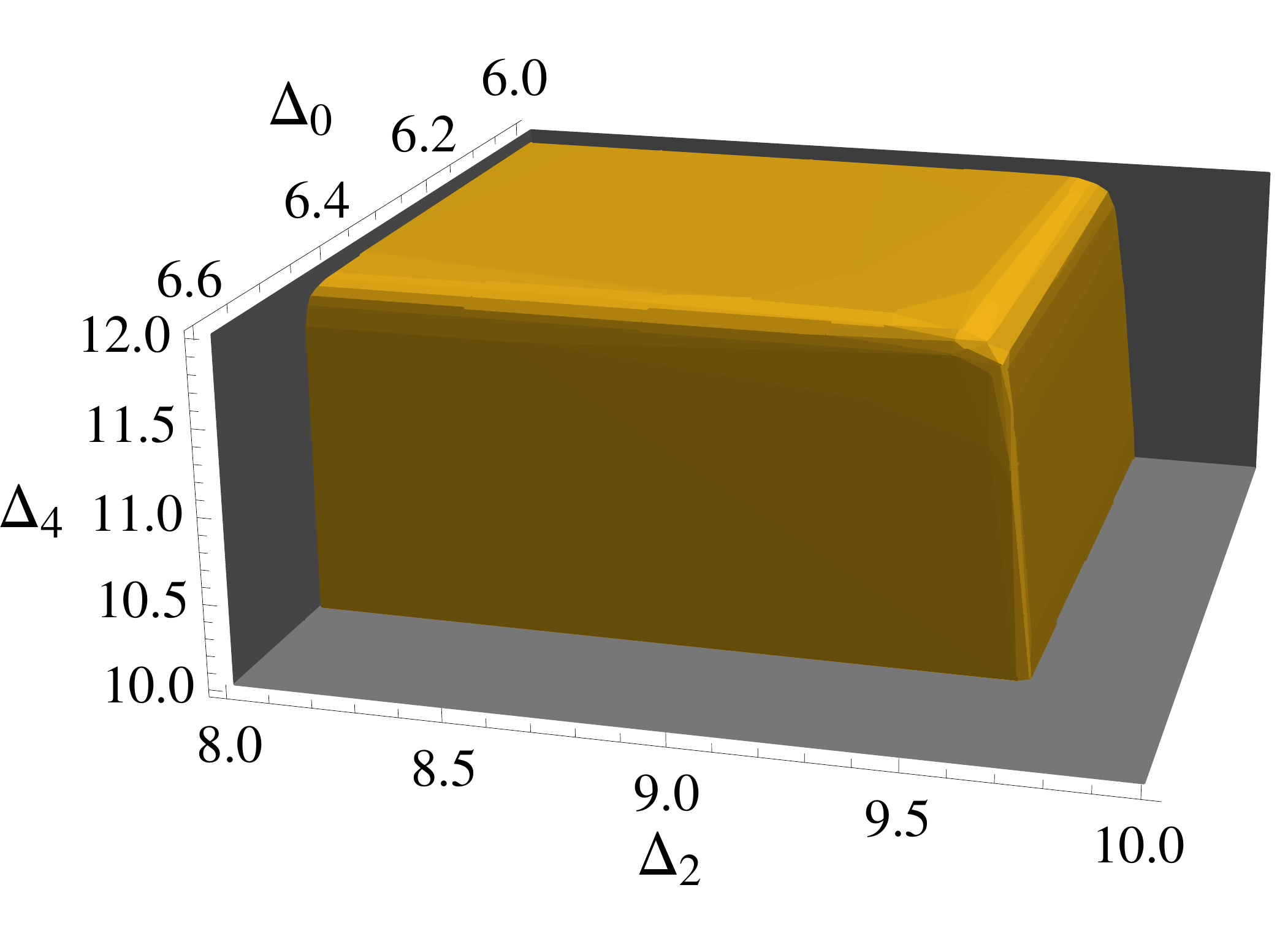}
              \caption{Simultaneous bounds on the spin $0,2,4$ superconformal primary dimensions when a gap is imposed in one of the other channels for a cutoff of $\Lambda=22$. These bounds are for the central charges corresponding to the $A_1$ theory ($c=25$), and are obtained with (left) and without (right) the addition of the short multiplet $\DD[0,4]$. The allowed region is inside of the region delimited by the golden surface.}
              	\label{Fig:cube}
            \end{center}
\end{figure}

In Fig.~\ref{Fig:cube} we impose simultaneous gaps in all three of the lowest spin operators, with (left) or without (right) the $\DD[0,4]$ multiplet. The region inside the approximate cuboid is allowed, the region outside of it is excluded. This being just the three-dimensional analogue of the two-dimensional plots shown in Figs.~\ref{Fig:square} and \ref{Fig:a1square_noD}, the allowed region in the space of these three dimensions should eventually converge to a perfect cuboid, which would once more demonstrate the uniqueness of the solution with scaling dimension determined by the location of its vertex. We ascribe the deviation from this cuboidal shape in Fig.~\ref{Fig:cube} to the finite value of $\Lambda$. Naturally, since we expect that with $\Lambda \to \infty$ the numerical bootstrap will show that this multiplet is absent at $c=25$, the cuboids on the left and on the right  
of Fig.~\ref{Fig:cube} should be converging to the same final bound. In this sense demanding the absence of the $\DD[0,4]$ multiplet is just a trick to overcome the slow convergence of our numerical results, leading to bounds closer to their $\Lambda \to \infty$ values.

%% file: sections/acknowledgments.tex

\acknowledgments
The authors have greatly benefited from discussions with Clay Cordova, Shiraz Minwalla, Nati Seiberg, and with
 many participants at the Bootstrap 2014 meeting in Porto and at the Bootstrap 2015 meeting at the Weizmann Institute, where preliminary versions of this work have been presented.
We are especially indebted to David Simmons-Duffin for many useful insights and for help with his \texttt{SDPB} software, and to Sheer El-Showk for sharing his \texttt{CPLEX} interface.
We are grateful to the Aspen Center for Physics (partially supported by NSF grant \#1066293) for hosting the summer workshop ``From scattering amplitudes to the conformal bootstrap'' and providing the perfect working environment  that allowed us to finalize this paper.
C. B. gratefully acknowledges support from the Frank and Peggy Taplin Fellowship at the IAS.
C. B. is also supported in part by the NSF through grant PHY-1314311.
M. L. is supported in part by FCT - Portugal, within the POPH/FSE programme, through grant SFRH/BD/70614/2010.
L. R. gratefully acknowledges the generosity of the Simons and Guggenheim foundations and the wonderful hospitality of the IAS, Princeton and of the KITP, Santa Barbara during his sabbatical year.
The work of M.L. and L.R. is supported in part by NSF Grant PHY-1316617.

%% file: appendices/A_representations.tex

\section{Unitary representations of \texorpdfstring{$\ospf(8^\star|4)$}{OSp(8*|4)}} 
\label{App:representations}

We recall the classification of unitarity irreducible representations of the $\ospf(8^\star|4)$ superalgebra. These have been described in \cite{Dobrev:1985qv,Dobrev:2002dt,Bhattacharya:2008zy}, and are reviewed in \cite{Beem:2014kka}. There are four linear relations at the level of quantum numbers that, if satisfied by the superconformal primary state in a representation, guarantee that the resulting representation is (semi-)short. We adopt the following notation for labelling these relations:
\begin{alignat}{3}
&\AA~&:&~ \Delta=\tfrac12 c_1+ c_2 + \tfrac{3}{2} c_3+2(d_1+d_2)+6~,\qquad&&\nn\\
&\BB~&:&~ \Delta=\tfrac12 c_1+ c_2 +2(d_1+d_2)+4~, && c_3=0~,\nn\\
&\CC~&:&~ \Delta=\tfrac12 c_1 +2(d_1+d_2)+2~, && c_2=c_3=0~,\nn\\
&\DD~&:&~ \Delta=2(d_1+d_2)~,	   && c_1=c_2=c_3=0~.
\label{eq:shortening}
\end{alignat}
The superconformal primaries of generic multiplets which obey no shortening condition, \ie, long multiplets, obey
\begin{equation}
\LL~:~ \Delta > \tfrac12 c_1 + c_2 + \tfrac{3}{2} c_3 +2(d_1+d_2)+6\,.
\end{equation}
Here $[d_1,d_2]$ are the Dynkin labels of the $\sof(5)_R$ representation of the superconformal primary\footnote{We use $\sof(5)$ conventions for the Dynkin labels, so the $\mathbf 5$ has Dynkin labels $[1,0]$. The authors of \cite{Dolan:2004mu} use $\uspf(4)$ conventions which means that the labels are interchanged.} and $[c_1,c_2,c_3]$ are the Dynkin labels of $\suf^\star(4)$, which are related to the orthogonal basis quantum numbers, $(h_1,h_2,h_3)$, by
\begin{equation}
\label{eq:so6_change_of_basis}
h_1=\tfrac12 c_1+ c_2+\tfrac12 c_3~,\quad h_2=\tfrac12 c_1+\tfrac12 c_3~,\quad h_3=\tfrac12 c_1-\tfrac12 c_3~.
\end{equation}
Short superconformal multiplets can be specified by the type of shortening condition they obey, together with the two $\sof(5)_R$ and the three $\suf^\star(4)$ Dynkin labels of the superconformal primary. For these multiplets the dimension of the superconformal primary is then fixed in terms of this information from \eqref{eq:shortening}. For long multiplets, $\LL$, one must specify the dimension of the superconformal primary in addition to the aforementioned quantum numbers.We will denote such a multiplet by
\begin{equation}
\XX[(\Delta);c_1,c_2,c_3;d_1,d_2]~,\qquad \XX=\AA,\BB,\CC,\DD,\LL~.
\end{equation}
In the special case of operators that transform as symmetric rank $\ell$ traceless tensors of $\sof(5,1)$ (which is the case for most operators discussed in this paper), we simplify the expression to
\begin{equation}
\XX[d_1,d_2]_{(\Delta),\ell}~.
\end{equation}

For a representation in any one of the classes listed in \eqref{eq:shortening}, the structure of null states in the Verma module built on the superconformal primary depends on the $\sof(5,1)$ representation of that primary. Every short representation possesses a single primary null state, with the additional null states being obtained by the action of additional raising operators on the null primary. Different locations for the primary null state lead to different multiplet structures, which we summarize in table \ref{Tab:shortenings}. In all cases, when some of the $c_i$ are written, the last one is necessarily nonzero. The quantum numbers $d_{1,2}$ in all cases are only restricted to be non-negative integers. The multiplets of the type $\BB[c_1,c_2,0;0,0]$, $\CC[c_1,0,0;d_1,d_2]$ with $d_1 + d_2 \leqslant 1$, and $\DD[0,0,0;d_1,d_2]$ with $d_1 + d_2 \leqslant 2$ contain conserved currents or free fields. In particular, the stress tensor multiplet is $\DD[0,0,0;2,0]$, which we denote simply by $\DD[2,0]$.

\begin{table}[ht!]
\centering
\renewcommand{\arraystretch}{1.3}
\begin{tabular}{|>{$}r<{$} @{[} >{$}c<{$} @{,} >{$}c<{$} @{,} >{$}c<{$} @{;} >{$}r<{$} @{,} >{$}r<{$} @{]\,\,\qquad} | @{\qquad}  >{$}r<{$} @{} >{$}l<{$}   @{\qquad} |@{\qquad\,\,[\,} r @{,\,} r @{,\,} r @{\,;\,} r @{,\,} r @{\,]\,\,\qquad}|}
\hline
\hline
\AA & c_1 & c_2 & c_3 & d_1 & d_2 & \QQ_{{\bf1}4}                                            & \, \psi = 0 & 0  & 0  &-1 & 0 & 1\\
\AA & c_1 & c_2 & 0   & d_1 & d_2 & \QQ_{{\bf1}3} \QQ_{{\bf1}4}                              & \, \psi = 0 & 0  & -1 & 0 & 0 & 2\\
\AA & c_1 & 0   & 0   & d_1 & d_2 & \QQ_{{\bf1}2} \QQ_{{\bf1}3} \QQ_{{\bf1}4}                & \, \psi = 0 & -1 & 0  & 0 & 0 & 3\\
\AA & 0   & 0   & 0   & d_1 & d_2 & \QQ_{{\bf1}1} \QQ_{{\bf1}2} \QQ_{{\bf1}3} \QQ_{{\bf1}4}  & \, \psi = 0 & 0  & 0  & 0 & 0 & 4\\
\hline
\BB & c_1 & c_2 & 0 & d_1 & d_2 & \QQ_{{\bf1}3}                             & \,\psi=0 & 0  & -1 & 1 & 0 & 1\\
\BB & c_1 & 0   & 0 & d_1 & d_2 & \QQ_{{\bf1}2} \QQ_{{\bf1}3}               & \,\psi=0 & -1 & 0  & 1 & 0 & 2\\
\BB & 0   & 0   & 0 & d_1 & d_2 & \QQ_{{\bf1}1} \QQ_{{\bf1}2} \QQ_{{\bf1}3} & \,\psi=0 & 0  & 0  & 1 & 0 & 3\\
\hline
\CC & c_1 & 0 & 0 & d_1 & d_2 & \QQ_{{\bf1}2}               & \,\psi=0 & -1 & 1 & 0 & 0 & 1\\
\CC & 0   & 0 & 0 & d_1 & d_2 & \QQ_{{\bf1}1} \QQ_{{\bf1}2} & \,\psi=0 &  0 & 1 & 0 & 0 & 2\\
\hline
\DD & 0 & 0 & 0 & d_1 & d_2 & \QQ_{{\bf1}1} & \,\psi=0 & 1 & 0 & 0 & 0 & 1\\
\hline
\hline
\end{tabular}
\caption{\label{Tab:shortenings}The primary null state for each of the shortened multiplets, expressed in terms of a combination of supercharges acting on the superconformal primary. The supercharges $\QQ_{{\bf A}a}$ transform in the $({\bf 4},{\bf 4})$ of $\uspf(4)_R\times\suf^\star(4)$. The expression in the second column is schematic -- the actual null state is a linear combination of this state with other descendants. The rightmost column contains the Dynkin labels corresponding to the combination of supercharges. Notice that Lorentz indices are implicitly antisymmetrized because of the identical $R$-symmetry indices on each supercharge.}
\end{table}

This structure of null states makes the decomposition rules for long multiplets transparent. Starting with a generic multiplet $\LL$ approaching the $\AA$-type bound for its dimension, the following decompositions take place (which decomposition occurs depends on the $\sof(5,1)$ representation of the long multiplet):
\begin{alignat}{3}
&\LL\lbrack \Delta^*+\delta;c_1,c_2,c_3;d_1,d_2\rbrack~
&\underset{\delta\to0}{\longrightarrow}
&~~\AA\lbrack c_1,c_2,c_3;d_1,d_2\rbrack&~\oplus~&\AA\lbrack c_1,c_2,c_3-1;d_1, d_2+1\rbrack~,\nn\\
&\LL\lbrack \Delta^*+\delta;c_1,c_2,0;d_1,d_2\rbrack~
&\underset{\delta\to0}{\longrightarrow}
&~~\AA\lbrack c_1,c_2,0;d_1,d_2\rbrack&~\oplus~&\BB\lbrack c_1,c_2-1,0;d_1,d_1+2\rbrack~,\nn\\
&\LL\lbrack \Delta^*+\delta;c_1,0,0;d_1,d_2\rbrack~
&\underset{\delta\to0}{\longrightarrow}
&~~\AA\lbrack c_1,0,0;d_1,d_2\rbrack&~\oplus~&\CC\lbrack c_1-1,0,0;d_1,d_2+3\rbrack~,\nn\\
&\LL\lbrack \Delta^*+\delta;0,0,0;d_1,d_2\rbrack~
&\underset{\delta\to0}{\longrightarrow}
&~~\AA\lbrack 0,0,0;d_1,d_2\rbrack&~\oplus~&\DD\lbrack 0,0,0;d_1,d_2+4\rbrack~.
\label{eq:decomposition}
\end{alignat}
In the partial wave analysis of the stress tensor multiplet four-point function the only recombinations that play a role are the second and last ones, where in both cases only the second multiplet appearing in the decomposition is allowed by selection rules in the OPE.
There is a relatively short list of multiplets that can never appear in a recombination rule:
\begin{eqnarray}
&&\BB[c_1,c_2,0;d_1,\{0,1\}]~,\nn\\
&&\CC[c_1,0,0;d_1,\{0,1,2\}]~,\\
&&\DD[0,0,0;d_1,\{0,1,2,3\}]~.\nn
\end{eqnarray}
Amusingly, the $\qq$-chiral operators that give rise to currents of the protected chiral algebra of \cite{Beem:2014kka} are all selected from among these non-recombinant representations. The operators in this list that make an appearance in the OPE of two stress tensors are precisely the ones whose OPE coefficients that were fixed in \eqref{eq:ashort}.

%% file: appendices/B_blocks.tex

\section{Superconformal blocks} 
\label{App:superblocks}

In this appendix we collect various expressions that are relevant for the decomposition into superconformal blocks of the stress tensor multiplet four-point function.

Let $\GG_{\D}^{(\ell)}(\D_{12},\D_{34};z,\bar z)$ with $\Delta_{ij} = \Delta_i - \Delta_j$ be the six-dimensional non-supersymmetric conformal blocks for a four-point function of scalar operators with conformal dimension $\D_i$, $i=1,\ldots,4$. These conformal blocks were given in closed form in \cite{Dolan:2003hv,Dolan:2011dv},
\begin{eqnarray}
\GG_{\D}^{(\ell)}(\D_{12},\D_{34};z,\bar z) &=& \FF_{00} - \frac{\ell+3}{\ell+1} \FF_{-1 1} + \frac{(\Delta-4)(\ell+3)}{16(\Delta-2)(\ell+1)} \nn\\
&& \frac{(\Delta-\ell-\Delta_{12}-4)(\Delta-\ell+\Delta_{12}-4)(\Delta-\ell+\Delta_{34}-4)(\Delta-\ell-\Delta_{34}-4)}{(\Delta-\ell-5)(\Delta-\ell-4)^2(\Delta-\ell-3)}\FF_{02} \nn\\
&-& \frac{\Delta -4 }{\Delta -2}\frac{(\Delta+ \ell-\Delta_{12})(\Delta+\ell+\Delta_{12})(\Delta+\ell+\Delta_{34})(\Delta+\ell-\Delta_{34})}{16 ( \Delta+ \ell-1)(\Delta+\ell)^2(\Delta+\ell+1)} \FF_{11} \nn\\
&+&\frac{2( \Delta-4)(\ell+3)\Delta_{12} \Delta_{34}}{(\Delta+\ell)(\Delta+\ell-2)(\Delta+\ell-4)(\Delta+\ell-6)} \FF_{01}~,
\label{eq:6dblock}
\end{eqnarray}
where
\begin{eqnarray}
\FF_{nm}(z, \bar{z}) &=& \frac{(z \bar z)^\frac{\Delta-\ell}{2}}{(z-\bar{z})^3}\left(\left(-\frac{z}{2}\right)^\ell z^{n+3} \bar{z}^{m} {}_2F_1\left(\frac{\Delta +\ell-\Delta_{12}}{2}+n,\frac{\Delta +\ell+\Delta_{34}}{2}+n,\Delta+\ell+2n,z \right) \right. \nn\\
&&\left. {}_2F_1\left(\frac{\Delta -\ell-\Delta_{12}}{2}-3+m,\frac{\Delta -\ell+\Delta_{34}}{2}-3+m,\Delta-\ell-6+2m,\bar z \right) - \left( z \longleftrightarrow \bar{z} \right)\right)~.\nn
\end{eqnarray}

The harmonic functions for the various $\sof(5)_R$ irreducible representation appearing in the decomposition of the four-point function \eqref{eq:decompG} are given by \cite{Nirschl:2004pa,Dolan:2004mu}
\begin{eqnarray}
Y^{[4,0]}(\alpha , \bar \alpha) &=& \sigma^2 +\tau ^2 + 4 \sigma  \tau -\frac{8 (\sigma +\tau )}{9}+\frac{8}{63}~,\nn\\
Y^{[2,2]}(\alpha , \bar \alpha) &=& \sigma^2 -\tau^2 -\frac{4 (\sigma -\tau )}{7}~, \nn\\
Y^{[0,4]}(\alpha , \bar \alpha) &=& \sigma^2 +\tau ^2 - 2 \sigma  \tau -\frac{2 (\sigma +\tau )}{3}+\frac{1}{6}~, \nn \\
Y^{[0,2]}(\alpha , \bar \alpha) &=& \sigma -\tau~,\nn\\
Y^{[2,0]}(\alpha , \bar \alpha) &=& \sigma +\tau -\frac{2}{5}~,\nn\\
Y^{[0,0]}(\alpha , \bar \alpha) &=& 1 ~,
\label{eq:Yprojectors}
\end{eqnarray}
where $\sigma= \alpha \bar \alpha$ and $\tau = (\alpha -1 )(\bar \alpha - 1)$. 

The superconformal blocks for the $\DD[2,0]$ four-point function decomposition were studied in \cite{Dolan:2004mu}. We quote here the results relevant for our purposes. For each $\sof(5)_R$ channel the superconformal blocks can be extracted from two functions $a(z,\bar z)$ and $h(z)$ as follows
\begin{eqnarray}
A_{[4,0]}(z,\bar z)&=& \frac{1}{6} u^{4} \Delta_{2}\left[ u^2 a(z,\bar z) \right]~,\nn\\
A_{[2,2]}(z,\bar z)&=& \frac{1}{2} u^{4} \Delta_{2}\left[ u(v-1) a(z,\bar z)\right]~,\nn\\
A_{[0,4]}(z,\bar z)&=& \frac{1}{6} u^{4} \Delta_{2}\left[ u(3(v+1)-u) a(z,\bar z)\right]~,\nn\\
A_{[0,2]}(z,\bar z)&=& \frac{1}{2} u^{4} \Delta_{2}\left[ (v-1)\left((v+1)-\frac{3}{7}u\right) a(z,\bar z) \right] \nn\\
&& - u^2 \left( \frac{ (z-2) z h'(z)+(\bar z-2) \bar z h'(\bar z)}{2 (z-\bar z)^2} +(z+ \bar z- z \bar z) \frac{ h(z) -  h(\bar z) }{ (z-\bar z)^3} \right)~,\nn \\
A_{[2,0]}(z,\bar z)&=& \frac{1}{2} u^{4} \Delta_{2}\left[ \left((v-1)^2 - \frac{1}{3}u(v+1)+\frac{2}{27}u^2\right) a(z,\bar z) \right] \nn\\
&& + u^2 \left(z \bar z  \frac{  h(z)-  h(\bar z)}{(z-\bar z)^3} -\frac{ z^2 h'(z)+\bar z^2 h'(\bar z)}{2 (z-\bar z)^2} \right)~,\nn\\
A_{[0,0]}(z,\bar z)&=& \frac{1}{4} u^{4} \Delta_{2}\left[ \left((v+1)^2- \frac{1}{5}(v-1)^2-\frac{3}{5}u (v+1)+\frac{3}{35}u^2\right) a(z,\bar z)\right] \nn\\
&& - u^2 \frac{ (5 (1-z) + z^2) h'(z)+(5(1-\bar z) + \bar z^2) h'(\bar z)}{5 (z-\bar z)^2}\nn\\
&& + u^2  \left(2 z \bar z +5(1- z)+ 5(1- \bar z)\right) \frac{ h(z) - h(\bar z) }{5 (z-\bar z)^3}~.
\label{eq:Ainah}
\end{eqnarray}
Each $A_{[i,j]}(z,\bar z)$ admits a decomposition in a finite sum of conformal blocks given in \eqref{eq:6dblock} with $\D_i=4$, with positive coefficients. As explained in Section \ref{Sec:blockdecomp}, the relative coefficients between conformal primaries belonging to the same superconformal multiplets are fixed, and there is only one unfixed OPE coefficient per superconformal multiplet. This is apparent from the form of \eqref{eq:Ainah}, where we see we only need to specify how each superconformal multiplet contributes to $a(z,\bar z)$ and $h(z)$. This information is summarized in table \ref{tab:superblocks}. To go from the contribution of each superconformal multiplet to $a(z,\bar z)$ and $h(z)$ to a finite sum over conformal blocks, which includes acting with the differential operator $\D_2$, one can make use of the recurrence relations given in Appendix D of \cite{Dolan:2004mu}, which were corrected in \cite{Chester:2014fya}.

%% file: appendices/C_lightcone.tex

\section{Lightcone limit}
\label{App:lightcone}

The lightcone limit of crossing equations has proved useful for studying the large spin asymptotics of the operator spectrum \cite{Fitzpatrick:2012yx,Komargodski:2012ek,Alday:2013cwa,Fitzpatrick:2014vua,Vos:2014pqa,Fitzpatrick:2015qma,Kaviraj:2015cxa,Alday:2015eya,Kaviraj:2015xsa,Alday:2015ota}. Here we analyze the crossing equation \eqref{eq:finalcrossingeq} in the lightcone limit. We show that the $\BB[0,2]_{\ell}$ operators with $\ell \gg 1$ are necessarily present in the theory, and obtain the large $\ell$ limit of their OPE coefficients. We also find the large spin limit of the the OPE coefficients and anomalous dimensions of twist eight long multiplets. The large $c$ limit of all these results (in addition to similar calculations not shown here for the twist ten and twelve long multiplets) match with expectations from eleven-dimensional supergravity.

\subsection{Lightcone crossing symmetry equation}
\label{subsubsec:lightcone_crossing_equation}

Our first order of business is to determined the lightcone limit of the building blocks of the main crossing equation \eqref{eq:finalcrossingeq}, which we recall takes the form
\begin{equation}
\begin{split}
&\underbrace{z \bar{z} a^{u}\left(z,\bar{z}\right)-(z-1) \left(\bar{z}-1\right) a^{u}\left(1-z,1-\bar{z}\right)}_{\CC^{u}(z,\bar z)} + \underbrace{z \bar{z} a^{\chi}\left(z,\bar{z}\right)-(z-1) \left(\bar{z}-1\right) a^{\chi}\left(1-z,1-\bar{z}\right)}_{\CC^{\chi}(z,\bar z)} \\ &\qquad = \underbrace{\frac{1}{(z - \bar z)^3} \left( \frac{h\left(1-\bar{z}\right)-h(1-z)}{(z-1) \left(\bar{z}-1\right)}+\frac{h\left(\bar{z}\right)-h(z)}{z \bar{z}} \right)}_{\CC^{h}(z,\bar z)}~.
\end{split}
\end{equation}
where we have explicitly separated out the protected and unprotected parts. The function $h(z)$ is given in \eqref{eq:h_sol} and it has the following asymptotic behavior
\begin{equation}
\label{eq:hasympt}
\begin{split}
h(z) &\underset{z \to 0}{=} \frac{1}{z} -\frac{1}{2} - \frac{8}{c} z + O(z^2)~,\\
h(z) &\underset{z \to 1}{=} \frac{- 1}{3 (1-z)^3} + \frac{1}{(1-z)^2} - \frac{1 + 8/c}{1-z} + O\left(\log(1-z)\right)~.
\end{split} 
\end{equation}
As in Section \ref{Sec:blockdecomp} we have set the integration constant $\b_5 = -1/6 + 8/c$ so the coefficient of $z^0$ in the expansion of $h(z)$ becomes $-1/2$, the same as the $z^0$ contribution of $h^{\text{at}}_{0}(z)$. The conformal block decomposition of $h(z)$ then takes the form
\begin{equation}
h(z) = \sum_{\ell = -4,\, \ell \text{ even}} b_\ell \, h^{\text{at}}_{\ell + 4}(z)~.
\end{equation}
The coefficients $b_\ell$ define the function $a^\chi(z,\bar z)$ as
\begin{equation}
a^{\chi}(z,\bar z) = \sum_{\ell=0,\, \ell \text{ even}}^\infty 2^{\ell} b_{\ell} \, a^\text{at}_{\ell+4,\ell}(z,\bar z)~.
\end{equation}
The atomic blocks $a^\text{at}_{\Delta,\ell}(z,\bar z)$ are defined in \eqref{eq:ahatom}. We will need their asymptotic behavior:
\begin{equation}
\label{eq:aatasympt}
\begin{split}
a^{\text{at}}_{\Delta,\ell}(z,\bar z) &\underset{z \to 0}{=} (z \bar z)^{(\Delta - \ell - 8)/2} \left( \frac{2^{2-\ell } \bar z^{\ell } \, _2F_1\left(\frac{1}{2} (\ell +\Delta +2),\frac{1}{2} (\ell +\Delta +4);\ell +\Delta +4;\bar z\right)}{(\Delta -\ell -2) (\Delta +\ell +2)} + O(z) \right)~,\\
a^{\text{at}}_{\Delta,\ell}	&\underset{z \to 1}{=} O(1)~.
\end{split}
\end{equation}

In the $z \to 1$ limit we find from \eqref{eq:hasympt} that the crossing symmetry becomes
\begin{equation}
\boxed{
\begin{aligned}
\CC^h(z,\bar z) \underset{z \to 1}{=}
&\phantom{+}\frac{1}{(1-z)^3}\left(\frac{1}{3 \bar z (1-\bar z)^3}\right)\\
&+\frac{1}{(1-z)^2}\left(\frac{1}{3 \bar z (1-\bar z)^3}\right)\\
&+ \frac{1}{1-z} \left( \frac{\bar z^5-2 \bar z^4+\bar z^3-2 \bar z+1}{3 (\bar z-1)^3 \bar z^3}+\frac{8 (\bar z-\log (\bar z)-1)}{c (\bar z-1)^4}\right)\\
&+ O\left(\log(1-z)\right)~.
\end{aligned}
}
\end{equation}
Replacing $z \to (1-z)$ in \eqref{eq:aatasympt} we find that in the same limit
\begin{equation}
\label{eq:alongcrossasympt}
(1-z)(1-\bar z) a^{u}(1-z,1-\bar z) \underset{z \to 1}{=} O\left((1-z)^{(\Delta_* - \ell - 6)/2}\right) = O(1)~,
\end{equation}
where in the last equality we have used $\Delta_* \geq \ell + 6$ with $\Delta_*$ the dimension of the operator(s) of lowest twist. This part of $\CC^u$ is therefore completely regular as $z \to 1$ and we find
\begin{equation}
\boxed{
\CC^u(z,\bar z) \underset{z \to 1}{=} \left( \lim_{z \to 1} z \bar z a^u(z,\bar z) \right) + O(1)~,}
\end{equation}
where the behavior of the first term is unknown since we cannot yet say much about $a^u(z,\bar z)$ in the relevant limit.

Finally for $\CC^\chi$ we observe the $\Delta = \ell + 4$ specialization of \eqref{eq:aatasympt} gives
\begin{equation}
\label{eq:aatasympttwist4}
a^{\text{at}}_{\ell + 4,\ell}(z,\bar z) \underset{z \to 0}{=}  \frac{- 2^{-\ell}}{z^2 \bar z^5} h^{\text{at}}_{\ell + 4}(\bar z) + O\left(\frac{1}{z}\right)~,
\end{equation}
and therefore
\begin{equation}
\label{eq:ashortasympt}
a^{\chi}(z,\bar z) \underset{z \to 0}{=} \frac{-1}{z^2 \bar z^5} \left(  h(\bar z) - b_{-4} h^{\text{at}}_0(\bar z) - b_{-2} h^{\text{at}}_2 (\bar z) \right) + O\left( \frac{1}{z} \right)~,
\end{equation}
which leads directly to
\begin{equation}
(1-z)(1-\bar z) a^{\chi}(1-z,1-\bar z) \underset{z \to 1}{=} \frac{-1}{(1-z) (1-\bar z)^4} \left(  h(1-\bar z) - b_{-4} h^{\text{at}}_0(1-\bar z) - b_{-2} h^{\text{at}}_2 (1-\bar z) \right) + O(1)~.
\end{equation}
We thus find that
\begin{equation}
\boxed{
\CC^\chi(z,\bar z) \underset{z \to 1}{=} \left( \lim_{z \to 1} z \bar z a^\chi(z,\bar z) \right) - \frac{-1}{(1-z) (1-\bar z)^4} \left(  h(1-\bar z) - b_{-4} h^{\text{at}}_0(1-\bar z) - b_{-2} h^{\text{at}}_2 (1-\bar z) \right) + O(1)~.
}
\end{equation}
Although $a^{\chi}(z,\bar z)$ is a known function, we do not have sufficient analytic control over it to find its $z \to 1$ behavior. We will make an estimate below.

Combining the three boxed equations above we have our final lightcone crossing symmetry equation
\begin{equation}
\boxed{
\label{eq:mainlightconeeqn}
\begin{aligned}
\lim_{z \to 1} \left( z \bar z a^{\chi}(z,\bar z) + z \bar z a^{u}(z,\bar z)\right) = 
&\phantom{+}\frac{1}{3 \bar z (1-\bar z)^3} \left(\frac{1}{(1-z)^3}+\frac{1}{(1-z)^2}+\frac{1}{1-z}\right)\\
&+\frac{8}{c(1-z)} \left(\frac{-2 \bar z^2-5 \bar z+1}{\bar z (1-\bar z)^5}-\frac{6 \bar z \log (\bar z)}{(1-\bar z)^6}\right)\\
&+ O\left(\log(1-z)\right)~.
\end{aligned}
}
\end{equation}
In the next subsection we will extract specific information about the spectrum and OPE coefficients from this equation. Before doing so, it is worthwhile to discuss it in slightly broader terms. To this end we compare \eqref{eq:mainlightconeeqn} with a specific solution to the crossing symmetry equations at large $c$. At strictly infinite $c$ the only known solution to the crossing symmetry equations, and therefore also to \eqref{eq:mainlightconeeqn}, is the so-called mean field solution $a^\text{MF}(z,\bar z)$. By definition this is the four-point function that is totally disconnected, \ie, it is a sum of products of two-point functions. At order $1/c$ we get corrections to the mean field solution from eleven-dimensional supergravity on ${\rm AdS}_7 \times S^4$. The resulting function $a^{\text{SG}}(z,\bar z)$ can be read off from the results in \cite{Arutyunov:2002ff,Heslop:2004du}.\footnote{As a check of our computations we have verified that the supergravity result satisfies \eqref{eq:mainlightconeeqn}. The relation between what is called $a(z,\bar z)$ and $h(z)$ here and the $F(z, \bar z) $ in \cite{Heslop:2004du} is  $F(z, \bar{z}) = - z \bar{z} (z - \bar z)^2 a(z, zb) -   \frac{1}{z \bar{z}} \frac{ h(z) - h(\bar{z})}{z - \bar{z}}$.}

What \eqref{eq:mainlightconeeqn} tells us is that as $z \to 1$, the leading behavior of the left hand side -- up to $O\left(\log(1-z)\right)$ -- is the \emph{same} as the mean field solution plus $1/c$ times the supergravity result, even for finite values of $c$. More precisely, for all $c$ we can write
\begin{equation}
\label{sugracomparison}
\lim_{z\to 1} z \bar z a(z,\bar z) = z \bar z  a^\text{MF}(z,\bar z) + \frac{1}{c} z \bar z  a^\text{SG}(z,\bar z) + O\left(\log(1-z)\right)~.
\end{equation}
This is why the mean field theory and the supergravity solutions will feature prominently in the rest of our discussion.

\subsection{OPE decomposition}
\label{subsec:lightcone_ope_decomposition}

To get a handle on \eqref{eq:mainlightconeeqn} we follow \cite{Fitzpatrick:2012yx,Komargodski:2012ek} and consider the series expansion around $\bar z = 0$. On the left-hand side we find a power of $\bar z^{(\Delta - \ell - 6)/2}$ for the leading term of a conformal block. In this sense powers of $\bar z$ count the twist of the operators. On the right-hand side we see that for each inverse power of $(1-z)$ we have an expansion in powers of $\bar z$ with the most singular term being $\bar z^{-1}$, plus, eventually, some logarithmic terms that we discuss below. These powers need to be matched on the left-hand side by operators of approximate twist four, six, eight, etc.. The first two leading orders in $(1-z)$ are the same as the mean field solution. Therefore, in a distributional sense the operators and OPE coefficients of the left-hand side must approximate the mean field solution at large $\ell$ for any even twist. The $1/c$ in the second subleading term in the $(1-z)$ expansion then implies certain corrections to the mean field behavior, which in turn are dictated by the supergravity solution.

\subsubsection{Mean field behavior at large $\ell$}
\label{subsubsec:larg_l_MFT}

Let us first consider the twist four operators. These are present only in $a^{\chi}(z,\bar z)$ and they need to reproduce the coefficient of $\bar z^{-1}$ on the right-hand side of \eqref{eq:mainlightconeeqn}. To see this we take the $\bar z \to 0$ limit of both sides, which allows us to use \eqref{eq:ashortasympt} but with $z$ and $\bar z$ interchanged. In this way we easily find a precise match of all the terms of order $\bar z^{-1}$ on both sides.

Next we can consider the twist six and eight operators, which should reproduce the coefficient of $\bar z^0$ and $\bar z^1$, respectively, on the right-hand side of \eqref{eq:mainlightconeeqn}. Here we face a small problem because the function $a^{\chi}(z,\bar z)$ also contains twist six and twist eight descendants. We therefore need to expand \eqref{eq:ashortasympt} to two orders higher in $z$ (and interchange $z$ and $\bar z$ for the problem at hand). We could in principle compute these corrections if we could resum the terms coming from the subleading corrections in \eqref{eq:aatasympttwist4} (again with $z$ and $\bar z$ interchanged). Unfortunately these terms do not appear to be easily summable. Things simplify if we only aim to reproduce the most singular term in \eqref{eq:mainlightconeeqn}, that is, the term of order $(1-z)^{-3}$ (and we still consider only the terms of order $\bar z^0$ and $\bar z^1$). In that case we can just take the large $\ell$ limit, and the corrections to \eqref{eq:aatasympttwist4} (with $z$ and $\bar z$ interchanged) take a simple factorized form
\begin{equation}
\label{eq:aatasympttwist4corr}
a^{\text{at}}_{\ell + 4,\ell}(z,\bar z) \underset{\substack{\bar z \to 0\\ \ell \to \infty}}{=}  \frac{- 2^{-\ell}}{z^5} h^{\text{at}}_{\ell + 4}(z) \left( \frac{1}{\bar z^2} + \frac{1}{\bar z} \left(\frac{2}{z} + \frac{1}{2}\right) + \frac{3}{z^2} + \frac{1}{z} + \frac{3}{10} +  O(\bar z,\ell^{-1}) \right)~.
\end{equation}
We can sum these corrections to find that
\begin{equation}
\begin{split}
a^{\chi}(z,\bar z) &\underset{\substack{\bar z \to 0\\ z \to 1}}{=} \frac{-1}{z^5} \left(  h(z) - b_{-4} h^{\text{at}}_0(z) - b_{-2} h^{\text{at}}_2 (z) \right)\left( \frac{1}{\bar z^2} + \frac{1}{\bar z} \left(\frac{2}{z} + \frac{1}{2}\right) + \frac{3}{z^2} + \frac{1}{z} + \frac{3}{10}+  O(\bar z) \right) \\
&\qquad + O\left( (1-z)^{-2}\right)\\
&= \frac{1}{(1-z)^3} \left( \frac{1}{3 \bar z^2} + \frac{5}{6 \bar z} + \frac{43}{30} + O(\bar z) \right) + O\left((1-z)^{-2}\right)~,
\end{split}
\end{equation}
and therefore, after substituting in \eqref{eq:mainlightconeeqn},
\begin{equation}
\lim_{z \to 1} z \bar z a^{u}(z,\bar z) =  \frac{1}{(1-z)^3} \left(\frac{1}{6} + \frac{17}{30} \bar z + O(\bar z^2) \right) + O\left( (1-z)^{-2}\right)\,.
\end{equation}
A closer examination of \eqref{eq:aatasympt} reveals a similar structure as for the twist four blocks, namely
\begin{equation}
a^{\text{at}}_{\tau + \ell,\ell} (z,\bar z) \underset{\bar z \to 0}{=} - \frac{2^{-\ell} \bar z^{\tau/2 - 4}}{z^5(\tau - 2)} h^{\text{at}}_{\tau/2 + \ell +2}(z) \left(2 +\bar{z} \frac{8+ z(\tau-2)}{2 z}\right) + O(\bar z^{\tau/2 - 2})~.
\end{equation}

This is useful because we already know how to reproduce a $(1-z)^{-3}$ singularity with $h^{\text{at}}_\beta(z)$, namely
\begin{equation}
\sum_{\ell = -4, \ell \text{ even}} \left( - \frac{3}{\b_1}\right) b_\ell h^{\text{at}}_{\ell + 4} (z) \underset{z \to 1}{=} \frac{1}{(1-z)^3} + O\left((1-z)^{-2}\right)~.
\end{equation}
In this sum $\ell$ is even, which means it is useful only for $\tau /2+\ell+2$ even, that is only when the twist is a multiple of four. To obtain results valid for all twists we must also be able to reproduce a $(1-z)^{-3}$ singularity as a sum over $h^{\text{at}}_{\ell } (z)$ with odd $\ell$.  Simply the term $(1-z)^{-3}$ does not have decomposition in these blocks, but just as the combination given in \eqref{eq:h_sol} had a decomposition in even blocks we can add sub-leading corrections to the $(1-z)^{-3}$ singularity such that it admits a block decomposition. The following function does admit a decomposition in odd $\ell$ blocks,
\begin{equation} 
\frac{z^3}{3}+\frac{1}{z-1}+\frac{1}{(z-1)^2}+\frac{1}{3 (z-1)^3}+\frac{1}{3} = \sum_{\ell =3}^{\infty } h^{\text{at}}_{\ell } d_{\ell }~,
\end{equation}
with the coefficients $d_\ell$ given by
\begin{equation} 
d_\ell= \frac{\left(\ell ^3-6 \ell ^2+11 \ell -6\right) \Gamma (\ell +3)}{9 \pi ^{-\frac{1}{2}} 2^{\frac{1}{4} (8 \ell -4)} \Gamma \left(\ell -\frac{1}{2}\right)}\,.
\end{equation}
Using that information it is not hard to find the leading behavior of the OPE coefficients of the twist six operators, which take the form\footnote{We refer to \cite{Fitzpatrick:2012yx,Komargodski:2012ek} for a more careful derivation which shows that the terms shown in this equation are all reliable at large $\ell$.}
\begin{equation}
\lambda^2_{6 + \ell, \ell} \underset{\ell \to \infty}{=} 2^\ell  d_{\ell +5} + \ldots = \frac{\sqrt{\pi } 2^{-\ell} \ell ^{13/2}}{4608}  + \ldots~.
\end{equation}
This proves that the $\BB[0,2]_{\ell-1}$ multiplets with asymptotically large $\ell$ are necessarily always present in a six-dimensional $(2,0)$ theory. For the twist eight operators we need to first subtract the descendants of the twist six operators, which we find account for $15/30$ of the $17/30$. Repeating the above procedure we find that
\begin{equation}
\label{eq:twist8coeff}
\lambda^2_{8 + \ell, \ell} \underset{\ell \to \infty}{=}  \frac{3 \cdot 2^{\ell}}{5} b_{\ell + 2} +\ldots = \frac{2^{-\ell} \sqrt{\pi } \ell ^{13/2}}{30720} + \ldots~.
\end{equation}
This equation may be a little misleading as it needs to be understood more in a distributional sense than as an exact equality. We will explain this in more detail below. We have checked that these OPE coefficients (as well as the twist ten and twelve ones) match the mean field theory OPE coefficients derived in \cite{Heslop:2004du} in the large $\ell$ limit.\footnote{The relation between the (square) of the OPE coefficients of \cite{Heslop:2004du} and here is $\lambda^{2,\text{here}}_{\tau+\ell,\ell}= 2^\ell (\ell+1) \lambda^{2,\text{there}}_{\tau + \ell, \ell}$.}

\subsubsection{\texorpdfstring{$1/c$}{1/c} corrections}
\label{subsubsec:finite_c_corrections}

We would now like to investigate the leading corrections $\delta a(z,\bar z)$ to the mean field solution. In the lightcone limit the manifest changes are due to $1/c$ corrections on the right-hand side of \eqref{eq:mainlightconeeqn} and in $a^{\chi}(z,\bar z)$. These will then induce some corrections to the very high spin operators in $a^{u}(z,\bar z)$ that we would like to compute. Our main equation therefore becomes
\begin{equation}
\label{eq:deltamainlightconeeqn}
\begin{split}
\lim_{z \to 1} \left( z \bar z \delta a^{\chi}(z,\bar z) + z \bar z \delta a^{u}(z,\bar z)\right) = &\phantom{+}\frac{8}{c(1-z)} \left(\frac{-2 \bar z^2-5 \bar z+1}{\bar z (1-\bar z)^5}-\frac{6 \bar z \log (\bar z)}{(1-\bar z)^6}\right)\\
&+ O\left(\log(1-z)\right)\,.
\end{split}
\end{equation}
Before we analyze this equation, we should note that the mean-field solution has non-zero OPE coefficients for all the operators of high spin and twist in $a^{u}(z,\bar z)$, and the corrections to these positive coefficients that we are about to compute are subleading (because they sum to a power $(1-z)^{-1}$ instead of the $(1-z)^{-3}$ for the mean field solution). They will therefore not be able to affect the positivity of the original number.

We again expand the right-hand side as a power series in $\bar z$. The leading order is $\bar z^{-1}$ which, as before, corresponds to twist four operators and is therefore captured by $a^{\chi}(z,\bar z)$. Indeed, using \eqref{eq:ashortasympt} and interchanging $z$ and $\bar z$ we find that
\begin{equation}
\label{eq:ashortasymptv2}
\begin{split}
\lim_{\substack{z \to 1\\ \bar z\to 0}} \left( z \bar z \delta a^{\chi}(z,\bar z)\right) &= \lim_{z \to 1} \left( \frac{-1}{\bar z  z^4} \left(  \delta h(z) - \delta b_{-4} h^{\text{at}}_0(z) - \delta b_{-2} h^{\text{at}}_2 (z) \right) + O\left( \frac{1}{\bar z} \right)\right)\\
&=  \left(\left(\frac{8}{c \bar z (1-z)} + O\left(\log(1-z)\right)\right) + O\left(\frac{1}{\bar z}\right)\right)~,
\end{split}
\end{equation}
which reproduces the coefficient of $\bar z^{-1}$ in \eqref{eq:deltamainlightconeeqn}. The contribution of the twist four operators is therefore consistent.

To say something about the higher twist operators we need to use \eqref{eq:aatasympttwist4corr} again. As usual we relate the large $\ell$ limit to the most singular term in $(1-z)$, and therefore the correction to \eqref{eq:ashortasymptv2} takes the form
\begin{equation}
\label{eq:ashortasymptv3}
\begin{split}
\lim_{\substack{z \to 1\\ \bar z\to 0}} \left( z \bar z \delta a^{\chi}(z,\bar z)\right) &= \frac{8}{c(1-z)} \left( \frac{1}{\bar z} + \frac{5}{2} + \frac{43}{10} \bar z + O(\bar z^2) \right) + O(\log(1-z))\,.
\end{split}
\end{equation}
Combining this equation with \eqref{eq:deltamainlightconeeqn} we now find 
\begin{equation}
\label{eq:alongasympt}
\lim_{z \to 1} \left(z \bar z\delta a^{u} (z,\bar z)\right) = \frac{8}{c(1-z)} \left( - \frac{5}{2} - \frac{163}{10} \bar z  - 6 \log(\bar z) \bar z + O(\bar z^2) \right) + O(\log(1-z))\,.
\end{equation}
The leading term is a correction to the OPE coefficients of the twist six operators in $a^{u}(z,\bar z)$. The subsequent two terms correct both the OPE coefficients and the scaling dimensions of the twist eight long multiplets. (The absence of a logarithm at the leading order is consistent with the fact that the twist six operators are protected.) We see that the corrections obtained from supergravity are in fact universal, \emph{i.e.} at large spin the corrections to OPE coefficients and operator dimensions have the same structure for all the $(2,0)$ theories, and only the prefactor $1/c$ is different. Notice that these are additional corrections that appear on top of $1/\ell$ corrections to the mean field solution.

Let us finally compute the anomalous dimensions at large spin and match them to the supergravity result. We write
\begin{equation}
a^{u} (z,\bar z) =  \sum_{\ell} \int d \tau \, c(\tau,\ell) a^{\text{at}}_{\tau + \ell,\ell}(z,\bar z)~,
\end{equation}
with a distribution $c(\tau,\ell)$ that takes the form
\begin{equation}
c(\tau,\ell) = \sum_n \delta(\tau - 2 \ell - 2 n - \gamma(n,\ell)) P_{n,\ell}~,
\end{equation}
and as $\ell \to \infty$ we have $\gamma(n,\ell) \to 0$ and $P_{n,\ell} \to P^{\text{mft}}_{n,\ell}$ which are the mean field theory values. The leading-order correction then takes the form
\begin{equation}
\delta a^{u}(z,\bar z) = \sum_{\ell} \sum_n \left( \delta P^{\text{MFT}}_{n,\ell} + \frac{1}{2} P^{\text{MFT}}(n,\ell) \gamma(n,\ell) \frac{d}{d n} \right) a^{\text{at}}_{2n + \ell, \ell}(z,\bar z)~.
\end{equation}
As before, we claim again that the right-hand side is an accurate representation of the summand for asymptotically large $\ell$, and as $z \to 1$ it reproduces \eqref{eq:alongasympt}. In the limit where $\bar z \to 0$ we can use the first line of \eqref{eq:aatasympt} (with $z$ and $\bar z$ interchanged) to see that the logarithm on the right-hand side of \eqref{eq:alongasympt} can only appear from the term involving $d/dn$ acting on the twist eight operators with $n = 4$. Isolating this term, and substituting \eqref{eq:aatasympt}, we find that
\begin{equation}
\lim_{z \to 1}  \sum_\ell P^{\text{MFT}}(4,\ell) \gamma(4,\ell) 
 2^{-\ell} h^{\text{at}}_{\ell + 6}(z) = \frac{ 288}{c(1-z)} + O(\log(1-z))~.
\end{equation}
From \eqref{eq:hasympt} we see that we can reproduce the $(1-z)^{-1}$ if we pick
\begin{equation}
P^{\text{MFT}}(4,\ell) \gamma(4,\ell)\underset{\ell \to \infty}{=} - 9 \cdot 2^{\ell + 2} \hat b_{\ell + 2} + \ldots~,
\end{equation}
where $\hat b_{\ell+2}$ is just the term proportional to $1/c$ in $b_{\ell +2}$. Using the coefficient \eqref{eq:twist8coeff}
\begin{equation}
P^{\text{MFT}}(4,\ell) \underset{\ell \to \infty}{=} \frac{3 \cdot 2^\ell}{5}b_{\ell + 2} + \ldots~,
\end{equation}
we find that
\begin{equation}
\gamma(4,\ell) \underset{\ell \to \infty}{=} - 60 \frac{\hat b_{\ell + 2}}{b_{\ell + 2}} = -\frac{17280}{c \ell^4} + O\left(\ell^{-3}\right)~,
\end{equation}
similarly we have computed the twist ten and twist twelve anomalous dimensions, and they agree precisely with the large $\ell$ limit of the conformal block decomposition of the supergravity solution of \cite{Heslop:2004du}.